\documentclass[11pt]{article}
\usepackage[margin=1in]{geometry}
\usepackage{graphicx}
\usepackage{placeins}
\usepackage{longtable}
\graphicspath{ {./images/} }
\usepackage{hyperref}
\usepackage{theapa}
\usepackage{longtable}
\usepackage{fourier}
\usepackage{array}
\usepackage[font={small}]{caption}
\makeatletter
\renewcommand{\maketitle}{\bgroup\setlength{\parindent}{0pt}
\begin{flushleft}
\fontsize{19}{24}\selectfont
  \textbf{\@title}
  
\end{flushleft}
\begin{flushleft}

  \@author
\end{flushleft}\egroup
}
\makeatother

\begin{document}

\title{Ethics and Governance of Artificial Intelligence: Evidence from a Survey of Machine Learning Researchers}

\author{%
{\bfseries
Baobao Zhang$^{1,*}$, Markus Anderljung$^{2,*}$, Lauren Kahn$^{3}$, Noemi Dreksler$^{2}$, Michael C. Horowitz$^{3}$ and Allan Dafoe$^{2}$} \\ \medskip \small
$^{1}$Department of Government, Cornell University\\ $^{2}$Centre for the Governance of AI, Future of Humanity Institute, University of Oxford\\
$^{3}$Perry World House, University of Pennsylvania\\
    $^{*}$ Corresponding authors, \href{mailto:baobaozhangresearch@gmail.com}{baobaozhangresearch@gmail.com}, \href{mailto:markus.anderljung@governance.ai}{markus.anderljung@governance.ai}
}


\date{}

\maketitle

\begin{abstract}
\noindent Machine learning (ML) and artificial intelligence (AI) researchers play an important role in the ethics and governance of AI, including taking action against what they perceive to be unethical uses of AI \cite{BelfieldActivism,van2020ethical}. Nevertheless, this influential group's attitudes are not well understood, which undermines our ability to discern consensuses or disagreements between AI/ML researchers. To examine these researchers' views, we conducted a survey of those who published in the top AI/ML conferences ($N = 524$). We compare these results with those from a 2016 survey of AI/ML researchers \cite{GraceEtAl} and a 2018 survey of the US public \cite{zhang2020us}. We find that AI/ML researchers place high levels of trust in international organizations and scientific organizations to shape the development and use of AI in the public interest; moderate trust in most Western tech companies; and low trust in national militaries, Chinese tech companies, and Facebook. While the respondents were overwhelmingly opposed to AI/ML researchers working on lethal autonomous weapons, they are less opposed to researchers working on other military applications of AI, particularly logistics algorithms. A strong majority of respondents think that AI safety research should be prioritized and that ML institutions should conduct pre-publication review to assess potential harms. Being closer to the technology itself, AI/ML researchers are well placed to highlight new risks and develop technical solutions, so this novel attempt to measure their attitudes has broad relevance. The findings should help to improve how researchers, private sector executives, and policymakers think about regulations, governance frameworks, guiding principles, and national and international governance strategies for AI.
\end{abstract}

\section{Introduction}
Tech companies and governments alike see the potential for AI and have moved to develop machine learning, particularly deep learning, applications across a variety of sectors --- from healthcare to national security \cite{T-Minus,horowitzTNSR}. Civil society groups, governments, and academic researchers have expressed concerns about AI related to safety \cite{amodei2016problemssafety,russell2019human}, discrimination and racial bias \cite{AlgorithmsOpression,barocas-hardt-narayanan}, and risks associated with uses of AI in a military and government context \cite{malicious2018,SpeedKills,StructuralRisks}.

``Narrow'' AI applications (such as self-driving cars, lethal autonomous weapons systems, and surveillance systems) have become an immediate cause for concern for AI/ML researchers, policymakers, and the public \cite{TrustLit}. Over the past two years, corporations, governments, civil society groups, and multi-stakeholder organizations have published dozens of high-level AI ethics principles \cite{fjeld2020principled}. Some early attempts at international governance include the OECD AI Principles adopted in May 2019 and the G20 Human-centered AI Principles adopted in June 2019 \cite{OECDPrinciples,g20report}.

Technical researchers are crucial in the formation of AI governance. Being close to the technology, AI/ML researchers are well placed to highlight new risks, develop technical solutions, and choose to work for organizations that align with their values.  Just as epistemic communities have developed norms to manage technologies that emerged in the 20th century, such as nuclear weapons and chlorofluorocarbons \cite{adler1992emergence,haas1992introduction,haas1992banning}, we expect AI/ML researchers to play a key role in AI governance.  For example, the Institute for Electrical and Electronics Engineers (IEEE) established the Global Initiative on Ethics of Autonomous and Intelligent Systems in 2016. Leading tech companies such as IBM, Google, and Microsoft have published frameworks and principles intended to guide how they deploy AI systems, and in several cases, have established research positions and units focused on AI ethics \cite{RUSIgovernance}. Individuals working within the AI/ML community have also begun to take an active role in directly shaping the societal and ethical implications of AI, by engaging with employers and governments \cite{BelfieldActivism}. For example, in summer 2018, over 3,000 Google employees signed a petition protesting Google’s involvement with Project Maven, a computer vision project run by the US Department of Defense \cite{ProjectMaven}. 

Through a survey of leading AI/ML researchers, we explore technical experts' attitudes about the governance of AI. We surveyed 524 AI/ML researchers in September and October 2019 who had a paper accepted at one of two leading AI research conferences: the Conference on Neural Information Processing Systems (NeurIPS) and the International Conference on Machine Learning (ICML). Our survey includes direct measures of trust, including attitudes about private and public sector actors. We then compare those results to a 2018 survey of AI attitudes among the US general public. This allows us to analyze attitudes on the current state of global AI governance: who are the most trusted actors to manage the technology, what AI governance challenges are perceived to be the most important, and which norms have already begun to shape AI development and deployment. 

There is a small but growing literature that surveys the AI/ML community. Most existing surveys focus on eliciting researcher forecasts on AI progress, such as when specific milestones will be reached or when AI will surpass human performance at nearly all tasks \cite{SandbergBostrom2011,BaumGoertzl2,MullerBostrom,GraceEtAl,walsh2018expert,gruetzemacher2019forecasting}. Others have focused on how computer scientists define AI \cite{krafft2020defining} or the impact of AI on society \cite{anderson2018artificial}. AI/ML professionals have also been surveyed in regard to their views on working on military-related projects \cite{CSETnewAIProfessionals} and their immigration pathways and intentions \cite{CSETimmigrationpathways}. A number of existing studies examine public opinion toward AI. Past survey research related to AI tends to focus on specific governance challenges, such as lethal autonomous weapons \cite{horowitz2016public}, algorithmic fairness \cite{saxena2019fairness}, or facial recognition technology \cite{smith2019facial,facevalue2019}. A few large-scale surveys have taken a more comprehensive approach by asking about a range of AI governance challenges \cite{eurobarometer460,smith2017,smith2018public,west2018,cave2019scary,eurobarometer2019,zhang2020us,ECAIwhitepaper,RSA2018}. While previous work has compared the public's and AI/ML researchers' forecast of AI development timelines \cite{walsh2018expert,zhang2019artificial}, little work compares the attitudes of AI/ML researchers and the public toward AI governance. 

Key results from our survey include:
\begin{itemize}
    \item Relative to the American public, AI/ML researchers place high levels of trust in international organizations (e.g., the UN, EU, etc.) to shape the development and use of AI in the public interest. While the American public rated the US military as one of the most trustworthy actors, AI/ML researchers place relatively low levels of trust in the militaries of countries where they do research.
    \item The majority of AI/ML researchers (68\%) indicate that AI safety, broadly defined, should be prioritized more than it is at present.
    \item While most researchers support openly sharing all aspects of research, a majority of AI/ML researchers (59\%) also support ``pre-publication review'' for ``work that has some chance of adverse impact.'' 
    Furthermore, there is considerable variation among the aspects of research that they feel must be shared: 84\% think that high-level description of the methods must be shared every time while only 22\% think that of the trained model.
    \item The respondents are wary of AI/ML researchers working on certain military applications of AI. Respondents are the most opposed to other researchers working on lethal autonomous weapons (58\% strongly oppose) but far fewer are opposed to others working on logistics algorithms (6\% strongly oppose) for the military. 31\% of researchers indicate that they would resign or threaten to resign from their jobs, and 25\% indicate that they would speak out publicly to the media or online, if their organization decided to work on lethal autonomous weapons.
\end{itemize}

\section{Methods}

To study attitudes about trust and governance in AI, we conducted a survey of AI/ML researchers between September 16 and October 13, 2019. The researchers were selected based on having papers accepted at two top AI research conferences, following the sampling frame of \cite{GraceEtAl}. One group of respondents had papers accepted to the 2018 NeurIPS conference and the other to the 2018 ICML conference. Another group had papers accepted at NeurIPS and ICML in 2015 and participated in a 2016 researcher survey on AI \cite{GraceEtAl}. Out of the 3,030 researchers who were contacted via email to complete our survey, 524 researchers (17\%) completed at least some part of the survey. To incentivize participation, we offered one in every ten respondents (via lottery) a gift card of \$250 USD.  The survey took a median 17.4 minutes to complete.

This paper presents the results from the component of the survey focused on AI governance. Other parts of the survey asked the respondents to forecast developments in AI research and about their immigration preferences. The full text of the survey questions reported in this paper can be found in the Supplementary Materials. We also collected relevant demographic data about the respondents (e.g., country of their undergraduate degree, workplace type, citation count, etc.) using publicly available information. For some questions, we compare responses from this survey with those from the US public. This public opinion data come from a representative national survey of 2,000 US adults conducted in 2018, in which similar questions were asked \cite{zhang2020us}.\footnote{For the public opinion results, we weighted the responses to be representative of the US adult population using weights provided to us by the survey firm YouGov that conducted the survey on our behalf.}

Our analysis is pre-registered using the Open Science Framework.\footnote{The project URL is \url{https://osf.io/fqz82/}.} Unless specified, we use multiple linear regression to analyze the associations between variables. For estimates of summary statistics or coefficients, ``don’t know'' or missing responses were re-coded to the weighted overall mean, unconditional on treatment conditions. Almost all questions had a ``don’t know'' option. If more than 10\% of the variable’s values were ``don’t know'' or missing, we included a (standardized) dummy variable for ``don’t know''/missing in the analysis. For survey experiment questions, we compared ``don’t know''/missing rates across experimental conditions. Our decision was informed by the Standard Operating Procedures for Don Green's Lab at Columbia University \cite{lin2016standard}. 

Heteroscedasticity-consistent standard errors were used to generate the margins of error at the 95\% confidence level. We report cluster-robust standard errors whenever there is clustering by respondent. In figures, each error bar shows the 95\% confidence intervals.

\section{Results}

\subsection{Evaluation of AI governance challenges}

To gauge their views on AI governance challenges, we asked our respondents: ``In the next 10 years, how important is it for tech companies and governments to carefully manage the following issues?'' Respondents were presented with a list of 5 randomly-selected items out of a total list of 13, that they then assigned a number value on a 4-point slider scale that allowed value input to the tenth decimal place. The scale ranged from 0 ``not at all important'' to 3 “very important.''\footnote{All the multiple-choice questions include an  ``I don't know'' answer option.}

\begin{figure}[htb]
    \centering
    \includegraphics[width=0.80\textwidth]{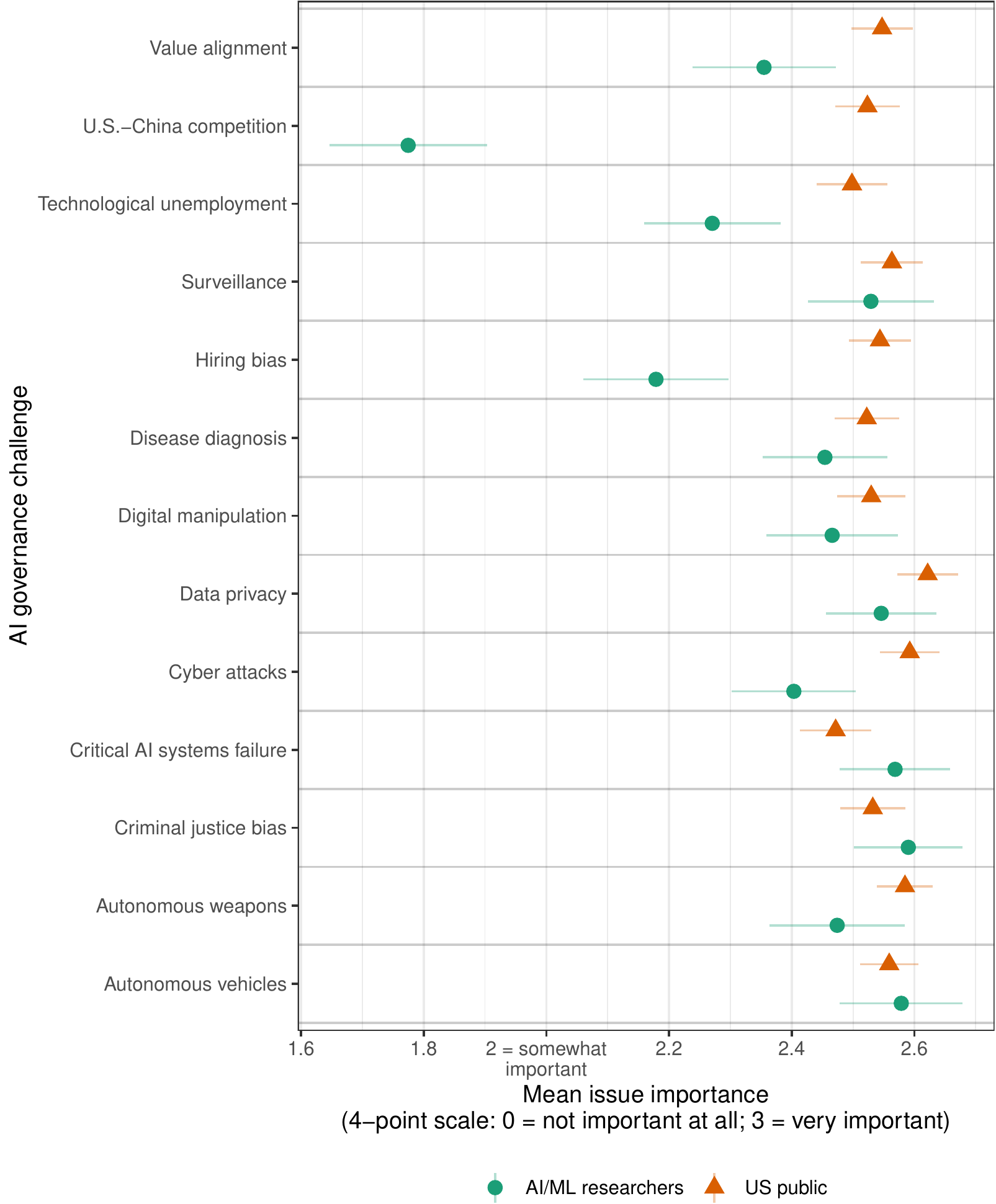}
    \label{fig1}
    \caption{Perceived issue importance of AI governance challenges (comparing AI/ML researchers' and the US public's responses). Each respondent was presented with 5 AI governance challenges randomly selected from a list of 13. Respondents were asked to evaluate the importance of each governance challenge using a four-point scale (the slider scale allows respondents to input values to the tenth decimal point): 0 = not important, 1 = not too important, 2 = somewhat important, 3 = very important. We present the mean responses for each governance challenge (by respondent type) along with the corresponding 95\% confidence intervals.}
\end{figure}

Figure 1 shows the mean importance of AI governance challenges, along with the corresponding 95\% confidence interval for both AI/ML researchers and the general public \cite{zhang2020us}. For the AI/ML researcher group, almost all issues were rated as having a mean importance of 2.5, between ``somewhat important'' and ``very important,'' with the top five issues including preventing criminal justice bias, ensuring autonomous vehicles are safe, preventing critical AI system failure, protecting data privacy, and preventing mass surveillance. Hiring bias and technological unemployment are rated slightly (about 0.3 points) lower than other issues. The one outlier is ``Reducing risks from US-China competition over AI,'' is rated significant below the other challenges at 1.8 (just below ``somewhat important''); this result may be an artifact of our question phrasing, in that AI/ML researchers may believe that risks from US-China competition are real, but not one that is helped by ``tech companies and governments'' trying to ``carefully manage'' them (see the Supplementary Materials for the text of the survey questions). As Table~S9 shows, AI/ML researchers who identified as male, compared with those who identify as female or other, tend to have lower issue importance scores across the board. 

There is considerable overlap between the assessment of AI governance challenges by AI/ML researchers and the US public (for details, see Table~S3). Both groups rate protecting data privacy, preventing mass surveillance, and ensuring that autonomous vehicles are safe as among the five most important governance challenges. AI/ML researchers placed significantly less importance on value alignment\footnote{Defined as ``AI systems are safe, trustworthy, and aligned with human values''}, technological unemployment, and hiring bias, and slightly more importance on critical AI systems failure, criminal justice bias, and autonomous vehicles, than the public. 

The gap between AI/ML researchers and the US public is particularly large when it comes to preventing the risks from US-China competition in AI. In contrast to AI/ML researchers' relatively low mean rating of 1.77 out of 3, the US public gave US-China competition a mean rating of 2.52 out of 3. One might think that breaking down the AI/ML researchers' responses by demographic subgroups (see Figure~S2 - S4), would reveal some potential explanations for the response pattern. However, the results are mixed. Respondents who attended undergraduate in China rate this issue relatively high (mean score of 2.26); in contrast, respondents who attended undergraduate in Europe give a mean score of only 1.59. While respondents who attended undergraduate in the US give a mean score of 1.90, the difference is not statistically significantly different to respondents with undergraduate degrees from China.

\FloatBarrier

\subsection{Trust in actors to shape the development and use of AI in the public interest}

Good governance requires understanding of what institutions and organizations AI/ML researchers (and other stakeholders) trust. To test AI/ML researcher trust in different governance options, we ask: ``Suppose the following organizations were in a position to strongly shape the development and use of advanced AI. How much trust do you have in each of these organizations to do so in the best interests of the public?'' Similar to the structure of the previous question, respondents were shown 5 randomly-selected actors. For each actor, they then assigned a number value on a 4-point scale ranging from 0 ``no trust at all'' to 3 ``a great deal of trust.''\footnote{The US military and the Chinese military were shown only to respondents who reported the US or China as the countries where they spend the most time doing research. These respondents had equal probability of being shown the US military or the Chinese military. Because very few responses came from respondents who do research in China, we dropped their responses in this figure. We break down responses to these two actors by the country where the respondents completed their undergraduate degree (US and China) in Figure~S6.}

Figure 2 shows the mean trust value for the actors, along with the corresponding 95\% confidence interval for both AI/ML researchers and the public. For AI/ML researchers, the most trusted actors, with a score above 2.0, were non-governmental scientific associations and intergovernmental research organizations. The Partnership on AI, a consortium of tech companies, academics, and civil society groups, is also rated relatively highly (mean score of 1.89). Out of the international institutions listed, the European Union (EU) is perceived to be the most trusted, with a mean score of 1.98, while the United Nations' (UN) mean trust rating is lower at 1.74 (two-sided $p$=0.010).\footnote{For comparing trust between actors, we use $F$-tests to test the equality of coefficients from the regression model presented in Table~S16.} It is noteworthy that these more neutral, scientific organizations received the highest trust ratings but currently play a relatively small role in AI development and management.

\begin{figure}[htb]
    \centering
    \includegraphics[width=0.75\textwidth]{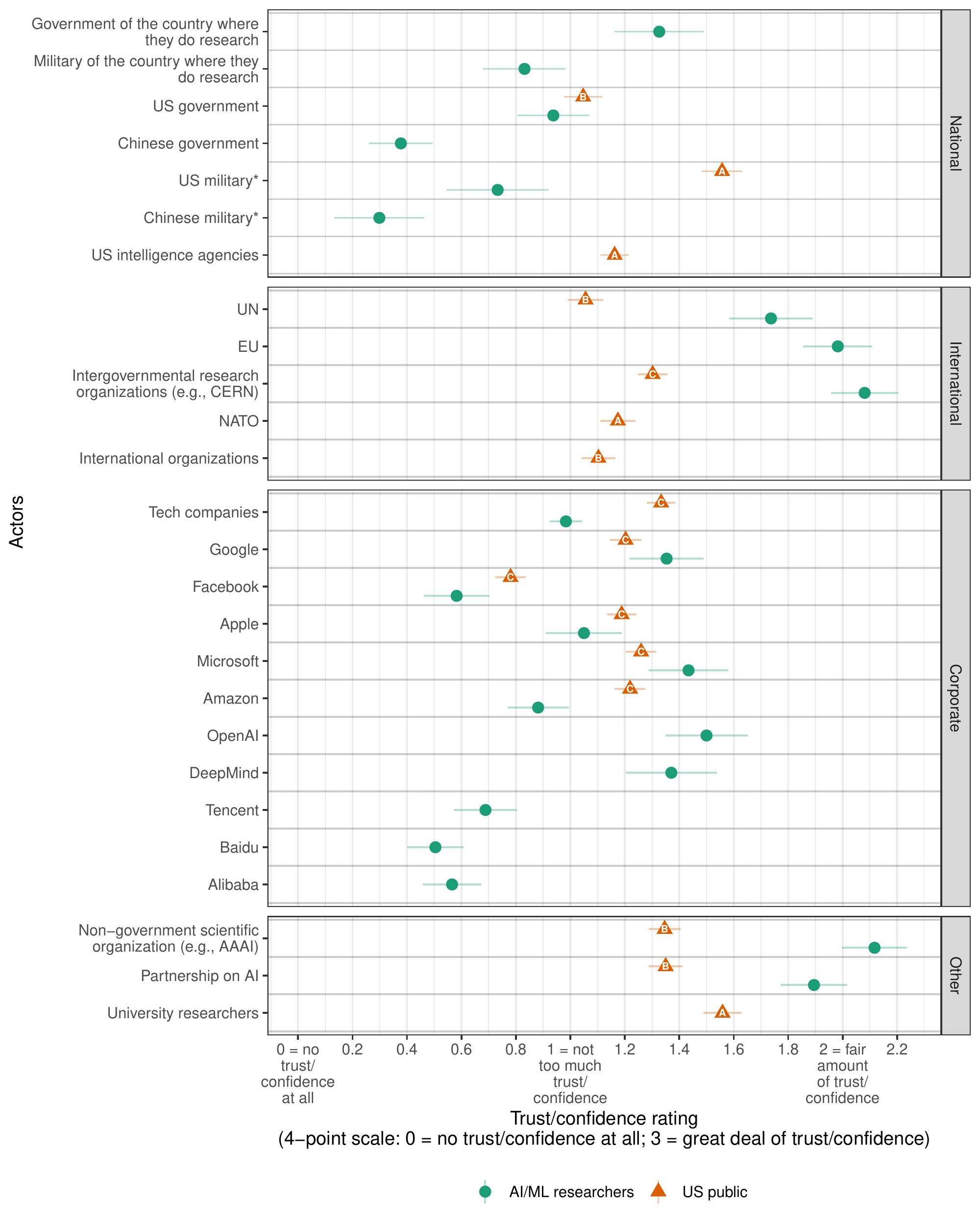}
    \label{fig2}
    \caption{Trust in actors to shape the development and use of AI in the public interest: comparing AI/ML researchers' and the US public's responses. AI/ML researchers were shown five randomly-selected actors and asked to evaluate how much they trust the actors using a four-point scale: 0 = no trust at all, 1 = not too much trust, 2 = a fair amount of trust, 3 = a great deal of trust. For the AI/ML researchers survey, the ``Tech companies'' result is the mean response across all corporate actors presented to respondents. *: The US military and the Chinese military responses are only for those who do research in the US. See Fig.~S5 to~S9 for further breakdown on this question by respondents' background. In the public opinion survey, respondents were asked about their confidence in the actors to develop AI (labeled ``A''), or to manage the development and use of AI (labeled ``B''), in the best interest of the public, using a similar four-point scale. For actors labeled ``C,'' both questions were asked; we averaged the responses to these two questions for each of these actors for clarity. For ``US intelligence agencies,'' we averaged across responses to the NSA, the FBI, and the CIA (which were similar). The circle and triangle shapes are the point estimates (mean responses) and the whiskers are the corresponding 95\% confidence intervals. The data for this figure, alongside more detailed breakdowns of the results, can be found in Tables~S10 - S15.}
    \end{figure}

Out of all the private tech companies listed, OpenAI\footnote{OpenAI announced in March 2019 that it would move from being a non-profit to being a ``capped-profit'' company, a for-profit and non-profit hybrid \cite{OpenAIannouncement}. The survey of the public was conducted before this change, whilst the survey of AI/ML researchers occurred afterwards.}, DeepMind, Google, and Microsoft are relatively more trusted. Facebook is ranked the least trustworthy of American tech companies, and the Chinese companies rated significantly less trustworthy than all listed US tech companies apart from Facebook. State actors, such as the US and Chinese governments or the militaries of the countries where the respondents do research, received relatively low trust scores from AI/ML researchers. In general, respondents trust the government of the country where they do research more than the military of that country (two-sided $p<$0.001).
As Figure~S6 shows, respondents who attended undergraduate in the US, compared with those who attended undergraduate in China, are significantly less trustful of the Chinese government and military, as well as the three Chinese tech companies presented to respondents (Tencent, Baidu, and Alibaba). The interaction plot in Figure~S10 shows that those who attended undergraduate in China trust both Chinese tech companies and Western tech companies more than those who attended undergraduate in the US. The difference in trust in Western versus Chinese tech companies is smaller for those who attended undergraduate in China than those who attended undergraduate in the US (two-sided $p<$0.001), as can be seen in Figure~S10.

AI/ML researchers, like the US public, as Figure 2 shows, distrust Facebook more than any other US tech company. A major difference between AI/ML researchers and the US public is their assessment of the military. Whereas surveyed AI/ML researchers, on average, do not have too much trust in the military of the countries where they do research, the US military is among the most trusted of institutions for the US public. Respondents who do research in the US gave the US military a mean rating of 0.73 (below ``1 - not too much trust''), whereas the US public gave their country's military a mean rating of 1.56 (see Tables~S10 - S11 and S15).

\FloatBarrier

In contrast, the US public, compared with AI/ML researchers, places much less trust in international institutions such as the UN. AI/ML researchers gave the UN a mean rating of 1.74 while the US public gave a mean rating of 1.06.

\begin{figure}[htb] 
    \centering
    \includegraphics[width=0.9\textwidth]{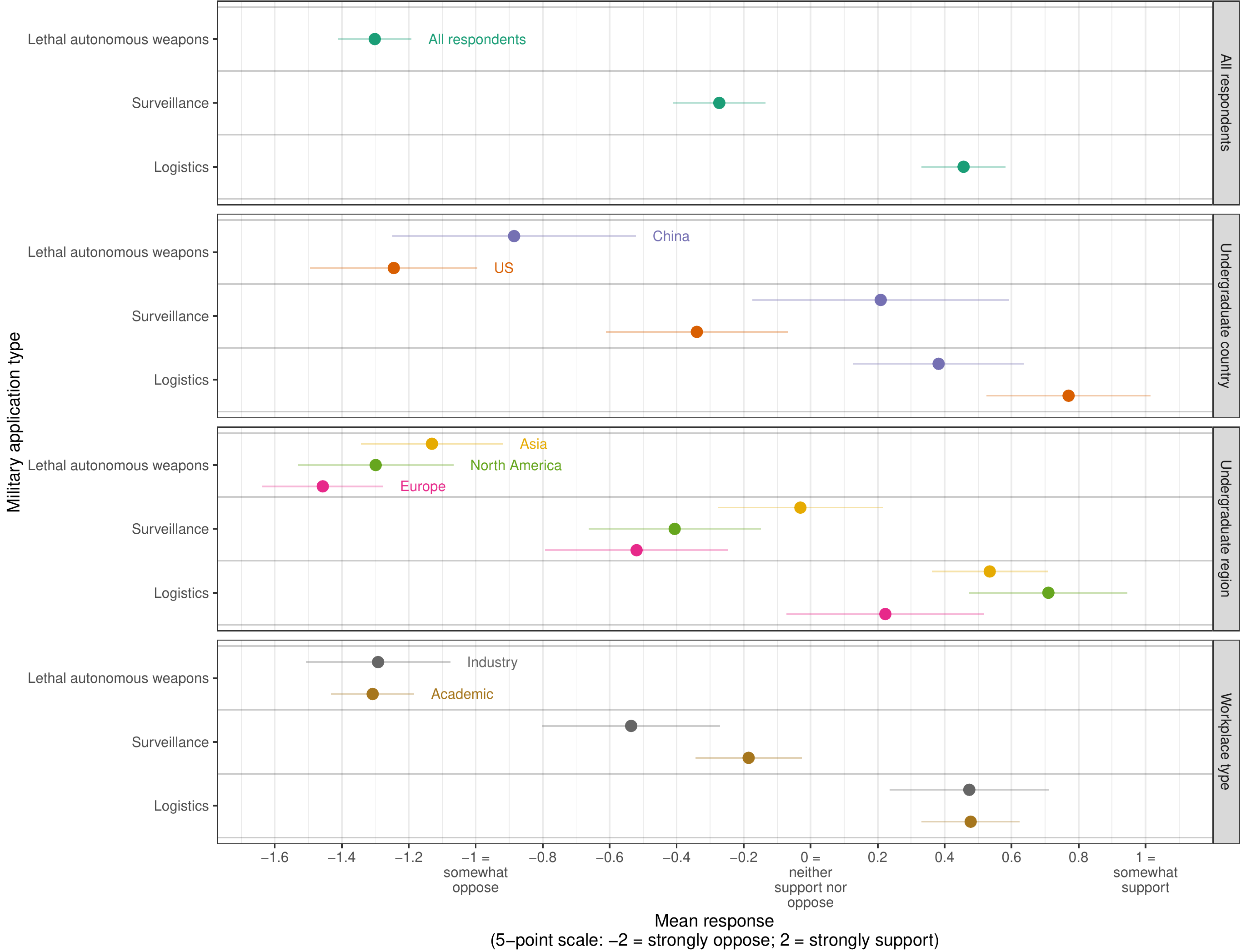}
    \label{fig4}
    \caption{Attitudes toward researchers working on military applications of AI. Respondents were presented with two applications randomly selected from the three and indicated how much they support or oppose other researchers working on those applications using a five-point scale: -2 = strongly oppose, -1 = somewhat oppose, 0 = neither support nor oppose, 1 = somewhat support, 2 = strongly support. We present the overall means and the demographic subgroup means with their corresponding 95\% confidence intervals. For the subgroup analysis, we broke down the responses by the respondents' undergraduate country, undergraduate region, and workplace type.}
\end{figure}

\FloatBarrier

\subsection{AI safety}

The safety of AI systems may be a critical factor in their development and adoption. We asked respondents about their familiarity with and prioritization of AI safety. We described AI safety in a broad way, as focused on ``making AI systems more robust, more trustworthy, and better at behaving in accordance with the operator’s intentions,'' and also provided examples (see Supplementary Materials). We first sought to understand how familiar  researchers were with AI safety research. We asked them to make a self-assessment using a five-point scale, ranging from 0 ``not familiar at all (first time hearing of the concept)'' to 4 ``very familiar (worked on the topic).'' To evaluate views about the value of AI safety research, we asked respondents, “How much should AI safety be prioritized relative to today?” Respondents selected answers on a 5-point Likert scale, ranging from -2 ``much less'' to 2 ``much more” with 0 meaning ``about the same.'' 

The AI/ML researchers we surveyed report, on average, moderate familiarity with AI safety as a concept (see Figure~S11). The distribution follows an approximately normal distribution, although it is right-skewed. 
 
3\% of respondents say that they are ``not familiar at all'' with AI safety while 15\% say that they are “very familiar.'' 

When asked about prioritizing AI safety, as Figure~S13 shows, an overwhelming majority of our respondents (68\%) say that the field should be prioritized more than at present. These results demonstrate significant growth in the reported prioritization of AI safety in the research community, though these differences may be driven by different definitions. In a similar survey of AI/ML researchers conducted in 2016, 47\% of respondents believed that AI safety should be prioritized more than it was at the time \cite{GraceEtAl}.\footnote{We updated the definition of AI safety research from \cite{GraceEtAl} after consultation with AI/ML researchers working in AI safety research. 

Contrasting with our definition (see Supplementary Materials), the 2016 definition of AI safety was as ``any AI-related research that, rather than being primarily aimed at improving the capabilities of AI systems, is instead primarily aimed at minimizing potential risks of AI systems (beyond what is already accomplished for those goals by increasing AI system capabilities).'' The examples provided in 2016 included: improving the human-interpretability of machine learning algorithms for the purpose of improving the safety and robustness of AI systems, not focused on improving AI capabilities; research on long-term existential risks from AI systems; AI-specific formal verification research; and policy research about how to maximize the public benefits of AI.}

\subsection{Publication norms}

\begin{figure}[htb] 
    \centering
    \includegraphics[width=0.9\textwidth]{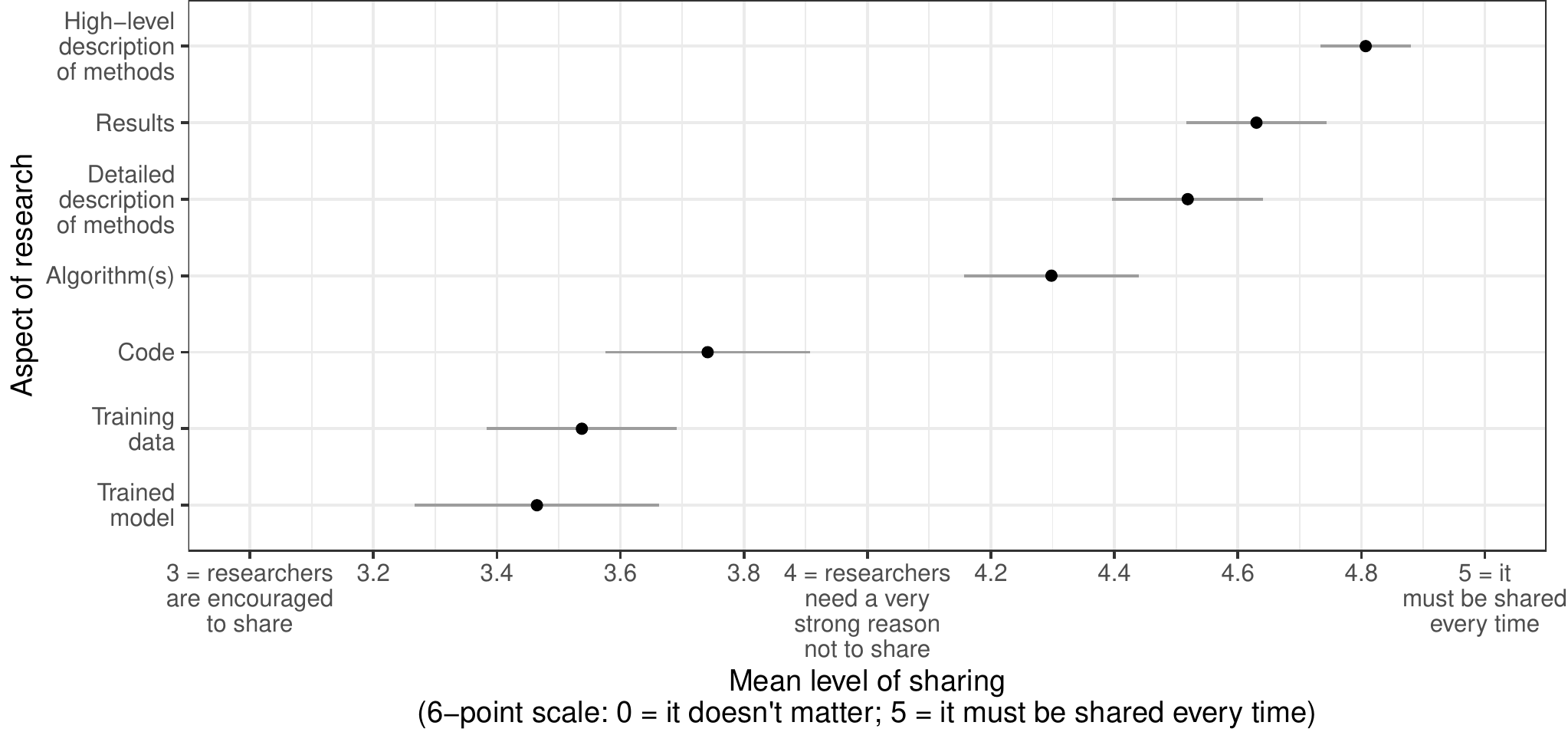}
    \label{fig3}
    \caption{Respondents' perceptions of how openly various aspects of research should be shared. Respondents were presented with three aspects of research, randomly selected from seven. They were asked how openly these aspects of research should be shared using a six-point scale: 0 = it doesn't matter, 1 = it's completely up to the researchers to share or not to share, 2 = it's preferred that researchers share but it's not paramount that they do, 3 = researchers are encouraged to share, 4 = researchers need a very strong reason not to share, 5 = it must be shared every time. We show the results (mean response and 95\% confidence intervals) for all respondents.}
\end{figure}

The AI/ML research community has recently seen innovation and subsequent controversy regarding publication norms, which also relate to questions of trust. Such norms concern when, how, and where research is published. OpenAI's release strategy for GPT-2, a text generation system, is a prime example. Citing concerns the system could be used for ``malicious purposes'', they employed a staged release strategy; the initial paper was accompanied by a smaller version of GPT-2, the full model only being released 8 months later \cite{GPT-2}. NeurIPS introduced further innovation, for the first time requiring researchers submit impact statements along with their papers to the 2020 conference \cite{NeurIPSImpact,NeurIPSImpactGuide}. The conference also employed a form of pre-publication review, rejecting four papers on ethical grounds after review from ethics advisors.  

We asked questions to generate insights into AI/ML researchers views on publication norms. First, we assessed how much they agree or disagree that ``machine learning research institutions (including firms, governments, and universities) should practice pre-publication review,'' which involves ``a strong norm or policy'' to have discussions which are ``informed, substantive, and serious'' about ``the ethical implications of publication'' (see Supplementary Materials). A majority of respondents agree (20\% strongly agree; 39\% somewhat agree) with the statement (see Table~S28). Additionally, as shown in Table~S30, both familiarity with AI safety and prioritization of AI safety significantly predict support for pre-publication review. These results speak to an interest amongst AI/ML researchers to take steps to address the risks of misuse of their work. 

Next, we asked respondents about the importance of sharing various aspects of AI/ML research. Respondents were shown three aspects of research, randomly selected from a list of seven (e.g. high-level description of methods, code, and training data). For each aspect of research, respondents could select from six levels of sharing, ranging from ``it doesn't matter'' to ``it must be shared every time.'' As Figure 3 shows, the respondents think that high-level descriptions of the methods, the results, and a detailed description of the methods should almost always be openly shared. However, support declines for requiring the sharing of other information that would be essential for replication, such as the code, the training data, or the trained model. Researchers felt that sharing these aspects of research should be encouraged, not required. On the high end, 84\% indicated that high-level description of methods must be shared every time; on the low end, only 22\% indicated that the trained model must be shared every time (see Figure~S18). We do not find significant differences in responses between researchers who work in academia versus in industry.

\subsection{Attitudes toward military applications of AI}

We also investigated researchers' views toward military applications of AI. Working on military uses of AI requires a great deal of trust in how they might be used, because of the central role that some think AI could play in the future of military power \cite{scharre2018army}. We asked about three areas of military applications of AI that have received public scrutiny: lethal autonomous weapon systems, surveillance technologies for intelligence agencies, and military logistics. Respondents were asked to evaluate two randomly-selected military applications out of the three. They were asked whether they would support or oppose researchers working on the application in the country where the respondent currently works or studies. Respondents selected answers on a Likert scale, ranging from -2 ``strongly oppose'' to 2 ``strongly support.'' Those who answered that they ``strongly oppose'' or ``somewhat oppose'' researchers working on the applications were asked what types of collective actions (e.g., signing a petition, or protesting) they would take if their organization decided to conduct such research. 

\FloatBarrier

Our results show researchers have substantial concerns regarding working on some military applications of AI. Nevertheless, there are nuances to their views. Figure 4 illustrates that researchers, on average, more than somewhat oppose work on lethal autonomous weapon systems (-1.3), very weakly oppose work on surveillance applications (-0.3), and very weakly support work on logistics applications (0.5). Additional detail in Figure~S15 demonstrates that 58\% strongly oppose other researchers working on lethal autonomous weapons, 20\% strongly oppose others working on surveillance tools, but only 6\% strongly oppose others working on military logistics. This is consistent with work by Aiken, Kagan, and Page \cite{CSETnewAIProfessionals}, which focuses just on US-based AI professionals and finds that US-based AI professionals are more opposed to working on battlefield applications of AI than other applications.

How would these AI/ML researcher attitudes translate into potential action? For each application (lethal autonomous weapon systems, surveillance, and logistics), the respondents who said that they strongly or somewhat opposed other researchers working on the application received a follow-up question asking if they would take action if their organization decided to work on the application. Figure~S16 shows the distribution of responses for each application. A majority of researchers who said they opposed others working on each application said that they would actively avoid working on the project, express their concern to a superior in their organization involved in the decision, and sign a petition against the decision. 75\% of researchers who said they opposed others working on lethal autonomous weapons said they would avoid working on lethal autonomous weapons themselves, and 42\% of those respondents said they would resign or threaten to resign from their jobs. In absolute terms, 31\% of researchers indicate that they would resign or threaten to resign from their jobs, and 25\% indicate that they would speak out publicly to the media or online if their organization decided to work on lethal autonomous weapons. Of those who say they oppose other researchers working on lethal autonomous weapons, less than 1\% said they would do nothing. The percentages for surveillance and logistical software are 4\% and 8\%, respectively (for further results see Figure~S16). 

A prominent conflict between the AI/ML community and a national military involved Google engineers protesting their company's participation in Project Maven in the US. In June 2018, some 3,000 Google employees signed a petition, voicing ethical concerns regarding the project \cite{GoogleLetter}. As a result, Google decided not to renew its Project Maven contract with the US Department of Defense.

Given the controversy over Google's participation in Project Maven, we asked respondents if they supported or opposed Google’s decision not to renew its contract, using a five-point Likert scale with -2 meaning ``strongly oppose” to 2 meaning ``strongly support.'' 
Figure~S17 details broad support within our AI/ML research respondents for Google's decision to withdraw from Project Maven. 38\% strongly support and 21\% support Google’s decision to pull out of Project Maven while only 9\% strongly or somewhat oppose the decision. 

The results are broadly consistent across demographic subgroups, as seen in Figure 4. Generally, across subgroups, respondents are the most opposed to working on lethal autonomous weapons and least opposed to working on military logistics.

\section{Conclusion}

It is important to recognize some of the limits to our findings, which future research can address. First, our sample strategy focused on those who publish in the top two AI/ML conferences; it thus may underweight the perspective of those subgroups of the AI/ML community who are less likely to publish there, such as perhaps product focused industry researchers. Second, this survey captures the views of the researchers at a particular point in time, while the norms around AI research and publishing continue to evolve, and significant shifts in the psychological,  political, and socioeconomic landscape continue to occur, for example, as a result of COVID-19. Future work could expand the sampling frame of respondents (e.g., to include more researchers who work in industry, and develop a more representative sample of the AI/ML community) and include panel studies that examine changes in respondents' attitudes over time. 

Another limitation might include demographic biases or response bias. Demographic characteristics of the respondents and non-respondents are found in Table~S1. A multiple regression that examines the association between demographic characteristics and response finds that respondents have lower h-indexes (a measure of productivity and citation impact of researchers) and are more likely to work in academia compared with non-respondents (see Table~S2). Overall, however, we do not see evidence of concerning levels of response bias. Compared with other work of its kind, our survey has more respondents, a higher response rate, and more global coverage than other surveys of AI/ML researchers we reviewed. Separately from response bias, there are other aspects of the population of AI/ML researchers worth keeping in mind, such as gender (91\% of our respondents and 89\% of non-respondents were male).

As institutions, regulations, and norms of AI governance are forming, this survey of AI/ML researchers provides insight into how this emerging epistemic community views the ethical and governance issues related to the technology. The respondents place relatively high levels of trust in international organizations to manage the development and use of AI in the public interest. But compared with the US public who place high levels of trust in the US military, AI/ML researchers are relatively distrustful of the military. Furthermore, the AI/ML researchers we surveyed are very opposed to working on certain military applications of AI, particularly lethal autonomous weapon systems. The respondents are also aware of potential adverse impacts of their research. Finally, a majority of respondents think that AI safety research should be prioritized and researchers should conduct pre-publication reviews to assess the potential harms their research could cause. This line of research could help guide policymakers, tech companies, civil society, and the AI/ML community in building and deploying safe and ethical AI systems.  

\newpage

\section*{Acknowledgements}

\noindent Funding: This research was supported by: the Ethics and Governance of AI Fund, the Open Philanthropy Project grant for ``Oxford University -- Research on the Global Politics of AI,'' and the Minerva Research Initiative under Grant \#FA9550-18-1-0194. The research reported here should solely be attributed to the authors; all errors are the responsibilities of the authors.

\

\noindent Authors contributions: A.D., B.Z., M.A., and M.H. (in alphabetical order) designed the research and provided the conceptual framing of the work. B.Z. and M.A. handled the data acquisition. B.Z. and N.D. analyzed the data. A.D., B.Z., L.K., M.A., M.H., and N.D. wrote the paper.

\

\noindent Competing interests: The authors declare no competing interests.

\

\noindent Data and materials availability: Due to the data privacy promised to respondents and that our IRB applications noted that we would not share individual-level results, we cannot release the data in full. We made this decision because the population we are sampling from is a relatively small group of individuals, which increases the likelihood of respondents being identifiable from individual-level data. Instead, we have opted to report detailed breakdowns of the data by key demographics in the Supplementary Materials.

\newpage
\section*{Supplementary Materials}

    \noindent Supplementary Text: Text of the Survey
    
    \noindent Table S1 - S31
    
    \noindent Fig S1 -  S19

\subsection*{Supplementary Text: Text of the Survey}
\label{appendix:survey-text}

Unless specified, the order of the items or scales presented here is the order presented to respondents in the survey.

\subsubsection*{AI governance challenges}

In the next 10 years, how important is it for tech companies and governments to carefully manage the following issues?\\

[Respondents were shown 5 randomly-selected items.]

\begin{itemize}
    \item Ensure fairness and transparency in AI used in hiring
    \item Ensure fairness and transparency in AI used in criminal justice
    \item Make AI used for medical diagnosis accurate and transparent
    \item Protect data privacy
    \item Ensure that autonomous vehicles are safe
    \item Prevent AI from being used to spread fake and harmful content online
    \item Prevent AI cyber attacks against governments, companies, organizations, and individuals
    \item Prevent AI-assisted surveillance from violating privacy and civil liberties
    \item Reducing risks from US-China competition over AI 
    \item Make sure AI systems are safe, trustworthy, and aligned with human values
    \item Develop treaties to prevent the misuse of lethal autonomous weapons
    \item Guarantee a good standard of living for those who lose their jobs to automation
    \item Prevent critical AI systems failures, such as a multi-day regional power outage or a trillion dollar market crash from automated algorithms
\end{itemize}

Answer choices: Slider that you can choose in between whole numbers (to 1 decimal point), marked 

\begin{itemize}
    \item 3 = Very important
    \item 2 = Somewhat important
    \item 1 = Not too important
    \item 0 = Not at all important
    \item I don't know
\end{itemize}

\subsubsection*{Trust in actors}

Suppose that the following organizations were in a position to strongly shape the development and use of advanced AI. How much trust do you have in each of these organizations to do so in the best interests of the public? 

[Respondents were shown 5 randomly-selected actors.]

Included if the person does not work in the US:
\begin{itemize}
    \item The government of \textless COUNTRY WHERE THEY DO RESEARCH\textgreater \footnote{Earlier in the survey, respondents were asked the following question: ``In which country do you spend the most time doing research?''. Respondents input the country from a drop-down menu. Those who did not input a country were assigned ``the country where you do research'' in questions that piped in the country where the respondent spent most of their time doing research.}
    \item The military of \textless COUNTRY WHERE THEY DO RESEARCH\textgreater 
\end{itemize}

Included if the person works in the US or China:
\begin{itemize}
    \item The US military 
    \item The Chinese military
\end{itemize}

To everyone else:
\begin{itemize}
    \item The US government
    \item The Chinese government 
    \item The United Nations (UN)
    \item The European Union (EU)
    \item An intergovernmental AI research organization (similar to CERN)
    \item Google 
    \item Facebook 
    \item Apple 
    \item Microsoft 
    \item Amazon
    \item OpenAI
    \item DeepMind
    \item Tencent
    \item Baidu
    \item Alibaba
    \item Non-governmental scientific organizations (e.g., AAAI)
    \item Partnership on AI, a consortium of tech companies, academics, and civil society groups
\end{itemize}

Answer choices:
\begin{itemize}
    \item A great deal of trust (3)
    \item A fair amount of trust (2)
    \item Not too much trust (1)
    \item No trust at all (0)
    \item I don’t know
\end{itemize}

\subsubsection*{AI safety}

\subsubsection*{AI safety introduction}
\label{AI safety}

AI safety research focuses on making AI systems more robust, more trustworthy, and better at behaving in accordance with the operator’s intentions.\medskip

Examples of such AI safety research include:
\begin{itemize}
    \item Making AI algorithms interpretable to humans
    \item Making sure that an AI system is robust to distributional shifts or adversarial inputs 
    \item Making sure that an AI system’s behavior aligns with the operator’s true intentions
\end{itemize}

\subsubsection*{Familiarity with AI safety research}

How familiar are you with AI safety research?\medskip

Use the slider to indicate your familiarity.

\begin{itemize}
    \item 0 means not familiar at all (e.g., this is the first time you’re hearing about the concept)
    \item 4 means very familiar (e.g., you have worked on the topic)
\end{itemize}

\subsubsection*{Prioritizing AI safety research} 

How much should AI safety research be prioritized -- by, for instance, the tech industry, the academic field, and governments -- relative to today?

Answer choices:

\begin{itemize}
    \item Much less (-2)	
\item Less (-1)	
\item About the same (0)	
\item More (1)	
\item Much more (2)
\item I don’t know
\end{itemize}

\newpage 

\subsubsection*{Attitudes toward military application of AI}
\subsubsection*{Support for others and themselves researching military technology}

[Respondents where shown 2 out of the 3 applications below; the order that the two questions were shown appear were randomized.] 

\noindent\rule{\textwidth}{1pt}

Do you support or oppose researchers in \textless COUNTRY WHERE THEY DO RESEARCH\textgreater working on the development of \textbf{lethal autonomous weapons} to be used by the military of \textless COUNTRY WHERE THEY DO RESEARCH\textgreater?

Lethal autonomous weapons are systems that, once activated by a human, are capable of targeting and firing on their own.

\noindent\rule{\textwidth}{1pt}

Do you support or oppose researchers in \textless COUNTRY WHERE THEY DO RESEARCH\textgreater  working on the development of \textbf{surveillance technologies} to be used by intelligence agencies of \textless COUNTRY WHERE THEY DO RESEARCH\textgreater ?

Intelligence agencies could use AI to expand their capacity to analyze image, video, sound, and text data.

\noindent\rule{\textwidth}{1pt}

Do you support or oppose researchers in \textless COUNTRY WHERE THEY DO RESEARCH\textgreater  working on the development of \textbf{logistics algorithms} to optimize storage and transportation for the military of \textless COUNTRY WHERE THEY DO RESEARCH\textgreater ?

The military could use machine learning algorithms to improve their logistics, such as the storage, purchasing and transportation of weapons and food.

\noindent\rule{\textwidth}{1pt}

Answer choices:

\begin{itemize}
    \item Strongly support (2)
    \item Somewhat support (1)
    \item Neither support nor oppose (0)
    \item Somewhat oppose (-1)
    \item Strongly oppose (-2)
    \item I don’t know
\end{itemize}

[For each of the questions above, if they selected ``somewhat oppose'' or ``strongly oppose'' above, the respondents were shown the respective question below.]

\noindent\rule{\textwidth}{1pt}

Suppose your organization has decided to research \textbf{lethal autonomous weapons} to be used by the military of \textless COUNTRY WHERE THEY DO RESEARCH\textgreater . Which, if any, of the following actions would you take?

\noindent\rule{\textwidth}{1pt}

Suppose your organization has decided to research \textbf{surveillance technologies} to be used by intelligence agencies of \textless COUNTRY WHERE THEY DO RESEARCH\textgreater . Which, if any, of the following actions would you take?

\noindent\rule{\textwidth}{1pt}

Suppose your organization has decided to research \textbf{logistics algorithms} to optimize storage and transportation for the military of \textless COUNTRY WHERE THEY DO RESEARCH\textgreater . Which, if any, of the following actions would you take?

\noindent\rule{\textwidth}{1pt}

\begin{itemize}
    \item Nothing
    \item Actively avoid working on the project
    \item Expressing your concern to a superior in your organization involved in the decision
    \item Sign a petition against the decision
    \item Participate in a public protest
    \item Speak out against the decision anonymously to the media or online
    \item Speak out against the decision publicly to the media or online
    \item Resign or threaten to resign from your job
    \item Other: [short textbox]
\end{itemize}

\subsubsection*{Project Maven}

Google had a contract to work on Project Maven, a US Department of Defense project that develops and integrates computer vision algorithms to support military operations. Some Google employees voiced ethical concerns regarding the project. Google decided not to renew its Project Maven contract with the US Department of Defense.
\smallskip

Do you support or oppose this decision by Google not to renew its contract? 
\smallskip

Answer choices:
\begin{itemize}
    \item Strongly support (2)
    \item Somewhat support (1)
    \item Neither support nor oppose (0)
    \item Somewhat oppose (-1)
    \item Strongly oppose (-2)
    \item I don’t know
\end{itemize}

[Optional question]: Would you like to elaborate on the reasoning behind your previous answer? 
[Text box]

\subsubsection*{Publication norms}

\subsubsection*{Pre-publication review}
\label{prepubtext}
Define ``\textbf{pre-publication review}'' as follows: For work that has some chance of adverse impacts, having a strong norm or policy to have discussions about the ethical implications of publication that are 

\begin{itemize}
\item Informed: the discussion includes the lead and senior authors
\item Substantive: the discussion lasts for at least an hour
\item Serious: the discussion can lead to real-world decisions (e.g., not to publish parts of the research in question)
\end{itemize}

Taking into account the cost (e.g., in terms of researcher time) to what extent do you agree or disagree with the following statement?
\medskip

Machine learning research institutions (including firms, governments, and universities) should practice pre-publication review.

\begin{itemize}
    \item Strongly agree (2)
    \item Somewhat agree (1)
    \item Somewhat disagree (-1)
    \item Strongly disagree (-2)
    \item I don't know
\end{itemize}

\subsubsection*{Sharing various aspects of research}

What is your view toward publicly sharing the following aspects of research, such as at conferences, in academic journals, or online?
\medskip

[Respondents were shown 3 aspects of research.]

\begin{itemize}
    \item High-level description of methods
    \item Detailed description of methods
    \item Results
    \item Code
    \item Training data
    \item Trained model
    \item Algorithm(s)
\end{itemize}

\noindent Answer choices:
\begin{itemize}
    \item It must be shared every time (5)
    \item Researchers need a very strong reason not to share (4)
    \item Researchers are encouraged to share (3)
	\item It’s preferred that researchers share but it’s not paramount that they do (2)
	\item It’s completely up to the researchers to share or not to share (1)
	\item It doesn't matter (0)
\end{itemize}

\clearpage

\subsection*{Demographics of survey respondents}

\begin{table}[htb]
\small
        \caption{Summary statistics of the non-respondents and respondents: binary differences. We collected demographic information for all our respondents and a random sample of 446 non-respondents using information publicly available online. The table presents the proportion of individuals in each demographic category for gender, region of undergraduate and PhD, region where the respondent works, and  the type of workplace for both non-respondents and respondents. The mean undergraduate graduation year and log citations are also shown. For each the difference between the non-respondents' and respondents' proportions is presented alongside the corresponding standard error. The Holm method was used to control the family-wise error rate.}\label{tab:summarystat-sample}

\centering
\renewcommand{\arraystretch}{1.2}
\begin{tabular}{|p{5.8cm}|r|r|l|r|}
\hline
Variable & Nonrespondent & Respondent & Difference (\textit{SE}) & Percent missing\\
\hline
Prop. male & 0.89 & 0.91 & 0.01 (0.02) & 0.01\\
\hline
Mean undergrad graduation year & 2007.62 & 2008.95 & 1.33 (0.47) & 0.21\\
\hline
Prop. undergrad region: North America & 0.25 & 0.27 & 0.02 (0.03) & 0.15\\
\hline
Prop. undergrad region: Europe & 0.26 & 0.29 & 0.02 (0.03) & 0.15\\
\hline
Prop. undergrad region: Asia & 0.43 & 0.39 & -0.04 (0.03) & 0.15\\
\hline
Prop. undergrad region: Other & 0.04 & 0.05 & 0.01 (0.01) & 0.15\\
\hline
Prop. PhD region: North America & 0.28 & 0.33 & 0.06 (0.03) & 0.08\\
\hline
Prop. PhD region: Europe & 0.59 & 0.53 & -0.07 (0.03) & 0.08\\
\hline
Prop. PhD region: Asia & 0.11 & 0.09 & -0.01 (0.02) & 0.08\\
\hline
Prop. PhD region: Other & 0.02 & 0.02 & 0.01 (0.01) & 0.08\\
\hline
Prop. currently enrolled in PhD & 0.20 & 0.33 & 0.12 (0.03)*** & 0.05\\
\hline
Mean log citations (all) & 6.75 & 6.26 & -0.49 (0.12)*** & 0.17\\
\hline
Mean h-index (all) & 19.68 & 14.42 & -5.26 (1.12)*** & 0.17\\
\hline
Prop. work region: Europe & 0.28 & 0.33 & 0.05 (0.03) & 0.01\\
\hline
Prop. work region: North America & 0.59 & 0.54 & -0.05 (0.03) & 0.01\\
\hline
Prop. work region: Asia & 0.12 & 0.12 & <0.01 (0.02) & 0.01\\
\hline
Prop. work region: Other & 0.01 & 0.02 & 0.01 (0.01) & 0.01\\
\hline
Prop. work in academia & 0.68 & 0.80 & 0.12 (0.03)*** & 0.00\\
\hline
Prop. work in industry & 0.36 & 0.35 & -0.01 (0.03) & 0.00\\
\hline
\end{tabular}
\end{table}

\clearpage

\begin{table}
\small
  \caption{Association between demographic characteristics and survey response: results from multiple regression model. We collected demographic information for all our respondents and a random sample of 446 non-respondents using information publicly available online. Here we use multiple linear regression to predict the response to the survey using the demographic variables that we collected.  The (arbitrarily-chosen) reference categories, the ones that are excluded from the list of coefficients, are female/other for gender, North America  for undergraduate, PhD, and work region, and industry for type of workplace. The \textit{F}-test of overall significance rejects the null hypothesis that respondents do not differ in whether they responded to the survey depending on demographic characteristics. The Holm method was used to control the family-wise error rate.} 
  \label{tab:associationresponse} 
\centering  
\begin{tabular}{ll}\hline
  &\hspace{-2.6em} Coefficient (\textit{SE}) \\
\hline
(Intercept) & 0.540$^{***}$ \\ 
  & (0.016) \\ 
  Male & 0.022 \\ 
  & (0.016) \\ 
  Undergrad graduation year \hspace{3em}& 0.002 \\ 
  & (0.021) \\ 
  Undergrad region: Europe & $-$0.030 \\ 
  & (0.022) \\ 
  Undergrad region: Asia & $-$0.037 \\ 
  & (0.020) \\ 
  Undergrad region: Other & $-$0.013 \\ 
  & (0.018) \\ 
  PhD region: Europe & 0.039 \\ 
  & (0.026) \\ 
  PhD region: Asia & $-$0.019 \\ 
  & (0.023) \\ 
  PhD region: Other & 0.009 \\ 
  & (0.021) \\ 
  Currently enrolled in PhD & 0.033 \\ 
  & (0.019) \\ 
  Log all citations & 0.022 \\ 
  & (0.028) \\ 
  All h-index & $-$0.083$^{*}$ \\ 
  & (0.026) \\ 
  Work region: Europe & 0.017 \\ 
  & (0.026) \\ 
  Work region: Asia & 0.034 \\ 
  & (0.024) \\ 
  Work region: Other & 0.014 \\ 
  & (0.021) \\ 
  Work in academia & 0.069$^{***}$ \\ 
  & (0.017) \\ 
  Missing: undergrad year & 0.001 \\ 
  & (0.028) \\ 
  Missing: undergrad region & $-$0.044 \\ 
  & (0.028) \\ 
  Missing: all citations & 0.009 \\ 
  & (0.016) \\ 
\hline
\textit{N} & 970\\
\hline
\multicolumn{2}{l}{\textsuperscript{} * \textit{p} $<$ 0.05, ** \textit{p} $<$ 0.01, *** \textit{p} $<$ 0.001}\\
\multicolumn{2}{l}{\textsuperscript{} \textit{F}(18, 951) = 4.196; \textit{p}-value: $<0.001$}\\
\end{tabular}
\end{table}
\clearpage

\subsection*{Additional figures} 

\renewcommand{\figurename}{Fig.}
\renewcommand{\thefigure}{S\arabic{figure}}

\subsubsection*{Evaluation of AI governance challenges \label{appendix:fig-issue-importance}}

\bigskip

\begin{figure} [ht]
    \centering
    \includegraphics[width=0.9\textwidth]{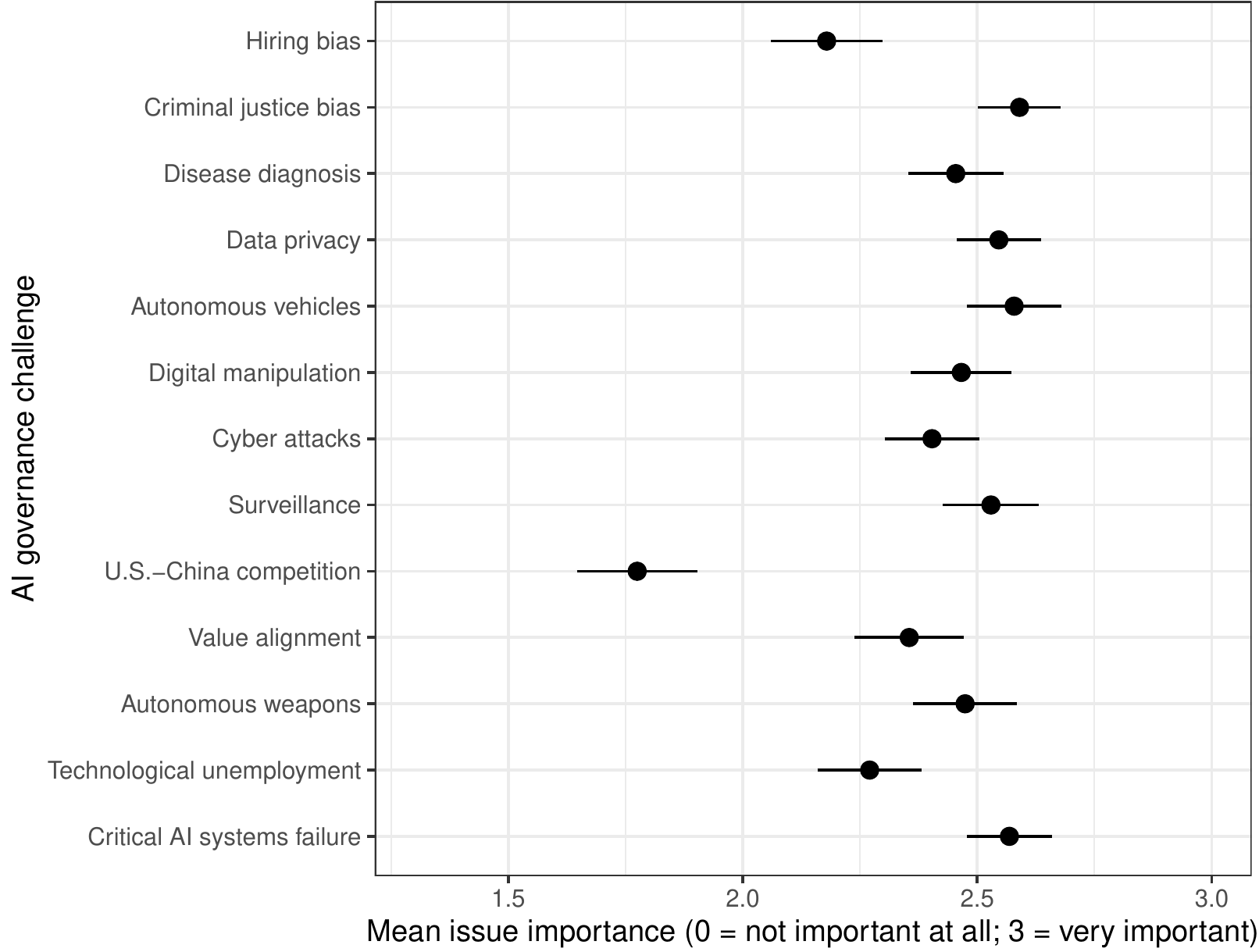}
    \caption{Perceived issue importance of AI governance challenges: all responses from the AI/ML researcher survey. Each respondent was presented with five AI governance challenges randomly selected from a list of 13. Respondents were asked to evaluate the importance of each governance challenge using a four-point scale (the slider scale allows respondents to input values to the tenth decimal point): 0 = not important, 1 = not too important, 2 = somewhat important, 3 = very important. We present the mean response for each governance challenge along with the corresponding 95\% confidence intervals.}
    \label{fig:issue-importance}
\end{figure}
\clearpage

\begin{figure}[ht]
    \centering
    \includegraphics[height=0.8\textheight]{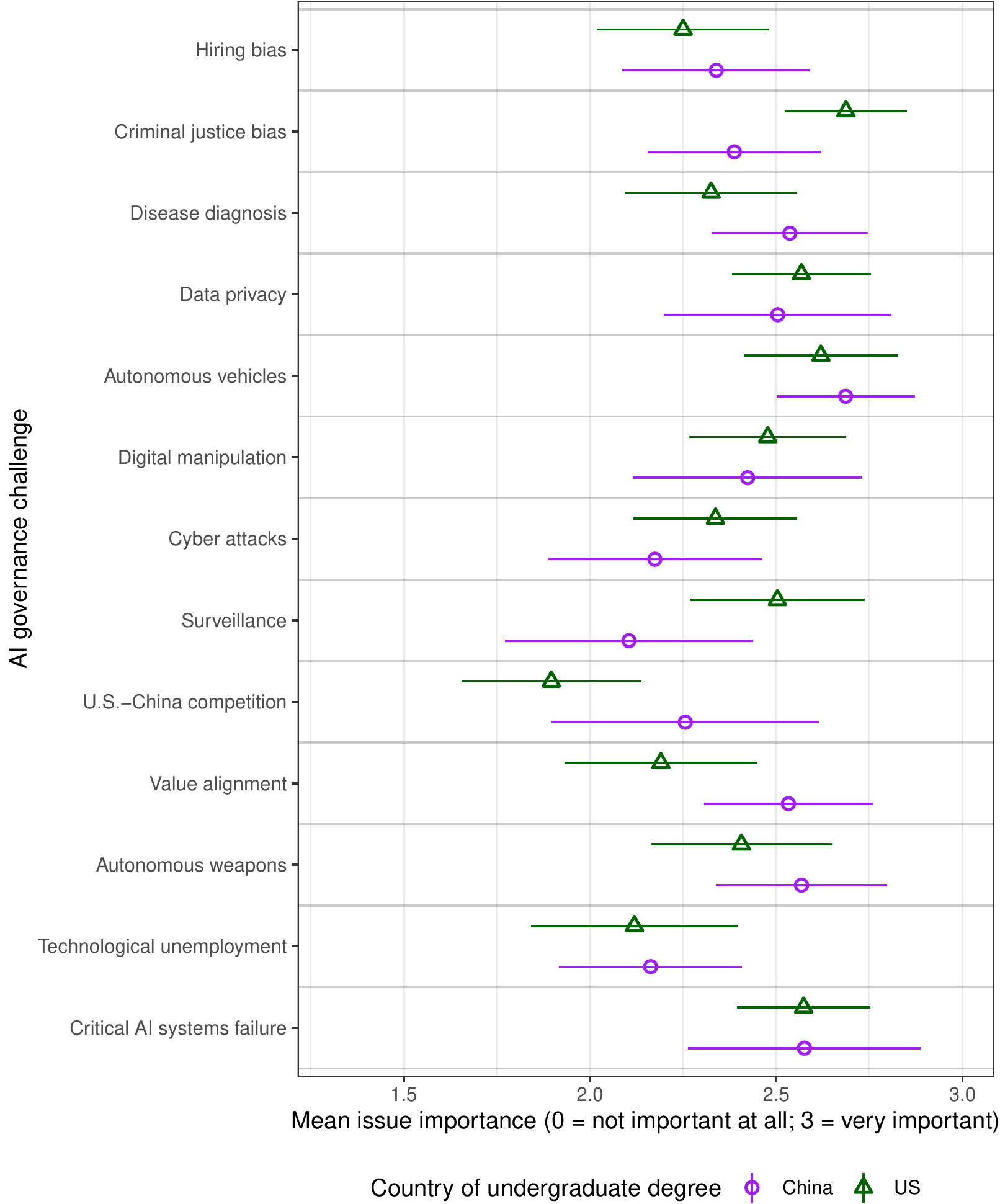}
    \caption{Perceived issue importance of AI governance challenges: by country of undergraduate degree (China and the US). Each respondent was presented with five AI governance challenges randomly selected from a list of 13. Respondents were asked to evaluate the importance of each governance challenge using a four-point scale (the slider scale allows respondents to input values to the tenth decimal point): 0 = not important, 1 = not too important, 2 = somewhat important, 3 = very important. We identified the country of respondents' undergraduate degrees using publicly available information on the internet. We present the mean response for each governance challenge (by country of undergraduate degree) along with the corresponding 95\% confidence intervals.}
    \label{fig:issue-importance-country}
\end{figure}
\clearpage

\begin{figure}[ht]
    \centering
    \includegraphics[height=0.78\textheight]{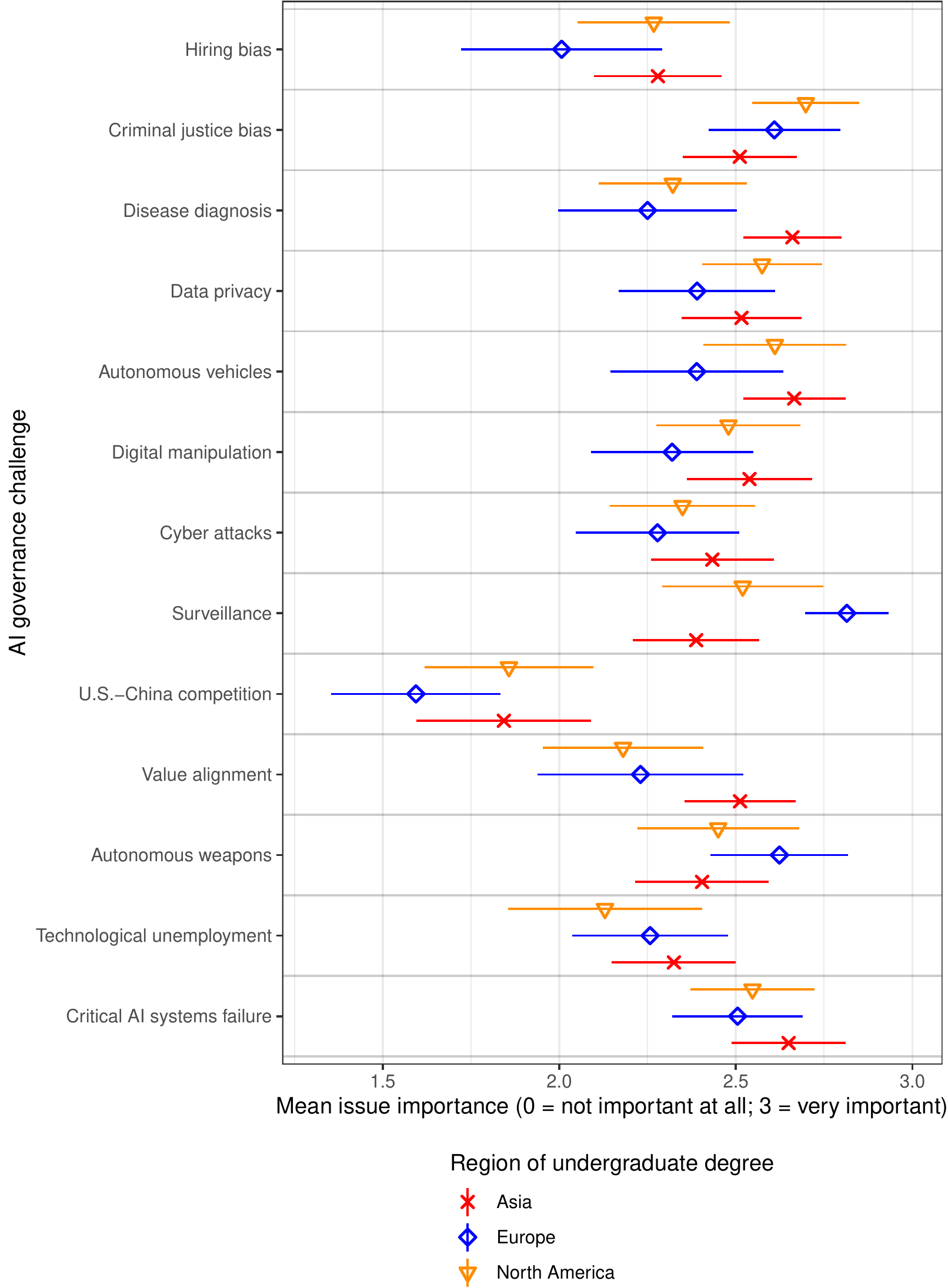}
    \caption{Perceived issue importance of AI governance challenges: by region of undergraduate degree (Asia, Europe, and North America). Each respondent was presented with five AI governance challenges randomly selected from a list of 13. Respondents were asked to evaluate the importance of each governance challenge using a four-point scale (the slider scale allows respondents to input values to the tenth decimal point): 0 = not important, 1 = not too important, 2 = somewhat important, 3 = very important. We present the mean response for each governance challenge (by region of undergraduate degree) along with the corresponding 95\% confidence intervals.}
    \label{fig:issue-importance-region}
\end{figure}
\clearpage

\begin{figure}[ht]
    \centering
    \includegraphics[width=0.9\textwidth]{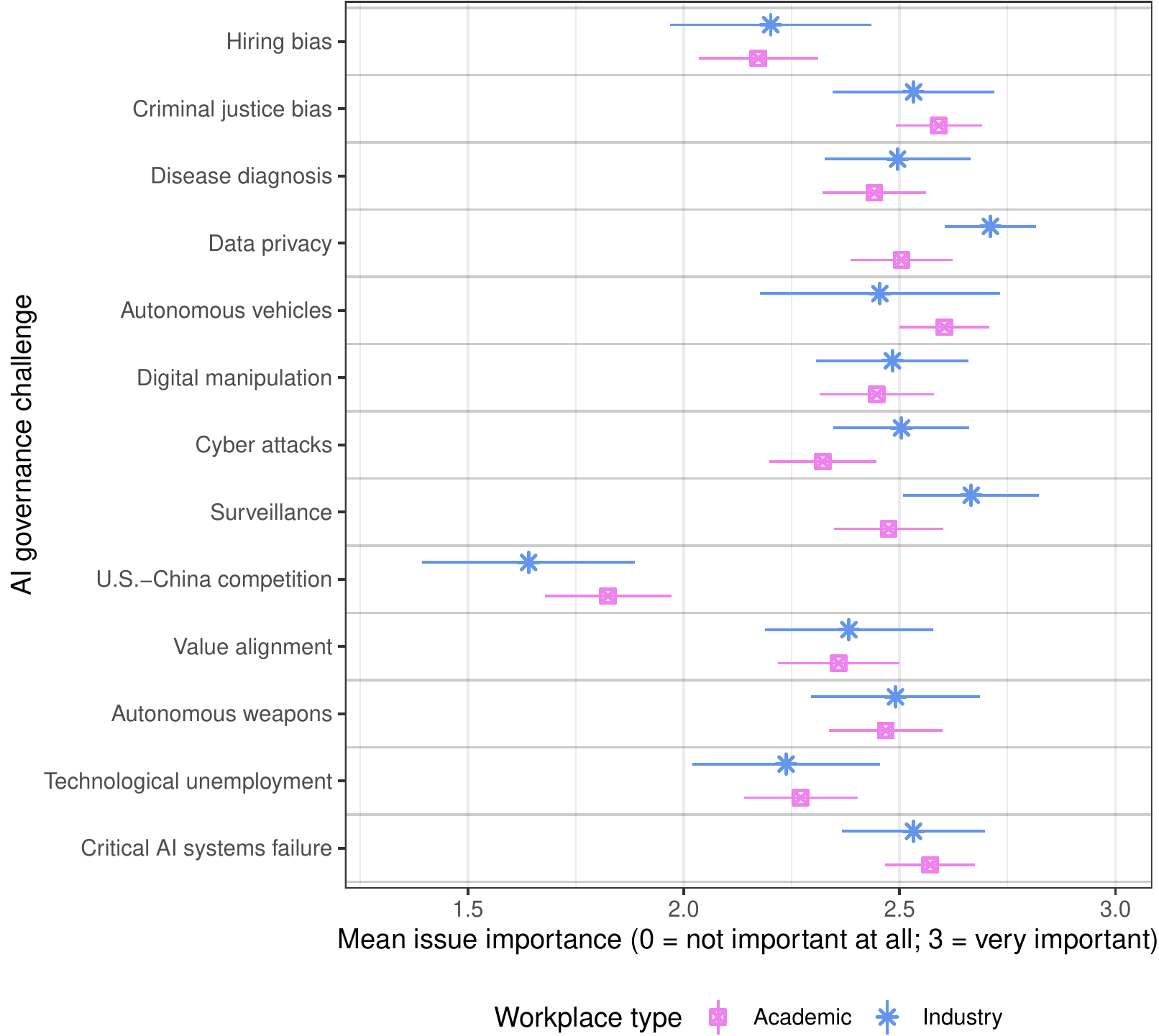}
    \caption{Perceived issue importance of AI governance challenges: by workplace type (academic and industry). Each respondent was presented with five AI governance challenges randomly selected from a list of 13. Respondents were asked to evaluate the importance of each governance challenge using a four-point scale (the slider scale allows respondents to input values to the tenth decimal point): 0 = not important, 1 = not too important, 2 = somewhat important, 3 = very important. We identified the respondents' workplace types using publicly available information on the internet. Note that a single respondent can work both in academia and industry. We present the mean response for each governance challenge (by workplace type) along with the corresponding 95\% confidence intervals.}
    \label{fig:issue-importance-workplace-type}
\end{figure}
\clearpage

\begin{figure}[htb]

\subsubsection*{Trust in actors to shape the development and use of AI in the public interest}

\medskip

    \centering
    \includegraphics[height=0.75\textheight]{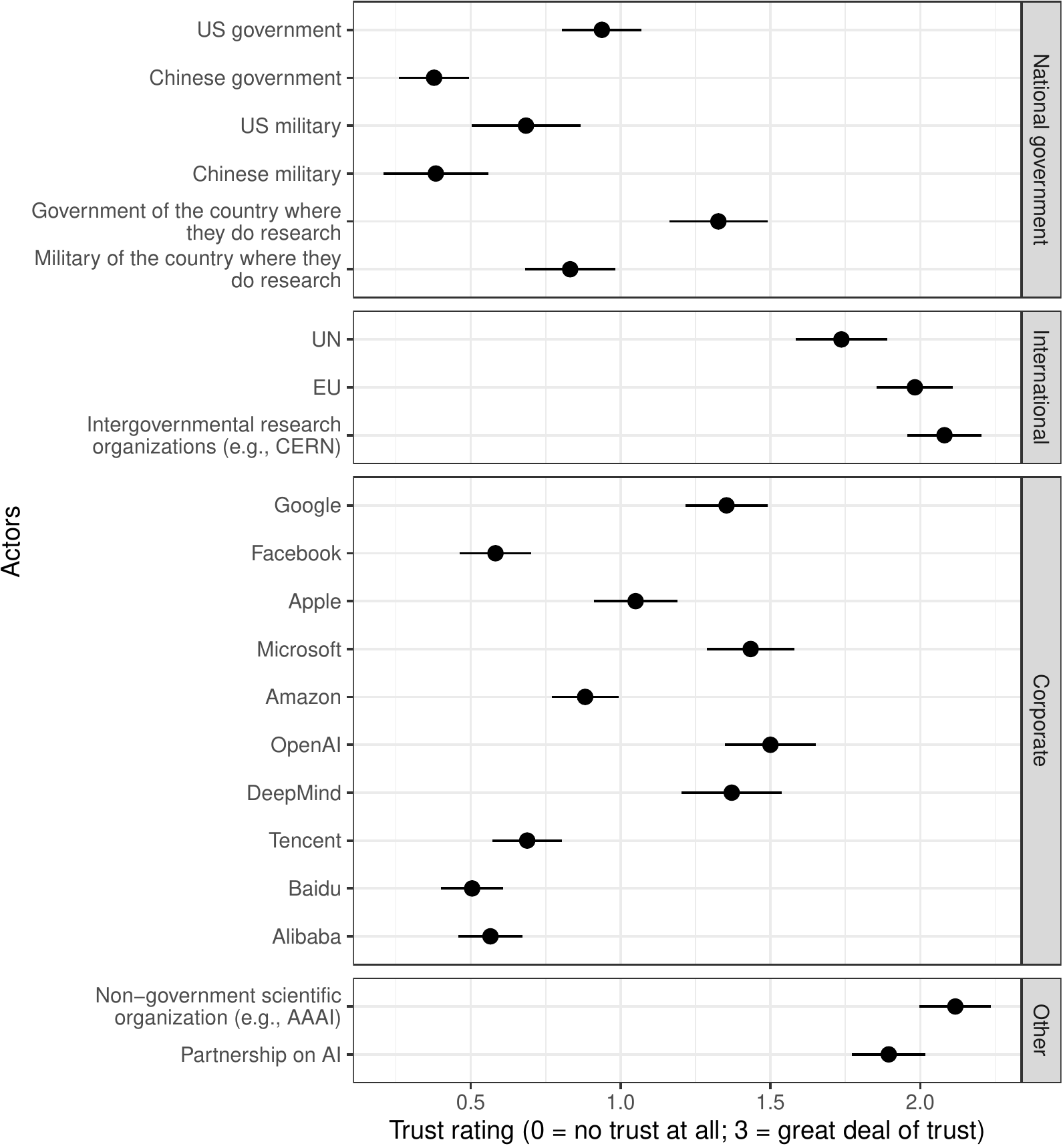}
    \caption{Trust in actors to shape the development and use of AI in the public interest: all responses from the AI/ML researcher survey. Respondents were shown five randomly-selected actors and asked to evaluate how much they trust the actors using a four-point scale: 0 = no trust at all, 1 = not too much trust, 2 = a fair amount of trust, 3 = a great deal of trust. The US military and the Chinese military were shown only to respondents who do research in the US or China; these respondents had equal probability of being shown the US military or the Chinese military.  Of the 60 responses to the US military, 56 came from those who do research in the US. Of the 66 responses to the Chinese military, 60 came from those who do research in the US. We present the mean response for each actor along with the corresponding 95\% confidence intervals.}
    \label{fig:trust-all}
\end{figure}
\clearpage

\begin{figure}[ht]
    \centering
    \includegraphics[height=0.8\textheight]{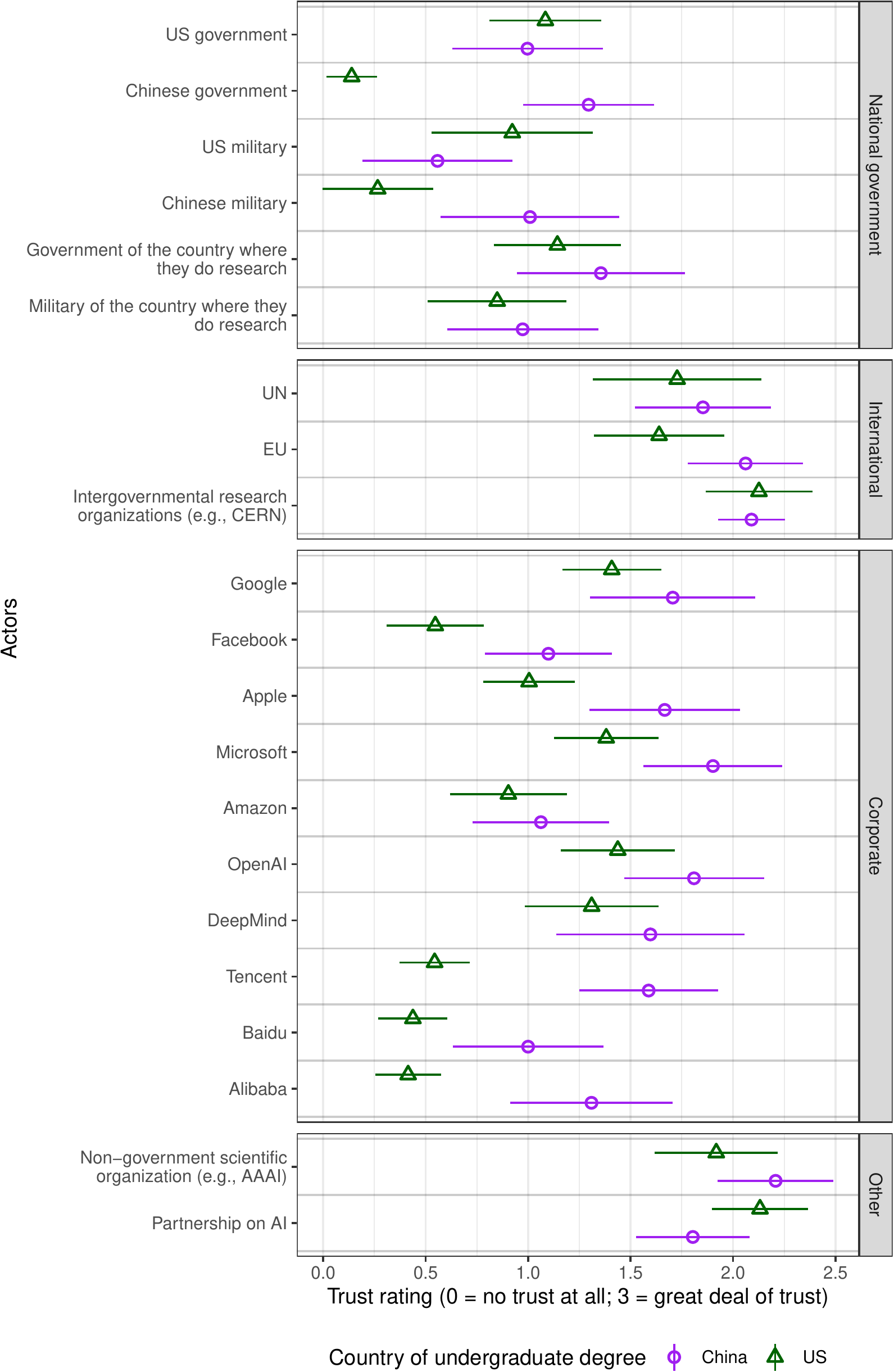}
    \caption{Trust in actors to shape the development and use of AI in the public interest: by country of undergraduate degree (China and the US). Respondents were shown five randomly-selected actors and asked to evaluate how much they trust the actors using a four-point scale: 0 = no trust at all, 1 = not too much trust, 2 = a fair amount of trust, 3 = a great deal of trust. The US military and the Chinese military were shown only to respondents who do research in the US or China; these respondents had equal probability of being shown the US military or the Chinese military. We present the mean response for each actor (by country of undergraduate degree) along with the corresponding 95\% confidence intervals.}
    \label{fig:trust-country}
\end{figure}

\begin{figure}[ht]
    \centering
    \includegraphics[height=0.8\textheight]{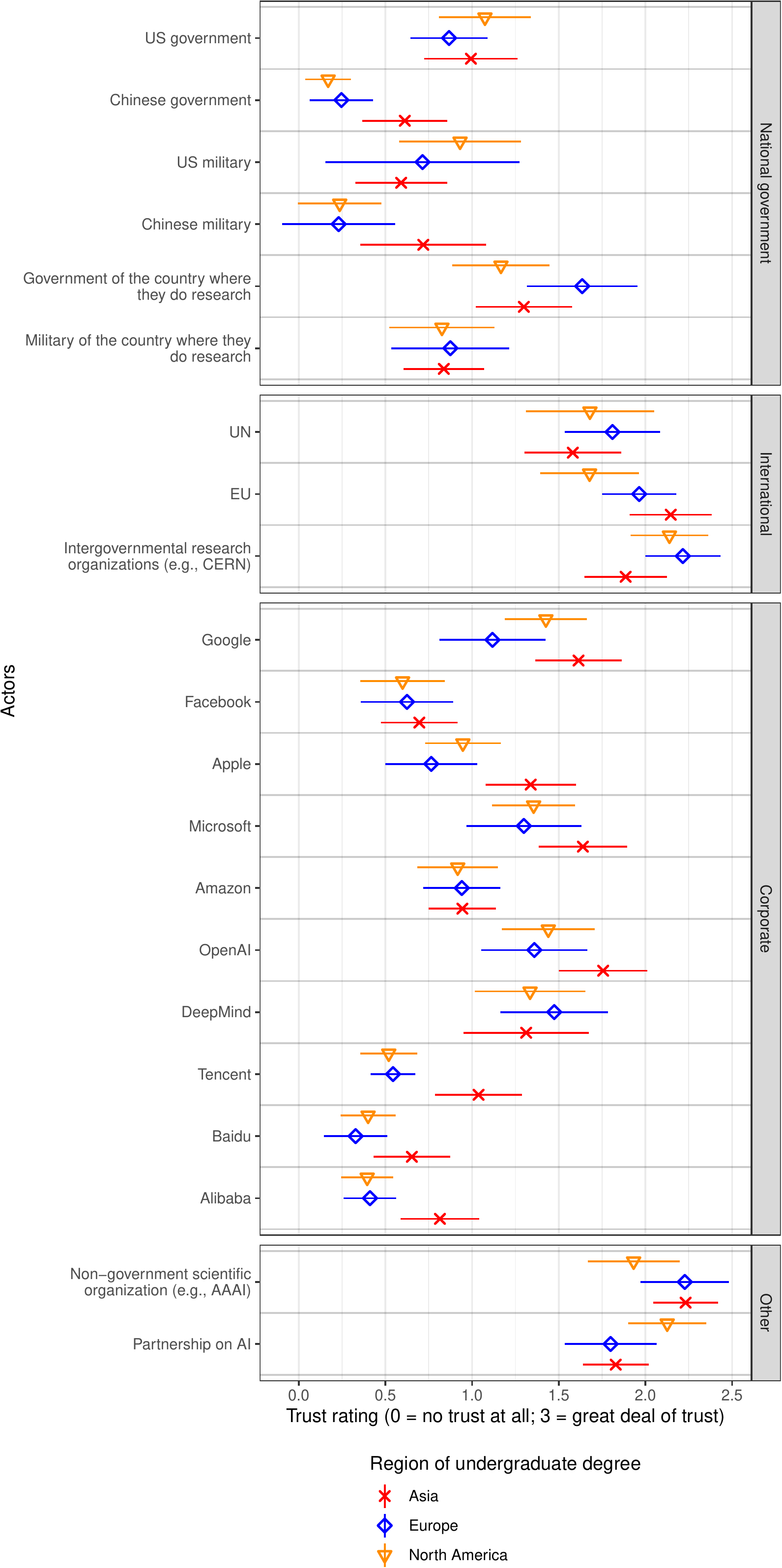}
    \caption{Trust in actors to shape the development and use of AI in the public interest: by region of undergraduate degree (Asia, Europe, and North America). Respondents were shown five randomly-selected actors and asked to evaluate how much they trust the actors using a four-point scale: 0 = no trust at all, 1 = not too much trust, 2 = a fair amount of trust, 3 = a great deal of trust. The US military and the Chinese military were shown only to respondents who do research in the US or China; these respondents had equal probability of being shown the US military or the Chinese military. We present the mean response for each actor (by region of undergraduate degree) along with the corresponding 95\% confidence intervals.}
    \label{fig:trust-region}
\end{figure}
\clearpage

\begin{figure}[ht]
    \centering
    \includegraphics[height=0.8\textheight]{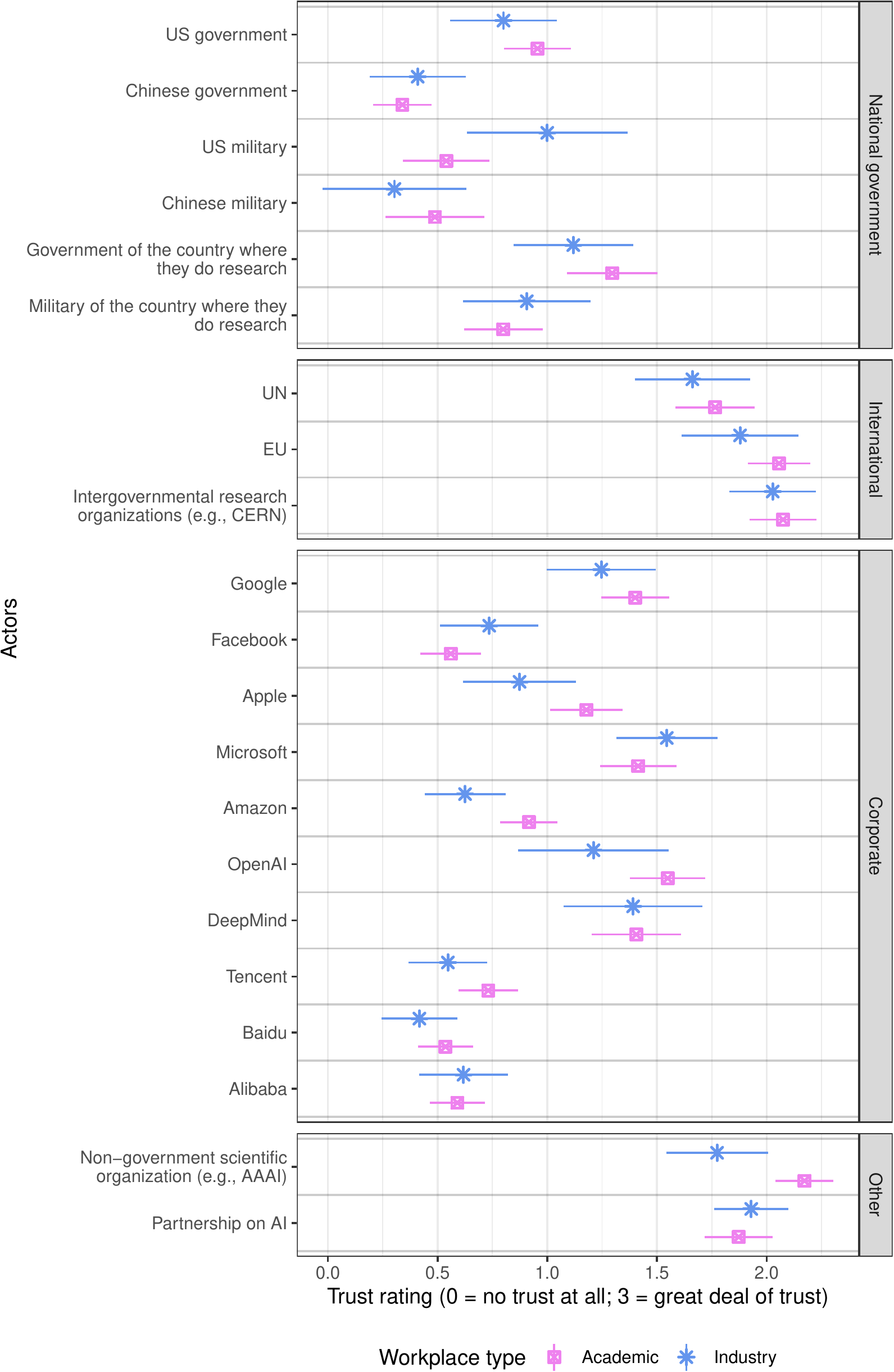}
    \caption{Trust in actors to shape the development and use of AI in the public interest: by workplace type (academic and industry). Respondents were shown five randomly-selected actors and asked to evaluate how much they trust the actors using a four-point scale: 0 = no trust at all, 1 = not too much trust, 2 = a fair amount of trust, 3 = a great deal of trust. The US military and the Chinese military were shown only to respondents who do research in the US or China; these respondents had equal probability of being shown the US military or the Chinese military. We present the mean response for each actor (by their workplace type) along with the corresponding 95\% confidence intervals.}
    \label{fig:trust-work}
\end{figure}
\clearpage

\begin{figure}[ht]
    \centering
    \includegraphics[height=0.75\textheight]{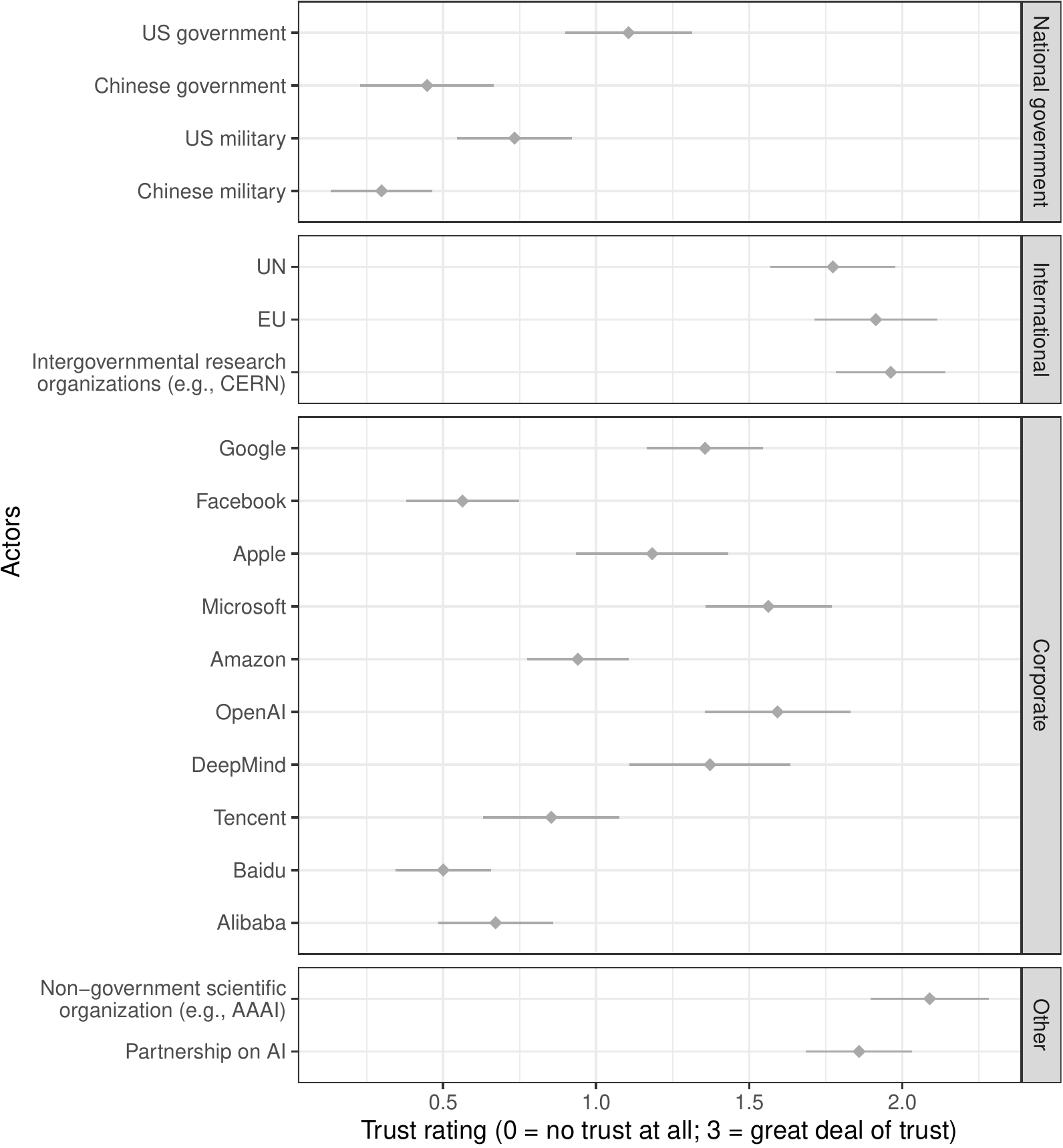}
    \caption{Trust in actors to shape the development and use of AI in the public interest: those who report spending most of their time doing research in the US. Respondents were shown five randomly-selected actors and asked to evaluate how much they trust the actors using a four-point scale: 0 = no trust at all, 1 = not too much trust, 2 = a fair amount of trust, 3 = a great deal of trust. The US military and the Chinese military were shown only to respondents who do research in the US or China; these respondents had equal probability of being shown the US military or the Chinese military. The country where each respondent spends the most time working or studying is self-reported in the survey. We present the mean responses for each actor along with the corresponding 95\% confidence intervals.}
    \label{fig:trust-work-US}
\end{figure}
\clearpage

\begin{figure} [ht]
    \centering
    \includegraphics[width=0.9\textwidth]{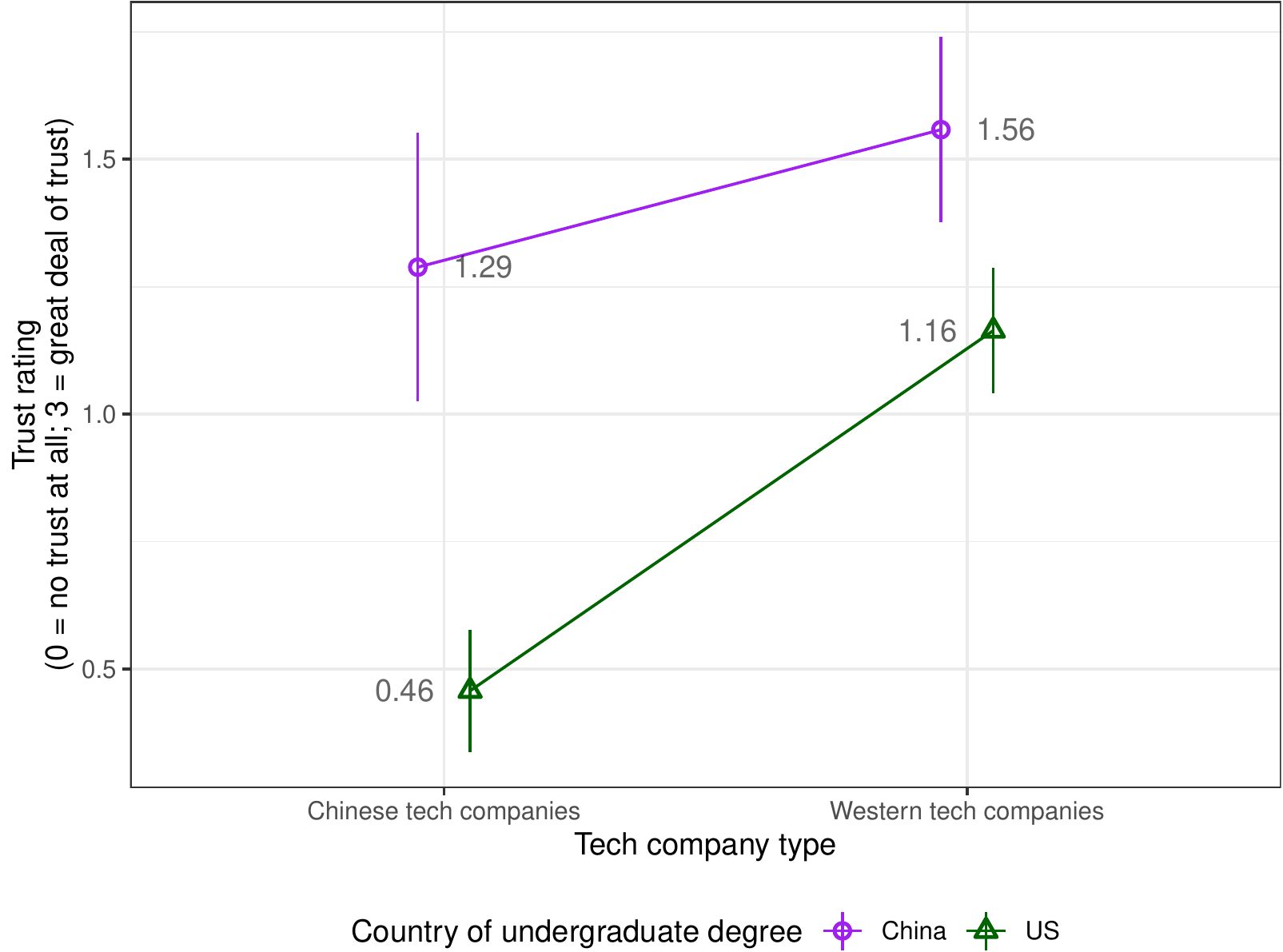}
    \caption{Interaction plot: how respondents who attended university in China versus the US rate trust in Chinese versus Western tech companies. Respondents were asked to evaluate how much they trust the companies using a four-point scale: 0 = no trust at all, 1 = not too much trust, 2 = a fair amount of trust, 3 = a great deal of trust. The figure is generated from a linear regression with a two-way interaction between Chinese versus Western tech companies and having an undergraduate degree from the US versus China. Only respondents who who received undergraduate degrees from the US and China are included in this analysis. See Table \ref{tab:interaction-tech-companies-us-china} for the regression output table.}
    \label{fig:us-china-tech-int}
\end{figure}
\clearpage

\subsubsection*{AI safety}

\bigskip

\begin{figure}[ht]
    \centering
    \advance\leftskip-0.3cm
    \includegraphics[height=0.4\textheight]{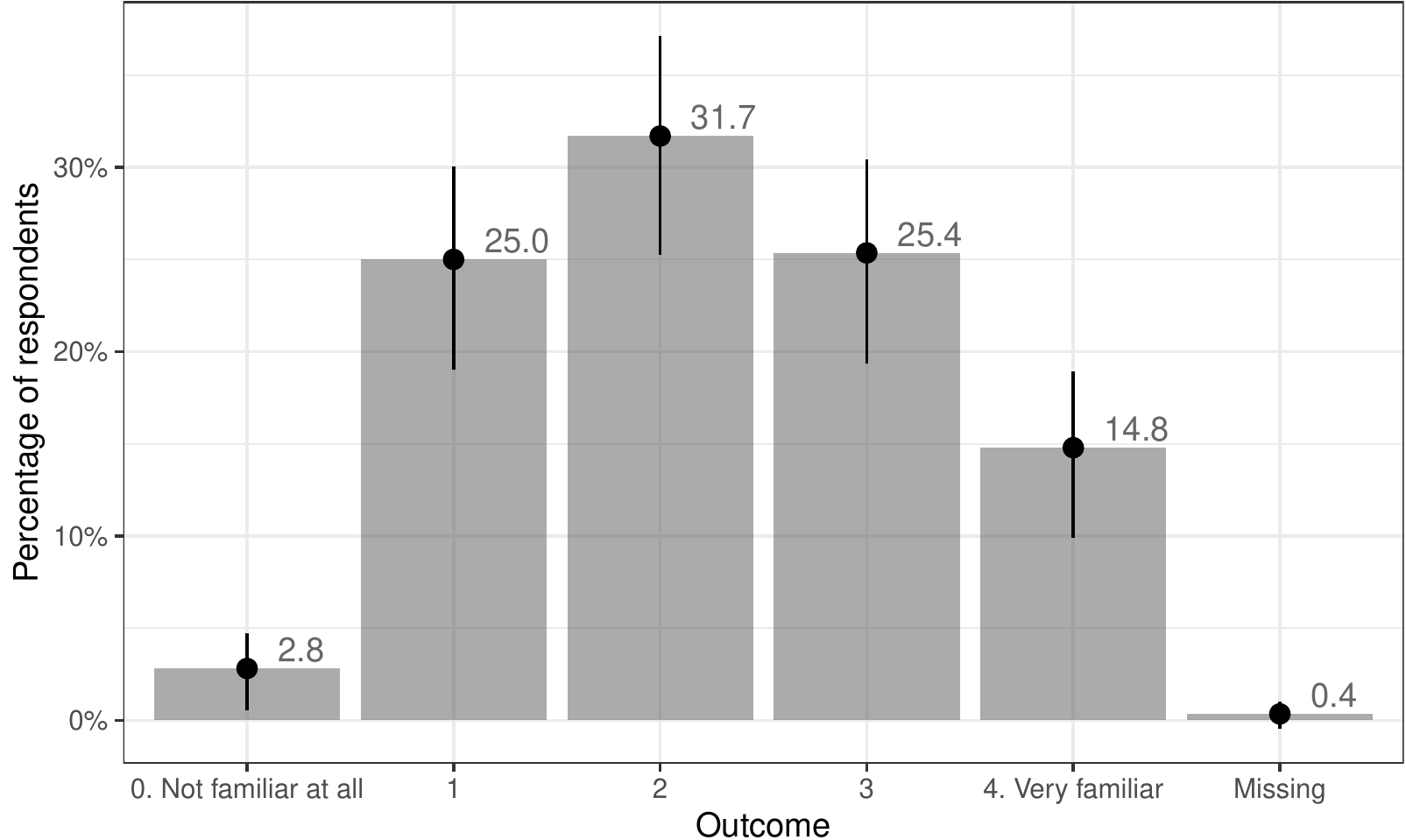}
    \caption{Familiarity with AI safety: distribution of responses. After reading a definition of AI safety (see survey text for the definition), respondents input their familiarity with AI safety using a 5-point slider (0 = not familiar at all; 4 = very familiar). We present the mean response at each level of familiarity with AI safety and for missing responses, along with the corresponding 95\% confidence intervals.}
    \label{fig:ai-safety-distribution}
\end{figure}
\clearpage

\begin{figure}[ht]
    \centering
    \includegraphics[height=0.65\textheight]{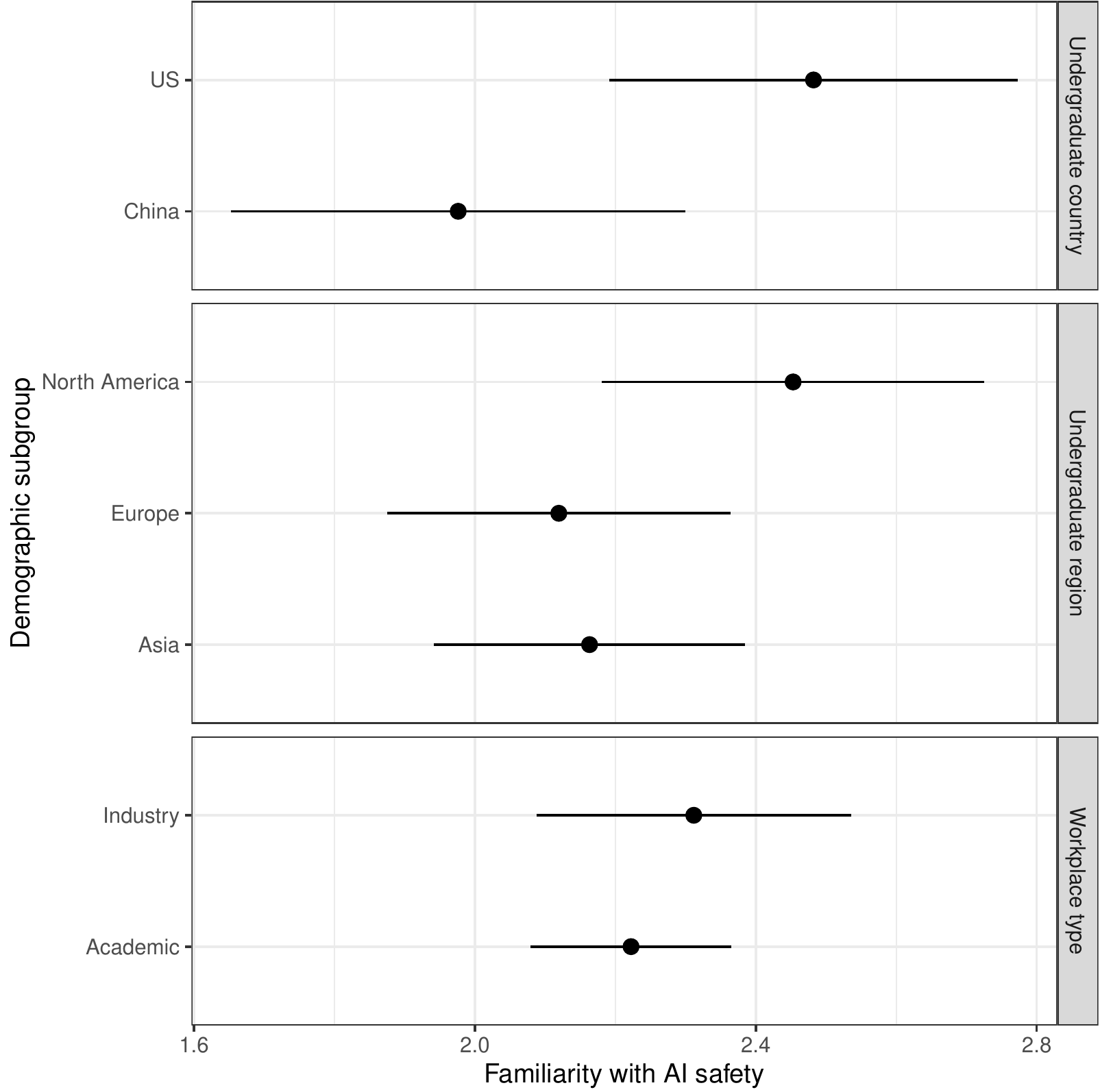}
    \caption{Familiarity with AI safety: mean response by demographic subgroups. After reading a definition of AI safety (see Survey text for the definition), respondents input their familiarity with AI safety using a 5-point slider (0 = not familiar at all; 4 = very familiar). We present the mean AI safety familiarity response by undergraduate country (US and China), undergraduate region (North America, Europe, and Asia), and workplace type (industry and academic), along with the corresponding 95\% confidence intervals.}
    \label{fig:ai-safety-subgroups}
\end{figure}
\clearpage

\begin{figure}[ht]
    \centering
    \includegraphics[height=0.43\textheight]{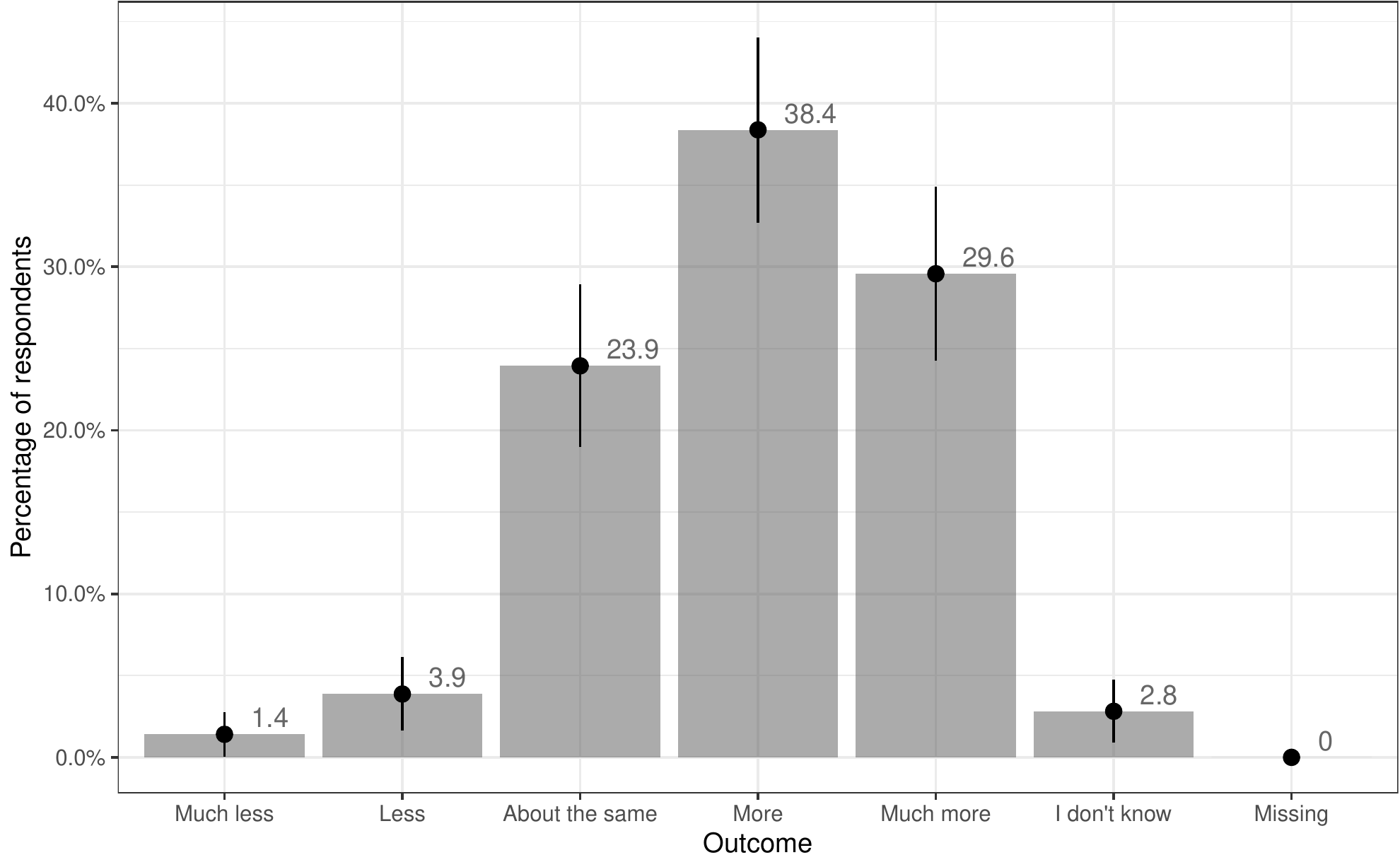}
    \caption{How much should AI safety be prioritized: distribution of responses. This question appears after the familiarity with AI safety question. Respondents were asked how much AI safety research should be prioritized relative to today. The answer choices are a Likert scale from -2 to 2: -2 = much less; -1 = less; 0 = about the same; 1 = more; 2 = much more. There was also an ``I don't know'' option. We present the mean response for each option along with the corresponding 95\% confidence intervals.}
    \label{fig:ai-safety-prioritized}
\end{figure}
\clearpage

\begin{figure}[ht]
    \centering
    \includegraphics[height=0.65\textheight]{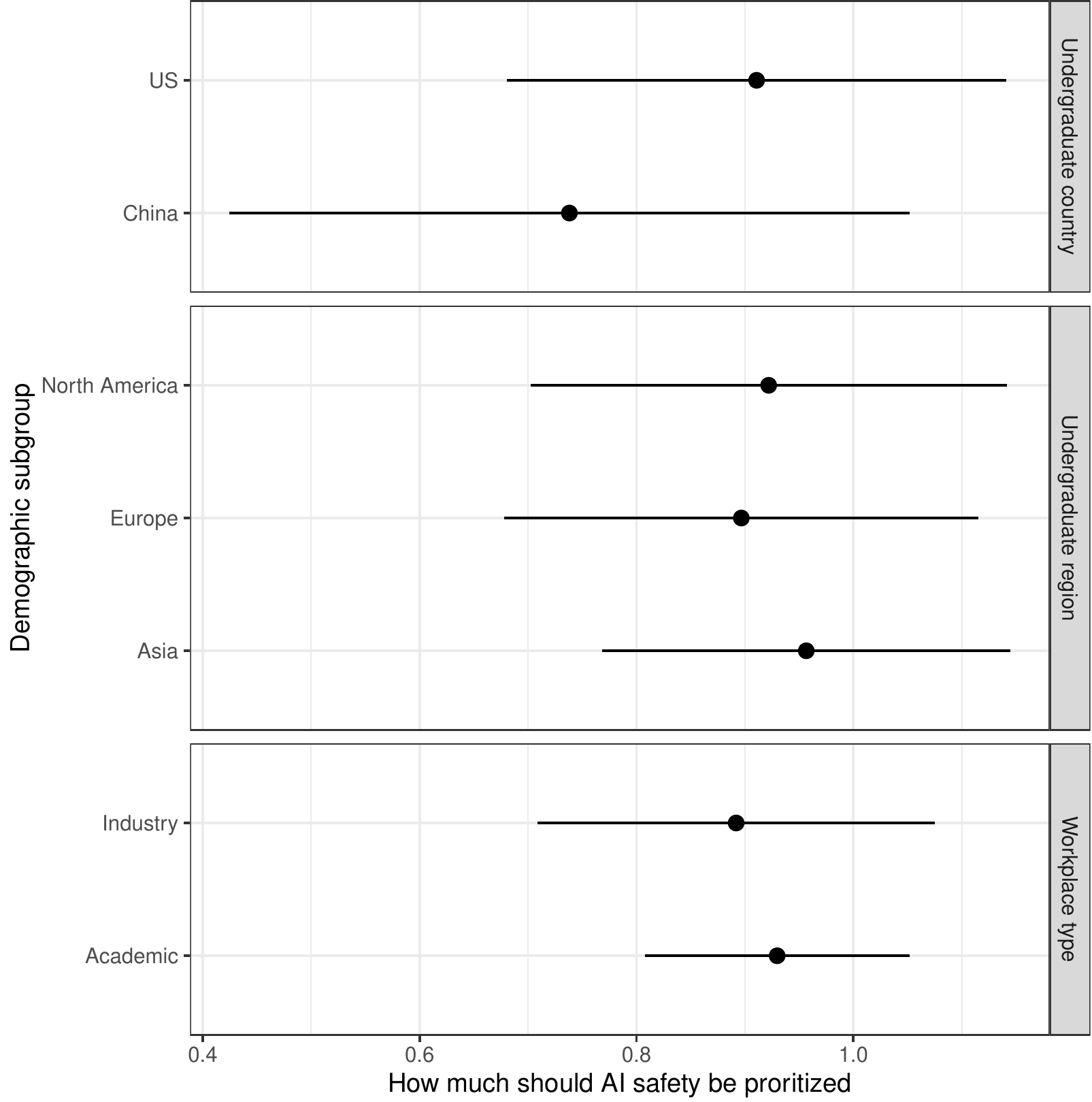}
    \caption{How much should AI safety be prioritized: mean response by demographic subgroups. Respondents were asked how much AI safety research should be prioritized relative to today. The answer choices are a Likert scale from -2 to 2: -2 = much less; -1 = less; 0 = about the same; 1 = more; 2 = much more. There was also an ``I don't know'' option. We present the mean response by undergraduate country (US and China), undergraduate region (North America, Europe, and Asia), and workplace type (industry and academic), along with the corresponding 95\% confidence intervals.}
    \label{fig:ai-safety-prioritized-subgroups}
\end{figure}
\clearpage

\subsubsection*{Attitudes toward military applications of AI}
\bigskip

\begin{figure}[h!]
    \centering
    \includegraphics[width=0.9\textwidth]{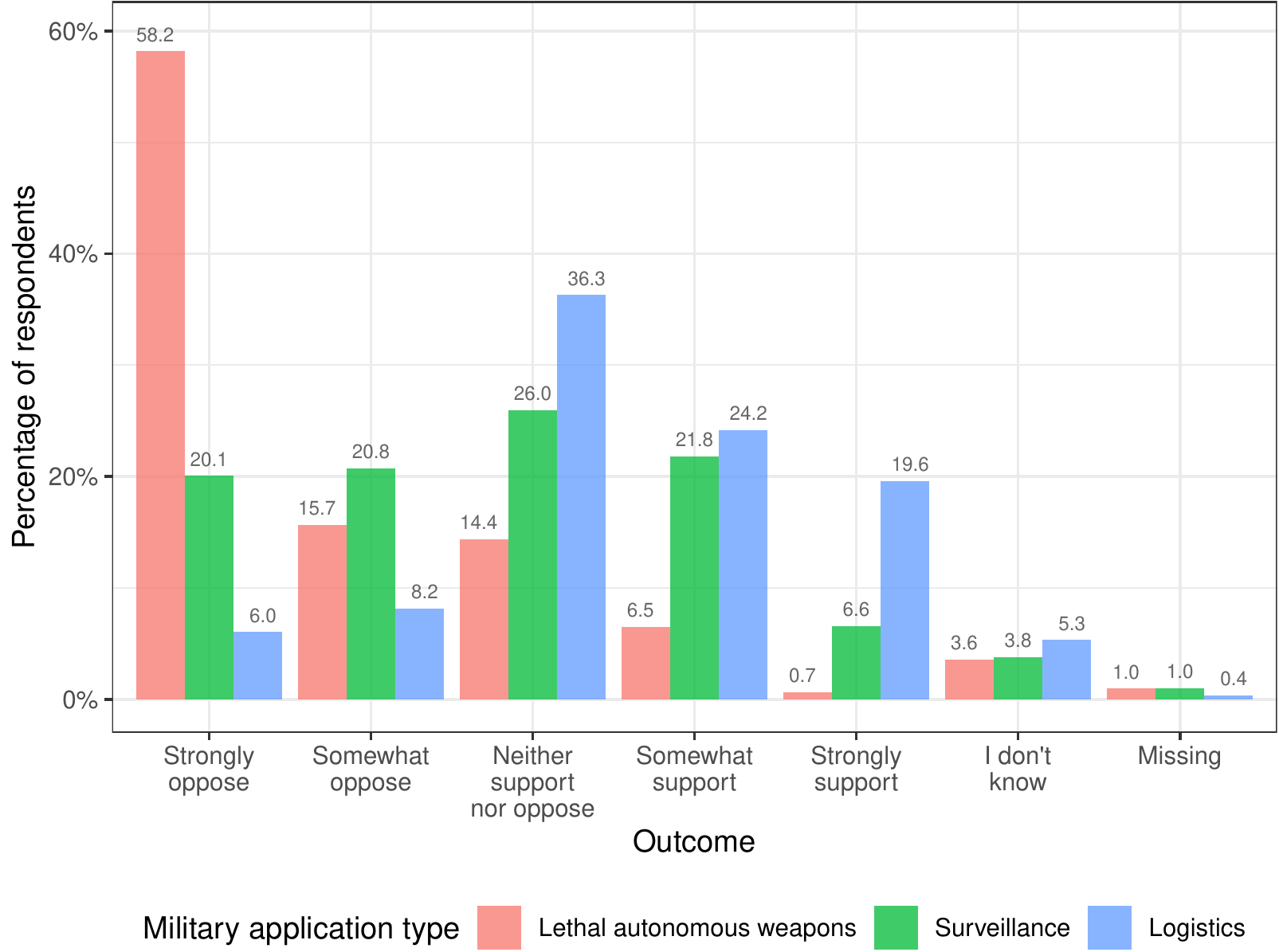}
    \caption{Attitudes toward researchers working on military applications of AI: distribution of responses. Respondents were asked to indicate their level of support for two of the three randomly presented military applications (lethal autonomous weapons, surveillance, and logistics) on a five-point scale from -2 to 2: -2 = strongly oppose, -1 = somewhat oppose, 0 = neither support nor oppose, 1 = somewhat support, 2 = strongly support. There was also an ``I don't know'' option. Each military application was defined when it was presented (see survey text for the definition). We present the percentage of respondents who chose each response as well as those who did not respond to the  question, along with the corresponding 95\% confidence intervals.}
    \label{fig:military-distribution}
\end{figure}
\clearpage

\begin{figure}[ht]
    \centering
    \includegraphics[height=0.4\textheight]{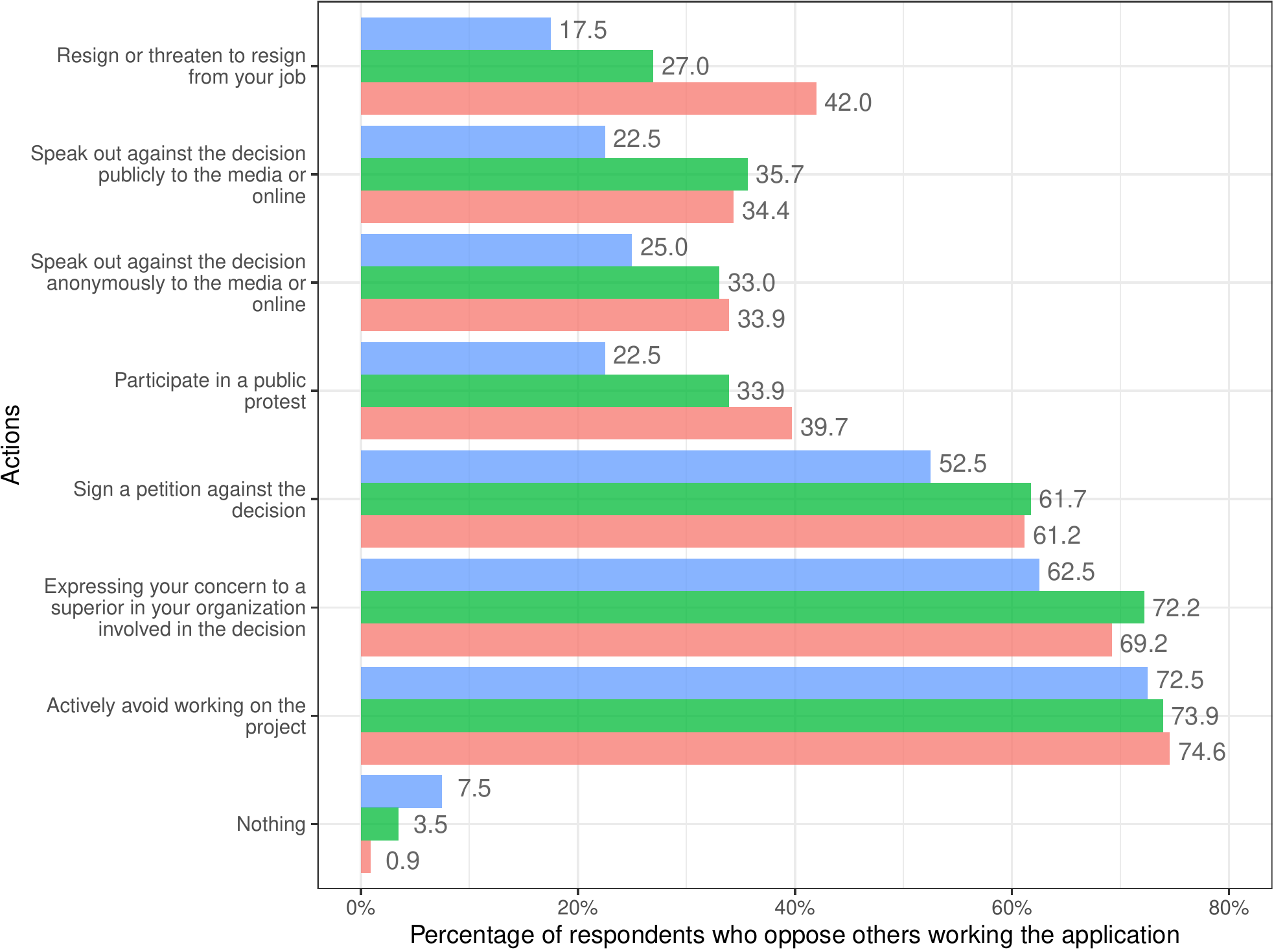}
    \includegraphics[height=0.4\textheight]{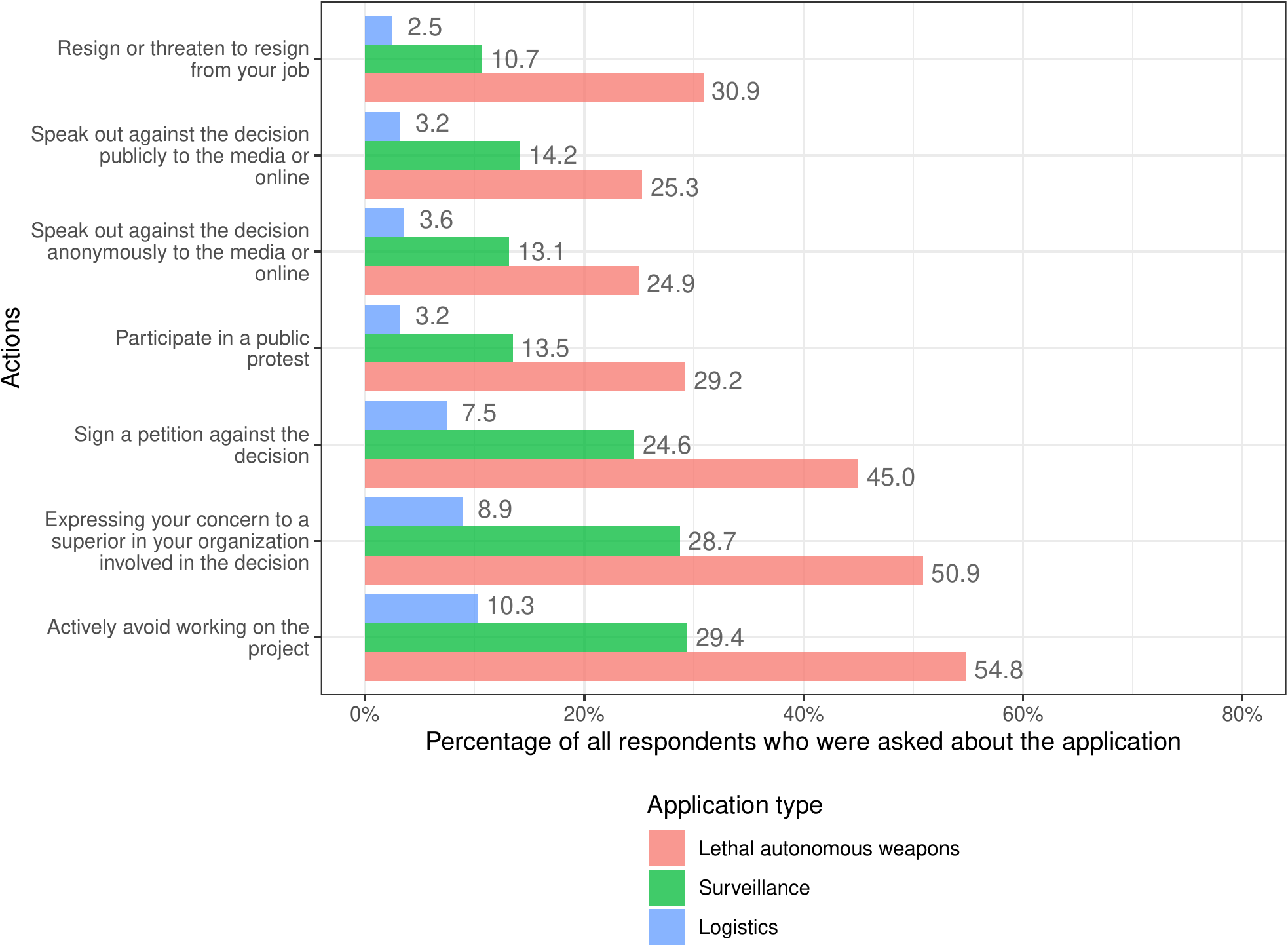}
    \caption{Support for collective action against research into military applications of AI: distribution of responses. Respondents were asked which actions, if any, they would take if their organization decided to research two randomly chosen applications from the following three: lethal autonomous weapons, surveillance, or logistics. The question text also highlighted that this was specific to where the respondent worked our studied. In the top panel, the x-axis is the percentage of respondents who were asked about the application \textit{and} indicated they ``oppose'' or ``strongly oppose'' researchers working on the application. In the bottom panel, the x-axis is the percentage of all respondents who were asked about the application. Recall that each respondent was asked about two applications randomly selected from the three.}
    \label{fig:collective-distribution}
\end{figure}
\clearpage

\begin{figure}[ht]
    \centering
    \advance\leftskip-0.45cm
    \includegraphics[height=0.45\textheight]{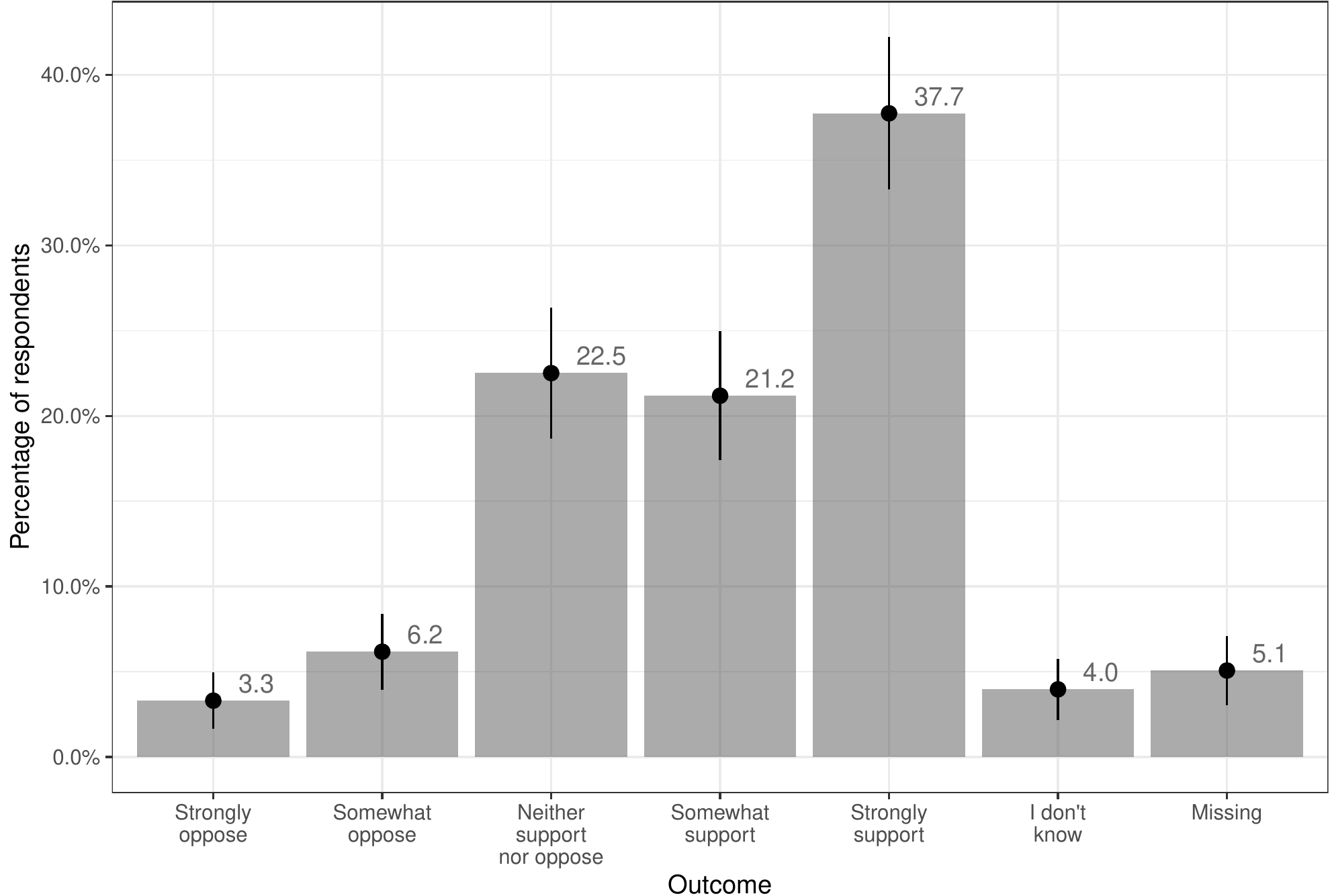}
    \caption{Attitudes toward Google pulling out of Project Maven: distribution of responses. Respondents were presented with a short description of the employees' reactions to Google's Project Maven and the following non-renewal of the contract (see survey text for the description) and were asked to indicate their support for the non-renewal decision on a five-point scale from -2 to 2: -2 = strongly oppose, -1 = somewhat oppose, 0 = neither support nor oppose, 1 = somewhat support, 2 = strongly support. There was also an ``I don't know'' option. We present the percentage of respondents choosing each option and who did not respond to the question, along with the corresponding 95\% confidence intervals.}
    \label{fig:maven-distribution}
\end{figure}
\clearpage

\subsubsection*{Publication norms}
\bigskip

\begin{figure}[ht]
    \centering
    \includegraphics[height=0.5\textheight]{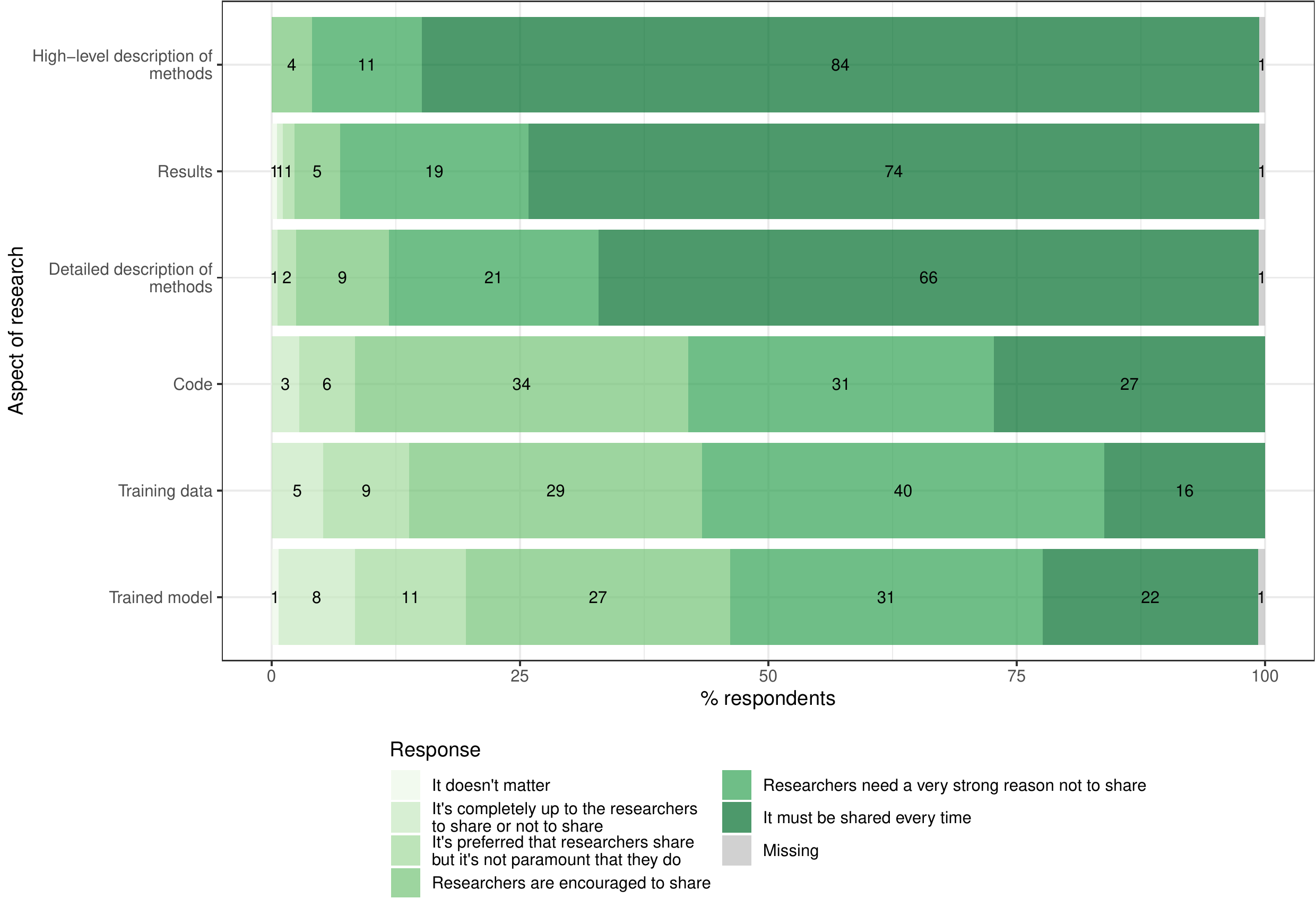}
    \caption{Sharing aspects of research: distribution of responses. Respondents were presented with three aspects of research randomly chosen from a list of six. For each aspect of research, they selected from six levels of openness (0 = it doesn't matter; 1 = it’s completely up to the researchers to share or not to share; 2 = it’s preferred that researchers share but it’s not paramount that they do; 3 = researchers are encouraged to share; 4 = researchers need a very strong reason not to share; 5 = it must be shared every time). We present the mean response for each level of openness for the different aspects of research, along with the corresponding 95\% confidence intervals.}
    \label{fig:sharing_bars}
\end{figure}
\clearpage

\begin{figure}[h!]
    \centering
    \includegraphics[height=0.878\textheight]{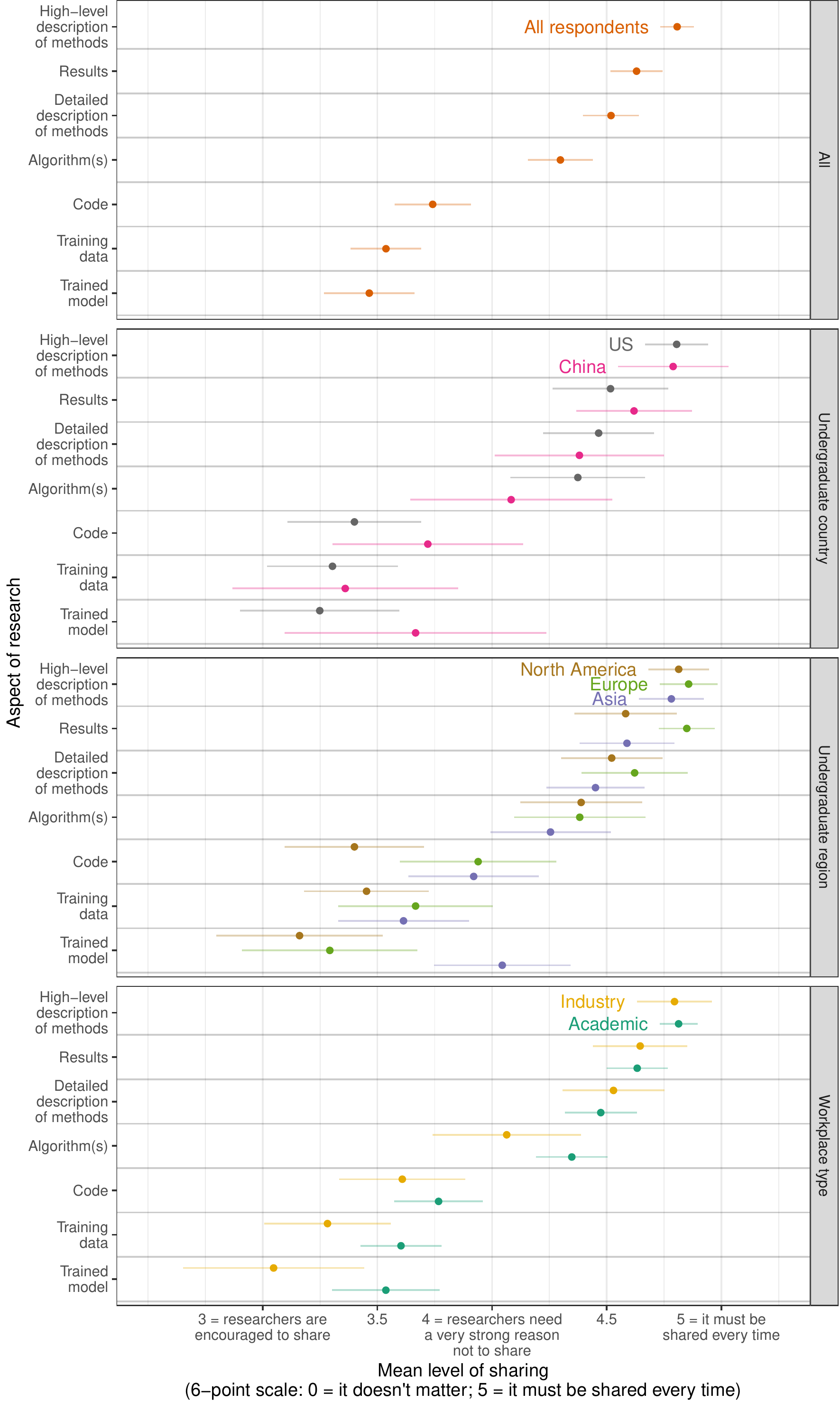}
    \caption{Sharing aspects of research: mean level of openness response for each aspect of research, by demographic subgroups. We present the mean openness response for each aspect  of research for all respondents as  well as by undergraduate country (US and China), undergraduate region (North America, Europe, and Asia), and workplace type (industry and academic). The corresponding 95\% confidence intervals are shown.}
    \label{fig:sharing_demo}
\end{figure}
\clearpage

\subsection*{Additional tables}
\subsubsection*{Evaluation of AI governance challenges \label{appendix:tab-issue-importance}}

\begin{table}[ht]
\small
\caption{\label{tab:issue-importance-vs-public} Perceived issue importance of AI governance challenges (comparing AI/ML researchers' and the US public's responses). The table presents the mean perceived importance of each of the AI governance challenges and the associated standard error and sample size for each type of respondent (AI/ML researcher and US public).}
\centering
\renewcommand{\arraystretch}{1.2}
\begin{tabular}[t]{|l|l|l|l|l|}
\hline
Governance challenge & Mean & \textit{SE} & \textit{N} & Respondent type\\
\hline
Hiring bias & 2.18 & 0.06 & 170 & AI/ML researchers\\
\hline
Criminal justice bias & 2.59 & 0.05 & 167 & AI/ML researchers\\
\hline
Disease diagnosis & 2.45 & 0.05 & 147 & AI/ML researchers\\
\hline
Data privacy & 2.55 & 0.05 & 177 & AI/ML researchers\\
\hline
Autonomous vehicles & 2.58 & 0.05 & 165 & AI/ML researchers\\
\hline
Digital manipulation & 2.47 & 0.05 & 157 & AI/ML researchers\\
\hline
Cyber attacks & 2.40 & 0.05 & 176 & AI/ML researchers\\
\hline
Surveillance & 2.53 & 0.05 & 172 & AI/ML researchers\\
\hline
US-China competition & 1.77 & 0.07 & 159 & AI/ML researchers\\
\hline
Value alignment & 2.35 & 0.06 & 156 & AI/ML researchers\\
\hline
Autonomous weapons & 2.47 & 0.06 & 157 & AI/ML researchers\\
\hline
Technological unemployment & 2.27 & 0.06 & 168 & AI/ML researchers\\
\hline
Critical AI systems failure & 2.57 & 0.05 & 179 & AI/ML researchers\\
\hline
Hiring bias & 2.54 & 0.03 & 760 & US Public\\
\hline
Criminal justice bias & 2.53 & 0.03 & 778 & US Public\\
\hline
Disease diagnosis & 2.52 & 0.03 & 767 & US Public\\
\hline
Data privacy & 2.62 & 0.03 & 807 & US Public\\
\hline
Autonomous vehicles & 2.56 & 0.02 & 796 & US Public\\
\hline
Digital manipulation & 2.53 & 0.03 & 741 & US Public\\
\hline
Cyber attacks & 2.59 & 0.02 & 745 & US Public\\
\hline
Surveillance & 2.56 & 0.03 & 784 & US Public\\
\hline
US-China competition & 2.52 & 0.03 & 766 & US Public\\
\hline
Value alignment & 2.55 & 0.03 & 783 & US Public\\
\hline
Autonomous weapons & 2.58 & 0.02 & 757 & US Public\\
\hline
Technological unemployment & 2.50 & 0.03 & 738 & US Public\\
\hline
Critical AI systems failure & 2.47 & 0.03 & 778 & US Public\\
\hline
\end{tabular}
\end{table}

\bigskip

\begin{table}[ht]
\small
\caption{\label{tab:top5-issue-importance}Top five most important AI governance challenges (AI/ML researchers versus US public). The table presents the five highest mean responses of perceived importance of the AI governance challenges ranked in descending order for AI/ML researchers and the US public.}
\centering
\renewcommand{\arraystretch}{1.2}
\begin{tabular}[t]{|l|l|l|}
\hline
Ranking & AI/ML researchers & US Public\\
\hline
1 & Criminal justice bias & Data privacy\\
\hline
2 & Autonomous vehicles & Cyber attacks\\
\hline
3 & Critical AI systems failure & Autonomous weapons\\
\hline
4 & Data privacy & Surveillance\\
\hline
5 & Surveillance & Autonomous vehicles\\
\hline
\end{tabular}
\end{table}

\clearpage

\begin{table}[ht]
\small
\caption{\label{tab:issue-importance-country}Rating issue importance of AI governance challenges (by undergraduate country). The table presents AI/ML researchers' mean perceived importance of each of the AI governance challenges and the associated standard error and sample size by country where the respondent completed their undergraduate degree (US and China).}
\centering
\renewcommand{\arraystretch}{1.2}
\begin{tabular}[t]{|l|l|l|l|l|}
\hline
Governance challenge & Mean & \textit{SE} & \textit{N} & Undergraduate country\\
\hline
Hiring bias & 2.25 & 0.12 & 36 & US\\
\hline
Criminal justice bias & 2.69 & 0.08 & 47 & US\\
\hline
Disease diagnosis & 2.33 & 0.12 & 34 & US\\
\hline
Data privacy & 2.57 & 0.10 & 42 & US\\
\hline
Autonomous vehicles & 2.62 & 0.11 & 35 & US\\
\hline
Digital manipulation & 2.48 & 0.11 & 37 & US\\
\hline
Cyber attacks & 2.34 & 0.11 & 38 & US\\
\hline
Surveillance & 2.50 & 0.12 & 30 & US\\
\hline
US-China competition & 1.90 & 0.12 & 39 & US\\
\hline
Value alignment & 2.19 & 0.13 & 31 & US\\
\hline
Autonomous weapons & 2.41 & 0.12 & 38 & US\\
\hline
Technological unemployment & 2.12 & 0.14 & 35 & US\\
\hline
Critical AI systems failure & 2.57 & 0.09 & 43 & US\\
\hline
Hiring bias & 2.34 & 0.13 & 23 & China\\
\hline
Criminal justice bias & 2.39 & 0.12 & 24 & China\\
\hline
Disease diagnosis & 2.54 & 0.11 & 26 & China\\
\hline
Data privacy & 2.50 & 0.16 & 24 & China\\
\hline
Autonomous vehicles & 2.69 & 0.09 & 31 & China\\
\hline
Digital manipulation & 2.42 & 0.16 & 20 & China\\
\hline
Cyber attacks & 2.17 & 0.15 & 23 & China\\
\hline
Surveillance & 2.10 & 0.17 & 27 & China\\
\hline
US-China competition & 2.26 & 0.18 & 17 & China\\
\hline
Value alignment & 2.53 & 0.12 & 23 & China\\
\hline
Autonomous weapons & 2.57 & 0.12 & 25 & China\\
\hline
Technological unemployment & 2.16 & 0.13 & 25 & China\\
\hline
Critical AI systems failure & 2.58 & 0.16 & 22 & China\\
\hline
\end{tabular}
\end{table}

\clearpage

\begin{table}[ht]
\small
\caption{\label{tab:issue-importance-region}Rating issue importance of AI governance challenges (by undergraduate region). The table presents AI/ML researchers' mean perceived importance of each of the AI governance challenges and the associated standard error and sample size by region where the respondent completed their undergraduate degree (Europe, North America, and Asia).}
\renewcommand{\arraystretch}{1.2}
\centering
\begin{tabular}{|l|l|l|l|l|}
\hline
Governance challenge & Mean & \textit{SE} & \textit{N} & Undergraduate region\\
\hline
Hiring bias & 2.01 & 0.15 & 43 & Europe\\
\hline
Criminal justice bias & 2.61 & 0.10 & 43 & Europe\\
\hline
Disease diagnosis & 2.25 & 0.13 & 30 & Europe\\
\hline
Data privacy & 2.39 & 0.11 & 41 & Europe\\
\hline
Autonomous vehicles & 2.39 & 0.12 & 37 & Europe\\
\hline
Digital manipulation & 2.32 & 0.12 & 41 & Europe\\
\hline
Cyber attacks & 2.28 & 0.12 & 41 & Europe\\
\hline
Surveillance & 2.81 & 0.06 & 42 & Europe\\
\hline
US-China competition & 1.59 & 0.12 & 42 & Europe\\
\hline
Value alignment & 2.23 & 0.15 & 42 & Europe\\
\hline
Autonomous weapons & 2.62 & 0.10 & 32 & Europe\\
\hline
Technological unemployment & 2.26 & 0.11 & 35 & Europe\\
\hline
Critical AI systems failure & 2.50 & 0.09 & 41 & Europe\\
\hline
Hiring bias & 2.27 & 0.11 & 40 & North America\\
\hline
Criminal justice bias & 2.70 & 0.08 & 52 & North America\\
\hline
Disease diagnosis & 2.32 & 0.11 & 40 & North America\\
\hline
Data privacy & 2.57 & 0.09 & 47 & North America\\
\hline
Autonomous vehicles & 2.61 & 0.10 & 38 & North America\\
\hline
Digital manipulation & 2.48 & 0.10 & 40 & North America\\
\hline
Cyber attacks & 2.35 & 0.11 & 45 & North America\\
\hline
Surveillance & 2.52 & 0.12 & 31 & North America\\
\hline
US-China competition & 1.86 & 0.12 & 44 & North America\\
\hline
Value alignment & 2.18 & 0.12 & 36 & North America\\
\hline
Autonomous weapons & 2.45 & 0.12 & 41 & North America\\
\hline
Technological unemployment & 2.13 & 0.14 & 40 & North America\\
\hline
Critical AI systems failure & 2.55 & 0.09 & 46 & North America\\
\hline
Hiring bias & 2.28 & 0.09 & 55 & Asia\\
\hline
Criminal justice bias & 2.51 & 0.08 & 44 & Asia\\
\hline
Disease diagnosis & 2.66 & 0.07 & 49 & Asia\\
\hline
Data privacy & 2.52 & 0.09 & 53 & Asia\\
\hline
Autonomous vehicles & 2.67 & 0.07 & 67 & Asia\\
\hline
Digital manipulation & 2.54 & 0.09 & 56 & Asia\\
\hline
Cyber attacks & 2.43 & 0.09 & 60 & Asia\\
\hline
Surveillance & 2.39 & 0.09 & 63 & Asia\\
\hline
US-China competition & 1.84 & 0.13 & 47 & Asia\\
\hline
Value alignment & 2.51 & 0.08 & 55 & Asia\\
\hline
Autonomous weapons & 2.40 & 0.10 & 57 & Asia\\
\hline
Technological unemployment & 2.32 & 0.09 & 68 & Asia\\
\hline
Critical AI systems failure & 2.65 & 0.08 & 56 & Asia\\
\hline
\end{tabular}
\end{table}

\clearpage

\begin{table}[ht]
\small
\caption{\label{tab:issue-importance-workplace}Rating issue importance of AI governance challenges (by workplace type). The table presents AI/ML researchers' mean perceived importance of each of the AI governance challenges and the associated standard error and sample size by workplace type (academic and industry).}
\centering
\renewcommand{\arraystretch}{1.2}
\begin{tabular}[t]{|l|l|l|l|l|}
\hline
Governance challenge & Mean & \textit{SE} & \textit{N} & Workplace type\\
\hline
Hiring bias & 2.17 & 0.07 & 131 & Academic\\
\hline
Criminal justice bias & 2.59 & 0.05 & 124 & Academic\\
\hline
Disease diagnosis & 2.44 & 0.06 & 113 & Academic\\
\hline
Data privacy & 2.50 & 0.06 & 125 & Academic\\
\hline
Autonomous vehicles & 2.60 & 0.05 & 131 & Academic\\
\hline
Digital manipulation & 2.45 & 0.07 & 113 & Academic\\
\hline
Cyber attacks & 2.32 & 0.06 & 132 & Academic\\
\hline
Surveillance & 2.47 & 0.06 & 127 & Academic\\
\hline
US-China competition & 1.82 & 0.07 & 123 & Academic\\
\hline
Value alignment & 2.36 & 0.07 & 111 & Academic\\
\hline
Autonomous weapons & 2.47 & 0.07 & 117 & Academic\\
\hline
Technological unemployment & 2.27 & 0.07 & 128 & Academic\\
\hline
Critical AI systems failure & 2.57 & 0.05 & 135 & Academic\\
\hline
Hiring bias & 2.20 & 0.12 & 42 & Industry\\
\hline
Criminal justice bias & 2.53 & 0.10 & 43 & Industry\\
\hline
Disease diagnosis & 2.50 & 0.09 & 42 & Industry\\
\hline
Data privacy & 2.71 & 0.05 & 53 & Industry\\
\hline
Autonomous vehicles & 2.45 & 0.14 & 35 & Industry\\
\hline
Digital manipulation & 2.48 & 0.09 & 46 & Industry\\
\hline
Cyber attacks & 2.50 & 0.08 & 48 & Industry\\
\hline
Surveillance & 2.67 & 0.08 & 52 & Industry\\
\hline
US-China competition & 1.64 & 0.13 & 40 & Industry\\
\hline
Value alignment & 2.38 & 0.10 & 51 & Industry\\
\hline
Autonomous weapons & 2.49 & 0.10 & 42 & Industry\\
\hline
Technological unemployment & 2.24 & 0.11 & 45 & Industry\\
\hline
Critical AI systems failure & 2.53 & 0.08 & 51 & Industry\\
\hline
\end{tabular}
\end{table}

\clearpage

\begin{table}[ht] \centering 
\small
  \caption{Association between perceived issue importance and different AI governance challenges. Output from the multiple linear regression used to compare differences in perceived issue importance between AI governance challenges. We regress perceived issue importance on all of the AI governance challenges. The (arbitrarily-chosen) reference category, the one that is excluded from the list of coefficients, is critical AI system failure. We clustered the standard errors by survey respondent because each respondent was presented with five AI governance challenges randomly chosen from the list of 13. The \textit{F}-test of overall significance rejects the null hypothesis that respondents perceive all the governance challenges to have equal issue importance. The Holm method was used to control the family-wise error rate.
  \label{tab:regression-issue-importance} }
\begin{tabular}{@{\extracolsep{3em}}ll} 
\\[-1.8ex]\hline &\hspace{-2em} Coefficient (\textit{SE}) \\
\hline \\[-1.8ex] 
 (Intercept) & 2.579$^{***}$ \\ 
  & (0.051) \\ 
  Autonomous weapons & $-$0.104 \\ 
  & (0.077) \\ 
  Criminal justice bias & 0.012 \\ 
  & (0.067) \\ 
  Critical AI systems failure & $-$0.010 \\ 
  & (0.065) \\ 
  Cyber attacks & $-$0.175 \\ 
  & (0.071) \\ 
  Data privacy & $-$0.033 \\ 
  & (0.066) \\ 
  Digital manipulation & $-$0.113 \\ 
  & (0.072) \\ 
  Disease diagnosis & $-$0.124 \\ 
  & (0.070) \\ 
  Hiring bias & $-$0.400$^{***}$ \\ 
  & (0.077) \\ 
  Surveillance & $-$0.049 \\ 
  & (0.071) \\ 
  Technological unemployment & $-$0.308$^{***}$ \\ 
  & (0.076) \\ 
  U.S.-China competition & $-$0.804$^{***}$ \\ 
  & (0.082) \\ 
  Value alignment & $-$0.224$^{*}$ \\ 
  & (0.075) \\ 
  \hline
 \textit{N} &  \hspace{-3em}2,150 responses; 430 unique respondents \\ 
\textit{F}-Statistic & \hspace{-3em}13.93$^{***}$ (\textit{df} = 12; 429) \\ 
\hline \\[-1.8ex] 
\multicolumn{2}{l}{$^{*}$p $<$ .05; $^{**}$p $<$ .01; $^{***}$p $<$ .001} \\ 
\end{tabular} 
\end{table} 


\begin{table}[ht] \centering 
\small
  \caption{Association between perceived issue importance and demographic variables controlling for issues. The table presents the output from the multiple linear regression used to compare differences in perceived issue importance between demographic subgroups whilst controlling for the different types of AI governance issues. We regressed perceived issue importance on the categorical variables of gender (female/other, male, and prefer not to say or missing response), location of undergraduate education (Europe, US, China, other, and missing response), location of work/study (Europe, US, China, and other), and type of workplace (industry and academic). The (arbitrarily-chosen) reference categories, the ones that are excluded from the list of coefficients, are female/other for gender, and China for location of undergraduate education and place of work/study. The \textit{F}-test of overall significance rejects the null hypothesis that respondents do not differ in their perceived importance of AI governance challenges (when controlling for issue) depending on demographic subgroups. We cluster the standard errors by respondents because each respondent was presented with five AI governance challenges. The Holm method was used to control the family-wise error rate.} 
   
  \label{tab:corr-gov-challenge-demo} 
\begin{tabular}{@{\extracolsep{3em}}ll} 
\\[-1.8ex]\hline  &\hspace{-2em} Coefficient (\textit{SE}) \\ 
\hline \\[-1.8ex] 
 (Intercept) & 3.058$^{***}$ \\ 
  & (0.135) \\ 
  Gender: male & $-$0.283$^{***}$ \\ 
  & (0.055) \\ 
  Gender: prefer not to say/NA & $-$0.311 \\ 
  & (0.172) \\ 
  Place of undergraduate degree: Europe & $-$0.005 \\ 
  & (0.091) \\ 
  Place of undergraduate degree: missing & 0.105 \\ 
  & (0.089) \\ 
  Place of undergraduate degree: other & 0.155 \\ 
  & (0.079) \\ 
  Place of undergraduate degree: US & 0.032 \\ 
  & (0.079) \\ 
  Place of work: Europe & $-$0.230 \\ 
  & (0.106) \\ 
  Place of work: other & $-$0.269 \\ 
  & (0.114) \\ 
  Place of work: US & $-$0.201 \\ 
  & (0.094) \\ 
  Job: industry & $-$0.021 \\ 
  & (0.071) \\ 
  Job: academic & $-$0.074 \\ 
  & (0.072) \\ 
  \hline
 \textit{N} &\hspace{-3em} 2,150 responses, 430 unique respondents \\ 
\textit{F}-Statistic &\hspace{-3em} 8.935$^{***}$ (\textit{df} = 23; 429) \\ 
\hline \\[-1.8ex] 
\multicolumn{2}{l}{$^{*}$p $<$ .05; $^{**}$p $<$ .01; $^{***}$p $<$ .001} \\ 
\end{tabular} 
\end{table} 

\clearpage

\subsubsection*{Trust in actors to shape the development and use of AI in the public interest}

{\renewcommand{\arraystretch}{1.3}\small
\begin{longtable}{|>{\raggedright\arraybackslash}m{2.9cm}|>{\centering\arraybackslash}m{1.5cm}|c|c|c|>{\centering\arraybackslash}m{3cm}|c|}
\caption{Trust in actors to shape the development and use of AI in the public interest: comparing AI/ML researchers' and the US public's responses. This table contains the data used to generate Fig. 2. AI/ML researchers were shown five randomly-selected actors and asked to evaluate how much they trust the actors using a four-point scale: 0 = no trust at all, 1 = not too much trust, 2 = a fair amount of trust, 3 = a great deal of trust. For the AI/ML researchers survey, the ``Tech companies'' result is the mean response across all corporate actors presented to respondents. The AI/ML researchers' responses to the US military and the Chinese military are denoted with $*$ because those two actors were shown only to those who do research in the US or China. These respondents had an equal probability of being shown the US military or the Chinese military. In the public opinion survey, respondents were asked about their confidence in the actors to develop AI (question type A) or manage the development and use of AI (question type B) in the best interest of the public using a similar four-point scale. For question type C, both questions were asked; we averaged the responses to these two questions for each of these actors for clarity. For ``US intelligence agencies,'' we averaged across responses to the NSA, the FBI, and the CIA. Table \ref{tab:trust-vs-public} contains the detailed breakdowns by actor and respondent type.}
\label{tab:figure2astable}\\\hline
Actor & Question type & Mean & SE & \textit{N} & Subgroup & Actor type\\
\hline
\endfirsthead
\multicolumn{7}{c}%
{\tablename\ \thetable\ -- \textit{Continued from previous page}} \\
\hline
Actor & Question type & Mean & SE & \textit{N} & Subgroup & Actor type\\
\hline
\endhead
\hline \multicolumn{7}{r}{\textit{Continued on next page}} \\
\endfoot
\hline
\endlastfoot
Government of the country where they do research &  & 1.33 & 0.08 & 100 & AI/ML researchers & National\\
\hline
Military of the country where they do research &  & 0.83 & 0.08 & 106 & AI/ML researchers & National\\
\hline
US government &  & 0.94 & 0.07 & 113 & AI/ML researchers & National\\
\hline
Chinese government &  & 0.38 & 0.06 & 115 & AI/ML researchers & National\\
\hline
US military* &  & 0.73 & 0.10 & 56 & AI/ML researchers (does research in the US only) & National\\
\hline
Chinese military* &  & 0.30 & 0.08 & 60 & AI/ML researchers (does research in the US only) & National\\
\hline
UN &  & 1.74 & 0.08 & 122 & AI/ML researchers & International\\
\hline
EU &  & 1.98 & 0.06 & 114 & AI/ML researchers & International\\
\hline
Intergovernmental research organizations (e.g., CERN) &  & 2.08 & 0.06 & 114 & AI/ML researchers & International\\
\hline
Tech companies &  & 0.98 & 0.03 & 1171 & AI/ML researchers & Corporate\\
\hline
Google &  & 1.35 & 0.07 & 131 & AI/ML researchers & Corporate\\
\hline
Facebook &  & 0.58 & 0.06 & 104 & AI/ML researchers & Corporate\\
\hline
Apple &  & 1.05 & 0.07 & 117 & AI/ML researchers & Corporate\\
\hline
Microsoft &  & 1.43 & 0.07 & 118 & AI/ML researchers & Corporate\\
\hline
Amazon &  & 0.88 & 0.06 & 108 & AI/ML researchers & Corporate\\
\hline
OpenAI &  & 1.50 & 0.08 & 113 & AI/ML researchers & Corporate\\
\hline
DeepMind &  & 1.37 & 0.09 & 99 & AI/ML researchers & Corporate\\
\hline
Tencent &  & 0.69 & 0.06 & 116 & AI/ML researchers & Corporate\\
\hline
Baidu &  & 0.50 & 0.05 & 134 & AI/ML researchers & Corporate\\
\hline
Alibaba &  & 0.57 & 0.05 & 131 & AI/ML researchers & Corporate\\
\hline
Non-government scientific organization (e.g., AAAI) &  & 2.12 & 0.06 & 104 & AI/ML researchers & Other\\
\hline
Partnership on AI &  & 1.89 & 0.06 & 103 & AI/ML researchers & Other\\
\hline
US government & B & 1.05 & 0.04 & 743 & US public & National\\
\hline
US military* & A & 1.56 & 0.04 & 638 & US public & National\\
\hline
US intelligence agencies & A & 1.16 & 0.03 & 2096 & US public & National\\
\hline
UN & B & 1.06 & 0.03 & 802 & US public & International\\
\hline
Intergovernmental research organizations (e.g., CERN) & C & 1.30 & 0.03 & 1392 & US public & International\\
\hline
NATO & A & 1.17 & 0.03 & 695 & US public & International\\
\hline
International organizations & B & 1.10 & 0.03 & 827 & US public & International\\
\hline
Tech companies & C & 1.33 & 0.03 & 1432 & US public & Corporate\\
\hline
Google & C & 1.20 & 0.03 & 1412 & US public & Corporate\\
\hline
Facebook & C & 0.78 & 0.03 & 1373 & US public & Corporate\\
\hline
Apple & C & 1.19 & 0.03 & 1367 & US public & Corporate\\
\hline
Microsoft & C & 1.26 & 0.03 & 1368 & US public & Corporate\\
\hline
Amazon & C & 1.22 & 0.03 & 1469 & US public & Corporate\\
\hline
Non-government scientific organization (e.g., AAAI) & B & 1.35 & 0.03 & 792 & US public & Other\\
\hline
Partnership on AI & B & 1.35 & 0.03 & 780 & US public & Other\\
\hline
University researchers & A & 1.56 & 0.03 & 666 & US public & Other\\
\hline
\end{longtable}}

\clearpage

{\renewcommand{\arraystretch}{1.3}\small
\begin{longtable}{|>{\raggedright\arraybackslash}m{3.5cm}|c|c|c|c|c|}
\caption{\label{tab:trust-vs-public} Trust in actors to shape the development and use of AI in the public interest: by respondent type. The table presents the mean trust in different actors and the associated standard error and sample size by type of actor (national government, international, corporate, and other) and type of respondent (AI/ML researchers and US public).}\\\hline
Actor & Mean & \textit{SE} & \textit{N} & Actor type & Respondent type\\
\hline
\endfirsthead
\multicolumn{6}{c}%
{\tablename\ \thetable\ -- \textit{Continued from previous page}} \\
\hline
Actor & Mean & \textit{SE} & \textit{N} & Actor type & Respondent type\\
\hline
\endhead
\hline \multicolumn{6}{r}{\textit{Continued on next page}} \\
\endfoot
\hline
\endlastfoot
US government & 0.94 & 0.07 & 113 & National & AI/ML researchers\\
\hline
Chinese government & 0.38 & 0.06 & 115 & National & AI/ML researchers\\
\hline
Government of the country where they do research & 1.33 & 0.08 & 100 & National & AI/ML researchers\\
\hline
Military of the country where they do research & 0.83 & 0.08 & 106 & National & AI/ML researchers\\
\hline
US military & 0.68 & 0.09 & 60 & National & AI/ML researchers\\
\hline
Chinese military & 0.38 & 0.09 & 66 & National & AI/ML researchers\\
\hline
UN & 1.74 & 0.08 & 122 & International & AI/ML researchers\\
\hline
EU & 1.98 & 0.06 & 114 & International & AI/ML researchers\\
\hline
Intergovernmental research organizations (e.g., CERN) & 2.08 & 0.06 & 114 & International & AI/ML researchers\\
\hline
Google & 1.35 & 0.07 & 131 & Corporate & AI/ML researchers\\
\hline
Facebook & 0.58 & 0.06 & 104 & Corporate & AI/ML researchers\\
\hline
Apple & 1.05 & 0.07 & 117 & Corporate & AI/ML researchers\\
\hline
Microsoft & 1.43 & 0.07 & 118 & Corporate & AI/ML researchers\\
\hline
Amazon & 0.88 & 0.06 & 108 & Corporate & AI/ML researchers\\
\hline
OpenAI & 1.50 & 0.08 & 113 & Corporate & AI/ML researchers\\
\hline
DeepMind & 1.37 & 0.09 & 99 & Corporate & AI/ML researchers\\
\hline
Tencent & 0.69 & 0.06 & 116 & Corporate & AI/ML researchers\\
\hline
Baidu & 0.50 & 0.05 & 134 & Corporate & AI/ML researchers\\
\hline
Alibaba & 0.57 & 0.05 & 131 & Corporate & AI/ML researchers\\
\hline
Non-government scientific organization (e.g., AAAI) & 2.12 & 0.06 & 104 & Other & AI/ML researchers\\
\hline
Partnership on AI & 1.89 & 0.06 & 103 & Other & AI/ML researchers\\
\hline
Amazon & 1.33 & 0.04 & 685 & Corporate & Public\\
\hline
Apple & 1.29 & 0.04 & 697 & Corporate & Public\\
\hline
CIA & 1.21 & 0.04 & 730 & National & Public\\
\hline
Facebook & 0.85 & 0.04 & 632 & Corporate & Public\\
\hline
FBI & 1.21 & 0.04 & 656 & National & Public\\
\hline
Google & 1.34 & 0.04 & 645 & Corporate & Public\\
\hline
Intergovernmental research organizations (e.g., CERN) & 1.42 & 0.04 & 645 & International & Public\\
\hline
Microsoft & 1.40 & 0.04 & 597 & Corporate & Public\\
\hline
NATO & 1.17 & 0.03 & 695 & International & Public\\
\hline
Non-profit (e.g., OpenAI) & 1.44 & 0.03 & 659 & Other & Public\\
\hline
NSA & 1.28 & 0.04 & 710 & National & Public\\
\hline
Tech companies & 1.44 & 0.03 & 674 & Corporate & Public\\
\hline
US civilian government & 1.16 & 0.03 & 671 & National & Public\\
\hline
US military & 1.56 & 0.04 & 638 & National & Public\\
\hline
University researchers & 1.56 & 0.03 & 666 & Other & Public\\
\hline
Amazon & 1.24 & 0.03 & 784 & Corporate & Public\\
\hline
Apple & 1.20 & 0.03 & 775 & Corporate & Public\\
\hline
Facebook & 0.91 & 0.03 & 741 & Corporate & Public\\
\hline
Google & 1.20 & 0.03 & 767 & Corporate & Public\\
\hline
Intergovernmental research organizations (e.g., CERN) & 1.27 & 0.03 & 747 & International & Public\\
\hline
International organizations & 1.10 & 0.03 & 827 & International & Public\\
\hline
Microsoft & 1.24 & 0.03 & 771 & Corporate & Public\\
\hline
Non-government scientific organization (e.g., AAAI) & 1.35 & 0.03 & 792 & Other & Public\\
\hline
Partnership on AI & 1.35 & 0.03 & 780 & Other & Public\\
\hline
Tech companies & 1.33 & 0.03 & 758 & Corporate & Public\\
\hline
US federal government & 1.05 & 0.04 & 743 & National & Public\\
\hline
US state governments & 1.05 & 0.03 & 713 & National & Public\\
\hline
UN & 1.06 & 0.03 & 802 & International & Public\\
\hline
\end{longtable}}

\clearpage

{\renewcommand{\arraystretch}{1.3}\small
\begin{longtable}{|>{\raggedright\arraybackslash}m{3.5cm}|c|c|c|>{\centering\arraybackslash}m{2.2cm}|c|}
\caption{Trust in actors to shape the development and use of AI in the public interest (by undergraduate country). The table presents the mean trust in different actors and the associated standard error and sample size by country of undergraduate education (US and China) and type of actor (national government, international, corporate, and other).}\label{tab:trust-country}\\\hline
Actor & Mean & \textit{SE} & \textit{N} & Undergraduate country & Actor type\\
\hline
\endfirsthead
\multicolumn{6}{c}%
{\tablename\ \thetable\ -- \textit{Continued from previous page}} \\
\hline
Actor & Mean & \textit{SE} & \textit{N} & Undergraduate country & Actor type\\
\hline
\endhead
\hline \multicolumn{6}{r}{\textit{Continued on next page}} \\
\endfoot
\hline
\endlastfoot
US government & 1.08 & 0.14 & 23 & US & National\\
\hline
Chinese government & 0.14 & 0.06 & 27 & US & National\\
\hline
Government of the country where they do research & 1.14 & 0.16 & 21 & US & National\\
\hline
Military of the country where they do research & 0.85 & 0.17 & 21 & US & National\\
\hline
US military & 0.92 & 0.20 & 17 & US & National\\
\hline
Chinese military & 0.27 & 0.14 & 24 & US & National\\
\hline
UN & 1.73 & 0.21 & 22 & US & International\\
\hline
EU & 1.64 & 0.16 & 25 & US & International\\
\hline
Intergovernmental research organizations (e.g., CERN) & 2.13 & 0.13 & 25 & US & International\\
\hline
Google & 1.41 & 0.12 & 32 & US & Corporate\\
\hline
Facebook & 0.55 & 0.12 & 23 & US & Corporate\\
\hline
Apple & 1.00 & 0.11 & 31 & US & Corporate\\
\hline
Microsoft & 1.38 & 0.13 & 30 & US & Corporate\\
\hline
Amazon & 0.90 & 0.15 & 22 & US & Corporate\\
\hline
OpenAI & 1.44 & 0.14 & 24 & US & Corporate\\
\hline
DeepMind & 1.31 & 0.17 & 27 & US & Corporate\\
\hline
Tencent & 0.54 & 0.09 & 23 & US & Corporate\\
\hline
Baidu & 0.44 & 0.09 & 32 & US & Corporate\\
\hline
Alibaba & 0.41 & 0.08 & 32 & US & Corporate\\
\hline
Non-government scientific organization (e.g., AAAI) & 1.92 & 0.15 & 23 & US & Other\\
\hline
Partnership on AI & 2.13 & 0.12 & 22 & US & Other\\
\hline
US government & 1.00 & 0.19 & 19 & China & National\\
\hline
Chinese government & 1.30 & 0.16 & 14 & China & National\\
\hline
Government of the country where they do research & 1.36 & 0.21 & 15 & China & National\\
\hline
Military of the country where they do research & 0.97 & 0.19 & 19 & China & National\\
\hline
US military & 0.56 & 0.19 & 15 & China & National\\
\hline
Chinese military & 1.01 & 0.22 & 16 & China & National\\
\hline
UN & 1.85 & 0.17 & 19 & China & International\\
\hline
EU & 2.06 & 0.14 & 16 & China & International\\
\hline
Intergovernmental research organizations (e.g., CERN) & 2.09 & 0.08 & 12 & China & International\\
\hline
Google & 1.71 & 0.21 & 17 & China & Corporate\\
\hline
Facebook & 1.10 & 0.16 & 16 & China & Corporate\\
\hline
Apple & 1.67 & 0.19 & 15 & China & Corporate\\
\hline
Microsoft & 1.90 & 0.17 & 16 & China & Corporate\\
\hline
Amazon & 1.06 & 0.17 & 16 & China & Corporate\\
\hline
OpenAI & 1.81 & 0.17 & 21 & China & Corporate\\
\hline
DeepMind & 1.60 & 0.23 & 9 & China & Corporate\\
\hline
Tencent & 1.59 & 0.17 & 17 & China & Corporate\\
\hline
Baidu & 1.00 & 0.19 & 19 & China & Corporate\\
\hline
Alibaba & 1.31 & 0.20 & 18 & China & Corporate\\
\hline
Non-government scientific organization (e.g., AAAI) & 2.21 & 0.14 & 15 & China & Other\\
\hline
Partnership on AI & 1.80 & 0.14 & 22 & China & Other\\
\hline
\end{longtable}}

\clearpage

{\renewcommand{\arraystretch}{1.3}\small
\begin{longtable}{|>{\raggedright\arraybackslash}m{3.5cm}|c|c|c|>{\centering\arraybackslash}m{2.2cm}|c|}
\caption{\label{tab:trust-region}Trust in actors to shape the development and use of AI in the public interest (by undergraduate region). The table presents the mean trust in different actors and the associated standard error and sample size by region of undergraduate education (Europe, North America, and Asia) and type of actor (national government, international, corporate, and other).}\\\hline
Actor & Mean & \textit{SE} & \textit{N} & Undergraduate region & Actor type\\
\hline
\endfirsthead
\multicolumn{6}{c}%
{\tablename\ \thetable\ -- \textit{Continued from previous page}} \\
\hline
Actor & Mean & \textit{SE} & \textit{N} & Undergraduate region & Actor type\\
\hline
\endhead
\hline \multicolumn{6}{r}{\textit{Continued on next page}} \\
\endfoot
\hline
\endlastfoot
US government & 0.87 & 0.11 & 33 & Europe & National\\
\hline
Chinese government & 0.25 & 0.09 & 31 & Europe & National\\
\hline
Government of the country where they do research & 1.64 & 0.16 & 22 & Europe & National\\
\hline
Military of the country where they do research & 0.88 & 0.17 & 24 & Europe & National\\
\hline
US military & 0.71 & 0.29 & 7 & Europe & National\\
\hline
Chinese military & 0.23 & 0.17 & 6 & Europe & National\\
\hline
UN & 1.81 & 0.14 & 33 & Europe & International\\
\hline
EU & 1.96 & 0.11 & 28 & Europe & International\\
\hline
Intergovernmental research organizations (e.g., CERN) & 2.22 & 0.11 & 30 & Europe & International\\
\hline
Google & 1.12 & 0.16 & 29 & Europe & Corporate\\
\hline
Facebook & 0.62 & 0.14 & 22 & Europe & Corporate\\
\hline
Apple & 0.76 & 0.14 & 24 & Europe & Corporate\\
\hline
Microsoft & 1.30 & 0.17 & 24 & Europe & Corporate\\
\hline
Amazon & 0.94 & 0.11 & 27 & Europe & Corporate\\
\hline
OpenAI & 1.36 & 0.16 & 32 & Europe & Corporate\\
\hline
DeepMind & 1.47 & 0.16 & 25 & Europe & Corporate\\
\hline
Tencent & 0.54 & 0.07 & 30 & Europe & Corporate\\
\hline
Baidu & 0.33 & 0.09 & 29 & Europe & Corporate\\
\hline
Alibaba & 0.41 & 0.08 & 37 & Europe & Corporate\\
\hline
Non-government scientific organization (e.g., AAAI) & 2.23 & 0.13 & 22 & Europe & Other\\
\hline
Partnership on AI & 1.80 & 0.14 & 16 & Europe & Other\\
\hline
US government & 1.07 & 0.14 & 26 & North America & National\\
\hline
Chinese government & 0.17 & 0.07 & 28 & North America & National\\
\hline
Government of the country where they do research & 1.17 & 0.14 & 24 & North America & National\\
\hline
Military of the country where they do research & 0.83 & 0.16 & 24 & North America & National\\
\hline
US military & 0.93 & 0.18 & 19 & North America & National\\
\hline
Chinese military & 0.24 & 0.12 & 27 & North America & National\\
\hline
UN & 1.68 & 0.19 & 25 & North America & International\\
\hline
EU & 1.68 & 0.15 & 28 & North America & International\\
\hline
Intergovernmental research organizations (e.g., CERN) & 2.14 & 0.11 & 30 & North America & International\\
\hline
Google & 1.43 & 0.12 & 33 & North America & Corporate\\
\hline
Facebook & 0.60 & 0.12 & 26 & North America & Corporate\\
\hline
Apple & 0.95 & 0.11 & 34 & North America & Corporate\\
\hline
Microsoft & 1.36 & 0.12 & 35 & North America & Corporate\\
\hline
Amazon & 0.92 & 0.12 & 27 & North America & Corporate\\
\hline
OpenAI & 1.44 & 0.14 & 25 & North America & Corporate\\
\hline
DeepMind & 1.33 & 0.16 & 28 & North America & Corporate\\
\hline
Tencent & 0.52 & 0.08 & 26 & North America & Corporate\\
\hline
Baidu & 0.40 & 0.08 & 35 & North America & Corporate\\
\hline
Alibaba & 0.40 & 0.08 & 35 & North America & Corporate\\
\hline
Non-government scientific organization (e.g., AAAI) & 1.93 & 0.14 & 26 & North America & Other\\
\hline
Partnership on AI & 2.13 & 0.11 & 23 & North America & Other\\
\hline
US government & 0.99 & 0.14 & 39 & Asia & National\\
\hline
Chinese government & 0.61 & 0.12 & 37 & Asia & National\\
\hline
Government of the country where they do research & 1.30 & 0.14 & 39 & Asia & National\\
\hline
Military of the country where they do research & 0.84 & 0.12 & 41 & Asia & National\\
\hline
US military & 0.59 & 0.14 & 26 & Asia & National\\
\hline
Chinese military & 0.72 & 0.19 & 23 & Asia & National\\
\hline
UN & 1.58 & 0.14 & 42 & Asia & International\\
\hline
EU & 2.15 & 0.12 & 34 & Asia & International\\
\hline
Intergovernmental research organizations (e.g., CERN) & 1.89 & 0.12 & 33 & Asia & International\\
\hline
Google & 1.61 & 0.13 & 43 & Asia & Corporate\\
\hline
Facebook & 0.70 & 0.11 & 37 & Asia & Corporate\\
\hline
Apple & 1.34 & 0.13 & 39 & Asia & Corporate\\
\hline
Microsoft & 1.64 & 0.13 & 38 & Asia & Corporate\\
\hline
Amazon & 0.94 & 0.10 & 38 & Asia & Corporate\\
\hline
OpenAI & 1.76 & 0.13 & 39 & Asia & Corporate\\
\hline
DeepMind & 1.31 & 0.18 & 28 & Asia & Corporate\\
\hline
Tencent & 1.04 & 0.13 & 40 & Asia & Corporate\\
\hline
Baidu & 0.65 & 0.11 & 46 & Asia & Corporate\\
\hline
Alibaba & 0.81 & 0.12 & 42 & Asia & Corporate\\
\hline
Non-government scientific organization (e.g., AAAI) & 2.23 & 0.10 & 37 & Asia & Other\\
\hline
Partnership on AI & 1.83 & 0.10 & 44 & Asia & Other\\
\hline
\end{longtable}}

\clearpage

{\renewcommand{\arraystretch}{1.3}\small
\begin{longtable}{|>{\raggedright\arraybackslash}m{3.5cm}|c|c|c|>{\centering\arraybackslash}m{2.2cm}|c|}
\caption{\label{tab:trust-workplace}Trust in actors to shape the development and use of AI in the public interest (by workplace and actor). The table presents the mean trust in different actors and the associated standard error and sample size by type of workplace (academic and industry) and type of actor (national government, international, corporate, and other).}\\\hline
Actor & Mean & \textit{SE} & \textit{N} & Workplace type & Actor type\\
\hline
\endfirsthead
\multicolumn{6}{c}%
{\tablename\ \thetable\ -- \textit{Continued from previous page}} \\
\hline
Actor & Mean & \textit{SE} & \textit{N} & Workplace type & Actor type\\
\hline
\endhead
\hline \multicolumn{6}{r}{\textit{Continued on next page}} \\
\endfoot
\hline
\endlastfoot
US government & 0.96 & 0.08 & 89 & Academic & National\\
\hline
Chinese government & 0.34 & 0.07 & 84 & Academic & National\\
\hline
Government of the country where they do research & 1.30 & 0.10 & 72 & Academic & National\\
\hline
Military of the country where they do research & 0.80 & 0.09 & 74 & Academic & National\\
\hline
US military & 0.54 & 0.10 & 39 & Academic & National\\
\hline
Chinese military & 0.49 & 0.11 & 49 & Academic & National\\
\hline
UN & 1.77 & 0.09 & 92 & Academic & International\\
\hline
EU & 2.06 & 0.07 & 88 & Academic & International\\
\hline
Intergovernmental research organizations (e.g., CERN) & 2.07 & 0.08 & 82 & Academic & International\\
\hline
Google & 1.40 & 0.08 & 97 & Academic & Corporate\\
\hline
Facebook & 0.56 & 0.07 & 80 & Academic & Corporate\\
\hline
Apple & 1.18 & 0.08 & 82 & Academic & Corporate\\
\hline
Microsoft & 1.41 & 0.09 & 89 & Academic & Corporate\\
\hline
Amazon & 0.92 & 0.07 & 81 & Academic & Corporate\\
\hline
OpenAI & 1.55 & 0.09 & 92 & Academic & Corporate\\
\hline
DeepMind & 1.41 & 0.10 & 68 & Academic & Corporate\\
\hline
Tencent & 0.73 & 0.07 & 94 & Academic & Corporate\\
\hline
Baidu & 0.54 & 0.06 & 99 & Academic & Corporate\\
\hline
Alibaba & 0.59 & 0.06 & 98 & Academic & Corporate\\
\hline
Non-government scientific organization (e.g., AAAI) & 2.17 & 0.07 & 80 & Academic & Other\\
\hline
Partnership on AI & 1.87 & 0.08 & 74 & Academic & Other\\
\hline
US government & 0.80 & 0.12 & 26 & Industry & National\\
\hline
Chinese government & 0.41 & 0.11 & 33 & Industry & National\\
\hline
Government of the country where they do research & 1.12 & 0.14 & 28 & Industry & National\\
\hline
Military of the country where they do research & 0.91 & 0.15 & 34 & Industry & National\\
\hline
US military & 1.00 & 0.19 & 19 & Industry & National\\
\hline
Chinese military & 0.30 & 0.17 & 19 & Industry & National\\
\hline
UN & 1.66 & 0.13 & 37 & Industry & International\\
\hline
EU & 1.88 & 0.14 & 33 & Industry & International\\
\hline
Intergovernmental research organizations (e.g., CERN) & 2.03 & 0.10 & 39 & Industry & International\\
\hline
Google & 1.25 & 0.13 & 37 & Industry & Corporate\\
\hline
Facebook & 0.74 & 0.11 & 29 & Industry & Corporate\\
\hline
Apple & 0.87 & 0.13 & 38 & Industry & Corporate\\
\hline
Microsoft & 1.54 & 0.12 & 32 & Industry & Corporate\\
\hline
Amazon & 0.63 & 0.09 & 27 & Industry & Corporate\\
\hline
OpenAI & 1.21 & 0.18 & 19 & Industry & Corporate\\
\hline
DeepMind & 1.39 & 0.16 & 30 & Industry & Corporate\\
\hline
Tencent & 0.55 & 0.09 & 24 & Industry & Corporate\\
\hline
Baidu & 0.42 & 0.09 & 36 & Industry & Corporate\\
\hline
Alibaba & 0.62 & 0.10 & 36 & Industry & Corporate\\
\hline
Non-government scientific organization (e.g., AAAI) & 1.77 & 0.12 & 25 & Industry & Other\\
\hline
Partnership on AI & 1.93 & 0.09 & 33 & Industry & Other\\
\hline
\end{longtable}}

\clearpage

\begin{table}[ht]
\small
\caption{\label{tab:trust-works-us}Trust in actors to shape the development and use of AI in the public interest (respondents who spend most of their time doing research in the US). The table presents the mean trust in different actors and the associated standard error and sample size by type of actor (national government, international, corporate, and other) for respondents who work in the US.}
\centering
\renewcommand{\arraystretch}{1.3}
\begin{tabular}[t]{|>{\raggedright\arraybackslash}m{3.5cm}|c|c|c|>{\centering\arraybackslash}m{2.2cm}|c|}
\hline
Actor & Mean & \textit{SE} & \textit{N} & Country where they do research & Actor type\\
\hline
US government & 1.11 & 0.11 & 55 & US & National\\
\hline
Chinese government & 0.45 & 0.11 & 45 & US & National\\
\hline
US military & 0.73 & 0.10 & 56 & US & National\\
\hline
Chinese military & 0.30 & 0.08 & 60 & US & National\\
\hline
UN & 1.77 & 0.10 & 52 & US & International\\
\hline
EU & 1.91 & 0.10 & 58 & US & International\\
\hline
Intergovernmental research organizations (e.g., CERN) & 1.96 & 0.09 & 47 & US & International\\
\hline
Google & 1.36 & 0.10 & 62 & US & Corporate\\
\hline
Facebook & 0.56 & 0.09 & 50 & US & Corporate\\
\hline
Apple & 1.18 & 0.13 & 44 & US & Corporate\\
\hline
Microsoft & 1.56 & 0.11 & 63 & US & Corporate\\
\hline
Amazon & 0.94 & 0.08 & 52 & US & Corporate\\
\hline
OpenAI & 1.59 & 0.12 & 54 & US & Corporate\\
\hline
DeepMind & 1.37 & 0.13 & 44 & US & Corporate\\
\hline
Tencent & 0.85 & 0.11 & 49 & US & Corporate\\
\hline
Baidu & 0.50 & 0.08 & 62 & US & Corporate\\
\hline
Alibaba & 0.67 & 0.10 & 61 & US & Corporate\\
\hline
Non-government scientific organization (e.g., AAAI) & 2.09 & 0.10 & 50 & US & Other\\
\hline
Partnership on AI & 1.86 & 0.09 & 51 & US & Other\\
\hline
\end{tabular}
\end{table}

\clearpage

\begin{table}  
\small
  \caption{Association between trust and all of the different actors. The table presents the output from the multiple linear regression used to compare differences in rated trust between all actors. We regressed trust on all actors. The (arbitrarily-chosen) reference category, the one that is excluded from the list of coefficients, is Alibaba. The \textit{F}-test of overall significance rejects the null hypothesis that trust does not differ between the actors. The Holm method was used to control the family-wise error rate.} 
  
  \label{tab:reg-trust-all-actors} 
  \renewcommand{\arraystretch}{1}
  \centering 
\begin{tabular}{@{\extracolsep{3em}}ll} 
\\[-1.8ex]\hline & \hspace{-2em} Coefficient (\textit{SE}) \\ 
\hline \\[-1.8ex] 
 (Intercept) & 0.566$^{***}$ \\ 
  & (0.054) \\ 
  Amazon & 0.316$^{***}$ \\ 
  & (0.078) \\ 
  Apple & 0.484$^{***}$ \\ 
  & (0.078) \\ 
  Baidu & $-$0.061 \\ 
  & (0.065) \\ 
  Chinese government & $-$0.188 \\ 
  & (0.075) \\ 
  Chinese military & $-$0.182 \\ 
  & (0.102) \\ 
  DeepMind & 0.805$^{***}$ \\ 
  & (0.102) \\ 
  EU & 1.416$^{***}$ \\ 
  & (0.084) \\ 
  Facebook & 0.017 \\ 
  & (0.076) \\ 
  Google & 0.788$^{***}$ \\ 
  & (0.086) \\ 
  Government of the country where they do research & 0.761$^{***}$ \\ 
  & (0.100) \\ 
  Intergovernmental research organizations (e.g., CERN) & 1.515$^{***}$ \\ 
  & (0.084) \\ 
  Microsoft & 0.868$^{***}$ \\ 
  & (0.087) \\ 
  Military of the country where they do research & 0.266$^{*}$ \\ 
  & (0.092) \\ 
  Non-government scientific organization (e.g., AAAI) & 1.551$^{***}$ \\ 
  & (0.078) \\ 
  OpenAI & 0.934$^{***}$ \\ 
  & (0.088) \\ 
  Partnership on AI & 1.329$^{***}$ \\ 
  & (0.082) \\ 
  Tencent & 0.123 \\ 
  & (0.064) \\ 
  U.S. government & 0.372$^{***}$ \\ 
  & (0.086) \\ 
  U.S. military & 0.119 \\ 
  & (0.105) \\ 
  UN & 1.171$^{***}$ \\ 
  & (0.092) \\ 
  \hline
 \textit{N} &\hspace{-3em} 2,288 responses, 434 unique respondents \\ 
\textit{F}-Statistic & \hspace{-3em} 70.07$^{***}$ (\textit{df} = 20; 433) \\ 
\hline \\[-1.8ex] 
\multicolumn{2}{l}{$^{*}$p $<$ .05; $^{**}$p $<$ .01; $^{***}$p $<$ .001} \\ 
\end{tabular} 
\end{table} 

\clearpage

\begin{table}
\small
  \caption{Association between trust in actors and demographic variables controlling for actors. The table presents the output from the multiple linear regression used to compare differences in trust in actors between demographic subgroups whilst controlling for the different individual actors. We regressed trust in actors on the categorical variables of gender (female/other, male, and prefer not to say or missing response), location of undergraduate education (Europe, the US, China, other, and missing response), location of work/study (Europe, the US, China, and other), and type of workplace (industry, academic, and other). The (arbitrarily-chosen) reference categories, the ones that are excluded from the list of coefficients, are female/other for gender, other for workplace type, and China for location of undergraduate education and place of work/study. The \textit{F}-test of overall significance rejects the null hypothesis that respondents do not differ in their trust of actors (when controlling for individual actors) between demographic subgroups. The Holm method was used to control the family-wise error rate.} 
  
  \label{tab:corr-trust-demo} 
  \renewcommand{\arraystretch}{1.25}
  \centering
\begin{tabular}{@{\extracolsep{3em}}ll} 
\\[-1.8ex]\hline & \hspace{-2em} Coefficient (\textit{SE}) \\ 
\hline \\[-1.8ex] 
 (Intercept) & 0.765$^{***}$ \\ 
  & (0.155) \\ 
  Gender: male & 0.122 \\ 
  & (0.081) \\ 
  Gender: prefer not to say/NA & 0.033 \\ 
  & (0.113) \\ 
  Place of undergraduate degree: Europe & $-$0.321$^{*}$ \\ 
  & (0.095) \\ 
  Place of undergraduate degree: missing & $-$0.342$^{*}$ \\ 
  & (0.103) \\ 
  Place of undergraduate degree: other & $-$0.402$^{***}$ \\ 
  & (0.087) \\ 
  Place of undergraduate degree: US & $-$0.359$^{***}$ \\ 
  & (0.083) \\ 
  Place of work: Europe & $-$0.046 \\ 
  & (0.134) \\ 
  Place of work: other & 0.054 \\ 
  & (0.137) \\ 
  Place of work: US & 0.005 \\ 
  & (0.125) \\ 
  Job: industry & $-$0.047 \\ 
  & (0.076) \\ 
  Job: academic & 0.023 \\ 
  & (0.079) \\ 
  \hline
 \textit{N} & \hspace{-3em}2,288 responses, 434 unique respondents \\ 
\textit{F}-Statistic & \hspace{-3em}55.41$^{***}$ (\textit{df} = 31; 433) \\ 
\hline \\[-1.8ex] 
\multicolumn{2}{l}{$^{*}$p $<$ .05; $^{**}$p $<$ .01; $^{***}$p $<$ .001} \\ 
\end{tabular} 
\end{table} 

\clearpage

\begin{table}[ht] \centering 
\small
  \caption{Interaction between country of undergraduate degree and trust of Western versus Chinese tech companies. For this regression analysis, we focus only on respondents who received their undergraduate degrees in the US or China and trust in tech companies. We regress trust on whether the tech company is Western or Chinese, the country of the respondent's undergraduate degree, and the interaction between the two. Google, Facebook, Apple, Microsoft, Amazon, OpenAI, and DeepMind are coded as Western tech companies. Tencent, Baidu, and Alibaba are coded as Chinese tech companies. The arbitrary reference group for tech company type is Chinese tech companies; the arbitrary reference group for country of undergraduate degree is the US. We cluster standard errors by respondent because each respondent evaluated multiple tech companies. The Holm method was used to control the family-wise error rate.} 
  \label{tab:interaction-tech-companies-us-china} 
  \renewcommand{\arraystretch}{1.15}
\begin{tabular}{@{\extracolsep{3em}}p{9.5cm}p{6cm}} 
\\[-1.8ex]\hline & \hspace{-2em}Coefficient (\textit{SE}) \\ 
\hline \\[-1.8ex] 
 (Intercept) & 0.457$^{***}$ \\ 
  & (0.061) \\ 
  Western tech companies & 0.707$^{***}$ \\ 
  & (0.075) \\ 
  Place of undergraduate degree: China & 0.831$^{***}$ \\ 
  & (0.148) \\ 
  Western tech companies: Place of undergraduate degree: China & $-$0.437$^{**}$ \\ 
  & (0.154) \\ 
  \hline
 \textit{N} &\hspace{-3em} 440 responses; 159 unique respondents \\ 
\textit{F}-Statistic &\hspace{-3em} 47.00$^{***}$ (\textit{df} = 3; 158) \\ 
\hline \\[-1.8ex] 
\multicolumn{2}{l}{$^{*}$p $<$ .05; $^{**}$p $<$ .01; $^{***}$p $<$ .001} \\ 
\end{tabular} 
\end{table} 
\clearpage 

\subsubsection*{AI safety}
\medskip

\begin{table}[ht]
\small
\caption{\label{tab:ai-safety-familiarity}Familiarity with AI safety. The table presents the raw frequency and proportion of respondents who indicated different levels of familiarity with the issue of AI safety. The standard errors of the proportions are also presented. For reference, after reading a definition of AI safety (see survey text for the definition), respondents input their familiarity with AI safety using a 5-point slider (0 = not familiar at all; 4 = very familiar).}
\centering
\renewcommand{\arraystretch}{1.2}
\begin{tabular}[t]{|l|l|l|l|}
\hline
AI safety familiarity level & Proportion & \textit{SE} & Frequency\\
\hline
Missing & $<0.001$ & 0.00 & 1\\
\hline
0 - Not familiar at all & 0.03 & 0.01 & 8\\
\hline
1 & 0.25 & 0.03 & 71\\
\hline
2 & 0.32 & 0.03 & 90\\
\hline
3 & 0.25 & 0.03 & 72\\
\hline
4 - Very familiar & 0.15 & 0.02 & 42\\
\hline
\end{tabular}
\end{table}

\medskip
\begin{table}[ht]
\small
\caption{\label{tab:ai-safety-familiarity-demo}Familiarity with AI safety (mean response by demographic subgroups). The table presents the AI/ML researchers’ mean familiarity with AI safety and the associated standard error and sample size, by demographic subgroup. For reference, after reading a definition of AI safety (see survey text for the definition), respondents input their familiarity with AI safety using a 5-point slider (0 = not familiar at all; 4 = very familiar).}
\centering
\renewcommand{\arraystretch}{1.2}
\begin{tabular}[t]{|l|l|l|l|l|}
\hline
Demographic subgroup & Demographic subgroup type & Mean & \textit{SE} & \textit{N}\\
\hline
US & Undergraduate country & 2.48 & 0.15 & 56\\
\hline
China & Undergraduate country & 1.98 & 0.17 & 42\\
\hline
Europe & Undergraduate region & 2.12 & 0.12 & 69\\
\hline
North America & Undergraduate region & 2.45 & 0.14 & 64\\
\hline
Asia & Undergraduate region & 2.16 & 0.11 & 98\\
\hline
Academic & Workplace type & 2.22 & 0.07 & 217\\
\hline
Industry & Workplace type & 2.31 & 0.11 & 77\\
\hline
\end{tabular}
\end{table}

\medskip

\begin{table}[ht]
\small
\caption{\label{tab:ai-safety-prioritized}How much should AI safety be prioritized? Respondents were asked how much AI safety research should be prioritized relative to today. The answer choices are a Likert scale from -2 to 2: -2 = much less; -1 = less; 0 = about the same; 1 = more; 2 = much more. There was also an “I don’t know” option. We present the proportion of respondents who chose each option or have a missing response, along with the associated standard error and raw frequency.}
\centering
\renewcommand{\arraystretch}{1.2}
\begin{tabular}[t]{|l|l|l|l|}
\hline
AI safety prioritization & Proportion & \textit{SE} & Frequency\\
\hline
-2: Much less & 0.01 & 0.01 & 4\\
\hline
-1: Less & 0.04 & 0.01 & 11\\
\hline
0: About the same & 0.24 & 0.03 & 68\\
\hline
1: More & 0.38 & 0.03 & 109\\
\hline
2: Much more & 0.30 & 0.03 & 84\\
\hline
Missing & 0.00 & 0.00 & 0\\
\hline
I don't know & 0.03 & 0.01 & 8\\
\hline
\end{tabular}
\end{table}

\clearpage

\begin{table}
\small
  \caption{Correlation between respondents' familiarity with AI safety and how much they think AI safety research should be prioritized. For both models, the outcome variable is how much the respondents think AI safety research should be prioritized. Model 1 looks at the bivariate relationship between these two variables. Model 2 includes demographic variables as controls, including gender (female/other, male, and prefer not to say or missing response), location of undergraduate education (Europe, US, Asia, other, and missing response), location of work (Europe, US, Asia, and other), and type of workplace (industry, academic, and other). The (arbitrarily-chosen) reference categories, the ones that are excluded from the list of coefficients, are female/other for gender, other for workplace type, and Asia for location of undergraduate education and place of work/study. The Holm method was used to control the family-wise error rate.} 
  \label{tab:regression-ai-safety} 
  \centering
\begin{tabular}{lcc}
\hline & \multicolumn{2}{c}{Coefficient (\textit{SE})} \\ 
\\[-1.8ex] & (1) & (2)\\ 
\hline \\[-1.8ex] 
 (Intercept) & 0.709$^{***}$ & 1.157 \\ 
  & (0.125) & (0.397) \\ 
  Familiarity with AI safety & 0.100 & 0.086 \\ 
  & (0.049) & (0.047) \\ 
  Gender: male &  & $-$0.484$^{*}$ \\ 
  &  & (0.156) \\ 
  Gender: prefer not to say/NA &  & $-$0.479 \\ 
  &  & (0.395) \\ 
  Place of undergraduate degree: Europe &  & 0.134 \\ 
  &  & (0.218) \\ 
  Place of undergraduate degree: missing &  & 0.206 \\ 
  &  & (0.224) \\ 
  Place of undergraduate degree: other &  & 0.421 \\ 
  &  & (0.205) \\ 
  Place of undergraduate degree: US &  & 0.086 \\ 
  &  & (0.203) \\ 
  Place of work: Europe &  & 0.004 \\ 
  &  & (0.340) \\ 
  Place of work: other &  & $-$0.229 \\ 
  &  & (0.374) \\ 
  Place of work: US &  & 0.066 \\ 
  &  & (0.335) \\ 
  Job: industry &  & $-$0.162 \\ 
  &  & (0.148) \\ 
  Job: academic &  & $-$0.154 \\ 
  &  & (0.165) \\ 
  \hline
 \textit{N} & 284 & 284 \\ 
\textit{F}-Statistic & 4.215 (\textit{df} = 1; 282) & 1.928 (\textit{df} = 12; 271) \\ 
\hline \\[-1.8ex] 
\multicolumn{3}{l}{$^{*}$p $<$ .05; $^{**}$p $<$ .01; $^{***}$p $<$ .001} \\ 
\end{tabular}
\end{table}

\clearpage

\subsubsection*{Attitudes toward military applications of AI}

\begin{table}[ht]
\small
\caption{\label{tab:military-distribution}Attitudes toward researchers working on military applications of AI: distribution of responses. Respondents were asked to indicate their level of support for two of the three randomly presented military applications (lethal autonomous weapons, surveillance, and logistics) on a five-point scale from $-2$ to 2. Each military application was defined when it was presented (see survey text for the definition). The table presents the proportion of respondents that indicated each level of response or answered ``I don't know'', along with the associated standard error and raw frequency, by military application type. The proportion of missing responses is also presented.}
\centering
\renewcommand{\arraystretch}{1.2}
\begin{tabular}[t]{|l|l|l|l|l|}
\hline
Military application type & Response & Proportion & \textit{SE} & Frequency\\
\hline
Lethal autonomous weapons & $-2$: Strongly oppose & 0.58 & 0.03 & 178\\
\hline
Lethal autonomous weapons & $-1$: Somewhat oppose & 0.16 & 0.02 & 48\\
\hline
Lethal autonomous weapons & 0: Neither support nor oppose & 0.14 & 0.02 & 44\\
\hline
Lethal autonomous weapons & 1: Somewhat support & 0.07 & 0.01 & 20\\
\hline
Lethal autonomous weapons & 2: Strongly support & 0.01 & 0.00 & 2\\
\hline
Lethal autonomous weapons & Missing & 0.01 & 0.01 & 3\\
\hline
Lethal autonomous weapons & I don’t know & 0.04 & 0.01 & 11\\
\hline
Surveillance & $-2$: Strongly oppose & 0.20 & 0.02 & 58\\
\hline
Surveillance & $-1$: Somewhat oppose & 0.21 & 0.02 & 60\\
\hline
Surveillance & 0: Neither support nor oppose & 0.26 & 0.03 & 75\\
\hline
Surveillance & 1: Somewhat support & 0.22 & 0.02 & 63\\
\hline
Surveillance & 2: Strongly support & 0.07 & 0.01 & 19\\
\hline
Surveillance & Missing & 0.01 & 0.01 & 3\\
\hline
Surveillance & I don’t know & 0.04 & 0.01 & 11\\
\hline
Logistics & $-2$: Strongly oppose & 0.06 & 0.01 & 17\\
\hline
Logistics & $-1$: Somewhat oppose & 0.08 & 0.02 & 23\\
\hline
Logistics & 0: Neither support nor oppose & 0.36 & 0.03 & 102\\
\hline
Logistics & 1: Somewhat support & 0.24 & 0.03 & 68\\
\hline
Logistics & 2: Strongly support & 0.20 & 0.02 & 55\\
\hline
Logistics & Missing & 0.00 & 0.00 & 1\\
\hline
Logistics & I don’t know & 0.05 & 0.01 & 15\\
\hline
\end{tabular}
\end{table}

\begin{table}
\small
\caption{\label{tab:military-mean-subgroup}Attitudes toward researchers working on military applications of AI (mean response by demographic subgroup). Respondents were asked to indicate their level of support for two of the three randomly presented military applications (lethal autonomous weapons, surveillance, and logistics) on a five-point scale from $-2$ to 2: $-2$ = strongly oppose, $-1$ = somewhat oppose, 0 = neither support nor oppose, 1 = somewhat support, 2 = strongly support. There was also an ``I don't know'' option. Each military application was defined when it was presented (see survey text for the definition). The table presents the proportion of respondents that indicated each level of response or answered ``I don't know'', along with the associated standard error and raw frequency, by military application and demographic subgroup. The proportion of missing responses is also presented.}
\centering
\renewcommand{\arraystretch}{1.1}
\begin{tabular}[t]{|l|l|l|l|p{2.5cm}|p{2.5cm}|}
\hline
Military application type & Mean & \textit{SE} & \textit{N} & Demographic subgroup & Demographic subgroup type\\
\hline
Lethal autonomous weapons & $-1.30$ & 0.06 & 306 & All respondents & All respondents\\
\hline
Surveillance & $-0.27$ & 0.07 & 289 & All respondents & All respondents\\
\hline
Logistics & 0.46 & 0.06 & 281 & All respondents & All respondents\\
\hline
Lethal autonomous weapons & $-1.24$ & 0.13 & 65 & US & Undergraduate country\\
\hline
Surveillance & $-0.34$ & 0.14 & 65 & US & Undergraduate country\\
\hline
Logistics & 0.77 & 0.13 & 68 & US & Undergraduate country\\
\hline
Lethal autonomous weapons & $-0.89$ & 0.19 & 41 & China & Undergraduate country\\
\hline
Surveillance & 0.21 & 0.20 & 43 & China & Undergraduate country\\
\hline
Logistics & 0.38 & 0.13 & 40 & China & Undergraduate country\\
\hline
Lethal autonomous weapons & $-1.46$ & 0.09 & 83 & Europe & Undergraduate region\\
\hline
Surveillance & $-0.52$ & 0.14 & 67 & Europe & Undergraduate region\\
\hline
Logistics & 0.22 & 0.15 & 62 & Europe & Undergraduate region\\
\hline
Lethal autonomous weapons & $-1.30$ & 0.12 & 71 & North America & Undergraduate region\\
\hline
Surveillance & $-0.41$ & 0.13 & 73 & North America & Undergraduate region\\
\hline
Logistics & 0.71 & 0.12 & 78 & North America & Undergraduate region\\
\hline
Lethal autonomous weapons & $-1.13$ & 0.11 & 101 & Asia & Undergraduate region\\
\hline
Surveillance & $-0.03$ & 0.13 & 101 & Asia & Undergraduate region\\
\hline
Logistics & 0.53 & 0.09 & 92 & Asia & Undergraduate region\\
\hline
Lethal autonomous weapons & $-1.31$ & 0.06 & 234 & Academic & Workplace type\\
\hline
Surveillance & $-0.19$ & 0.08 & 216 & Academic & Workplace type\\
\hline
Logistics & 0.48 & 0.07 & 204 & Academic & Workplace type\\
\hline
Lethal autonomous weapons & $-1.29$ & 0.11 & 82 & Industry & Workplace type\\
\hline
Surveillance & $-0.54$ & 0.14 & 77 & Industry & Workplace type\\
\hline
Logistics & 0.47 & 0.12 & 81 & Industry & Workplace type\\
\hline
\end{tabular}
\end{table}

\clearpage

{\small
\renewcommand{\arraystretch}{1.25}
\begin{longtable}{|p{1.9cm}|p{3.3cm}|p{2.2cm}|p{2.25cm}|p{2cm}|p{2cm}|}
\caption{\label{tab:collective-distribution}Support for collective action against research into military applications of AI: distribution of responses. The table presents the proportion and number of respondents who oppose others working on the different military applications of AI (lethal autonomous weapons, logistics, and surveillance) and who were asked about each application in the survey, broken down by type of collective action}\\\hline
\begin{tabular}[t]{c} \\[-6pt] Military \\ application \\ type\end{tabular} & \begin{tabular}[t]{c} \\ Collective action \\ \end{tabular} & Proportion of respondents who oppose others working on the application & Proportion of all respondents who were asked about the application & Number of respondents who oppose others working on the application & Number of respondents asked about the application\\
\hline
\endfirsthead
\multicolumn{6}{c}%
{\tablename\ \thetable\ -- \textit{Continued from previous page}} \\
\hline
\begin{tabular}[t]{c} \\[-6pt] Military \\ application \\ type\end{tabular} & \begin{tabular}[t]{c} \\ Collective action \\ \end{tabular} & Proportion of respondents who oppose others working on the application & Proportion of all respondents who were asked about the application & Number of respondents who oppose others working on the application & Number of respondents asked about the application\\
\hline
\endhead
\hline \multicolumn{6}{r}{\textit{Continued on next page}} \\
\endfoot
\hline
\endlastfoot
\hline
Lethal \mbox{autonomous} weapons & Nothing & 0.01 & --- & 225 & 306\\
\hline
Lethal \mbox{autonomous} weapons & Actively avoid working on the project & 0.75 & 0.55 & 225 & 306\\
\hline
Lethal \mbox{autonomous} weapons & Expressing your concern to a superior in your organization involved in the decision & 0.69 & 0.51 & 225 & 306\\
\hline
Lethal \mbox{autonomous} weapons & Sign a petition against the decision & 0.61 & 0.45 & 225 & 306\\
\hline
Lethal \mbox{autonomous} weapons & Participate in a public protest & 0.40 & 0.29 & 225 & 306\\
\hline
Lethal \mbox{autonomous} weapons & Speak out against the decision anonymously to the media or online & 0.34 & 0.25 & 225 & 306\\
\hline
Lethal \mbox{autonomous} weapons & Speak out against the decision publicly to the media or online & 0.34 & 0.25 & 225 & 306\\
\hline
Lethal \mbox{autonomous} weapons & Resign or threaten to resign from your job & 0.42 & 0.31 & 225 & 306\\
\hline
Logistics & Nothing & 0.07 & --- & 40 & 281\\
\hline
Logistics & Actively avoid working on the project & 0.72 & 0.10 & 40 & 281\\
\hline
Logistics & Expressing your concern to a superior in your organization involved in the decision & 0.62 & 0.09 & 40 & 281\\
\hline
Logistics & Sign a petition against the decision & 0.52 & 0.07 & 40 & 281\\
\hline
Logistics & Participate in a public protest & 0.22 & 0.03 & 40 & 281\\
\hline
Logistics & Speak out against the decision anonymously to the media or online & 0.25 & 0.04 & 40 & 281\\
\hline
Logistics & Speak out against the decision publicly to the media or online & 0.22 & 0.03 & 40 & 281\\
\hline
Logistics & Resign or threaten to resign from your job & 0.18 & 0.02 & 40 & 281\\
\hline
Surveillance & Nothing & 0.03 & --- & 115 & 289\\
\hline
Surveillance & Actively avoid working on the project & 0.74 & 0.29 & 115 & 289\\
\hline
Surveillance & Expressing your concern to a superior in your organization involved in the decision & 0.72 & 0.29 & 115 & 289\\
\hline
Surveillance & Sign a petition against the decision & 0.62 & 0.25 & 115 & 289\\
\hline
Surveillance & Participate in a public protest & 0.34 & 0.13 & 115 & 289\\
\hline
Surveillance & Speak out against the decision anonymously to the media or online & 0.33 & 0.13 & 115 & 289\\
\hline
Surveillance & Speak out against the decision publicly to the media or online & 0.36 & 0.14 & 115 & 289\\
\hline
Surveillance & Resign or threaten to resign from your job & 0.27 & 0.11 & 115 & 289\\
\hline
\end{longtable}
}
\clearpage

\begin{table}[ht]
\small
\caption{\label{tab:maven-distribution}Attitudes toward Google pulling out of Project Maven. Respondents were presented with a short description of the employees' reactions to Google's Project Maven and the following non-renewal of the contract (see survey text for the description) and were asked to indicate their support for the non-renewal decision on a five-point scale from $-2$ to 2. There was also an ``I don't know'' option. The table presents the proportion of respondents choosing each option and who did not respond to the question, along with the associated standard error and raw frequency.}
\centering
\renewcommand{\arraystretch}{1.2}
\begin{tabular}[t]{|l|l|l|l|}
\hline
Response & Proportion & \textit{SE} & Frequency\\
\hline
$-2$: Strongly oppose & 0.03 & 0.01 & 15\\
\hline
$-1$: Somewhat oppose & 0.06 & 0.01 & 28\\
\hline
0: Neither support nor oppose & 0.23 & 0.02 & 102\\
\hline
1: Somewhat support & 0.21 & 0.02 & 96\\
\hline
2: Strongly support & 0.38 & 0.02 & 171\\
\hline
Missing & 0.05 & 0.01 & 23\\
\hline
I don’t know & 0.04 & 0.01 & 18\\
\hline
\end{tabular}
\end{table}

\clearpage

\begin{table}
\small
  \caption{Correlation between support for researchers working on lethal autonomous weapons attitude and support for Google pulling out of Project Maven. In both models, the outcome variable is support for Google pulling out of Project Maven. Model 1 looks at the bivariate relationship between these two variables. Model 2 includes demographic variables as controls, including gender (female/other, male, and prefer not to say or missing response), location of undergraduate education (Europe, US, Asia, other, and missing response), location of work (Europe, US, Asia, and other), and type of workplace (industry, academic, and other). The (arbitrarily-chosen) reference categories, the ones that are excluded from the list of coefficients, are female/other for gender, other for workplace type, and Asia for location of undergraduate education and place of work/study. The Holm method was used to control the family-wise error rate.} 
  \label{tab:corr-maven-laws-regression} 
  \renewcommand{\arraystretch}{1.2}
  \centering
\begin{tabular}{p{8cm}cc}
\hline & \multicolumn{2}{c}{Coefficient (\textit{SE})} \\ 
\\[-1.8ex] & (1) & (2)\\ 
\hline \\[-1.8ex] 
 (Intercept) & 0.388$^{**}$ & 0.324 \\ 
  & (0.108) & (0.393) \\ 
  Support for researchers working on lethal autonomous weapons & $-$0.469$^{***}$ & $-$0.441$^{***}$ \\ 
  & (0.067) & (0.070) \\ 
  Gender: male &  & $-$0.030 \\ 
  &  & (0.185) \\ 
  Gender: prefer not to say/NA &  & $-$0.428 \\ 
  &  & (0.472) \\ 
  Place of undergraduate degree: Europe &  & $-$0.129 \\ 
  &  & (0.242) \\ 
  Place of undergraduate degree: missing &  & $-$0.221 \\ 
  &  & (0.277) \\ 
  Place of undergraduate degree: other &  & 0.246 \\ 
  &  & (0.241) \\ 
  Place of undergraduate degree: US &  & 0.237 \\ 
  &  & (0.224) \\ 
  Place of work: Europe &  & 0.736 \\ 
  &  & (0.451) \\ 
  Place of work: missing &  & 0.393 \\ 
  &  & (0.647) \\ 
  Place of work: other &  & $-$0.092 \\ 
  &  & (0.480) \\ 
  Place of work: US &  & 0.090 \\ 
  &  & (0.429) \\ 
  Job: industry &  & $-$0.209 \\ 
  &  & (0.182) \\ 
  Job: academic &  & $-$0.186 \\ 
  &  & (0.179) \\ 
  \hline
\textit{N} & 306 & 306 \\ 
\textit{F}-Statistic & 49.41$^{***}$ (\textit{df} = 1; 304) & 10.36$^{***}$ (\textit{df} = 13; 292) \\ 
\hline \\[-1.8ex] 
\multicolumn{3}{l}{$^{*}$p $<$ .05; $^{**}$p $<$ .01; $^{***}$p $<$ .001} \\ 
\end{tabular} 
\end{table} 

\clearpage

\subsubsection*{Publication norms}

\begin{table}[ht]
\small
\caption{\label{tab:prepub_review}Responses to statement on pre-publication review. After seeing a definition of ``pre-publication review'' (see survey text for the definition), respondents were asked to indicate their level of agreement with the statement: ``Machine learning research institutions (including firms, governments, and universities) should practice pre-publication review.'' The respondents could choose responses from a four-point scale. There was also an ``I don't know'' option. The table presents the proportion of respondents who indicated each option or had a missing response, along with the associated standard error and raw frequency.}
\centering
\renewcommand{\arraystretch}{1.2}
\begin{tabular}[t]{|l|l|l|l|}
\hline
Response & Proportion & \textit{SE} & Frequency\\
\hline
$-2$: Strongly disagree & 0.14 & 0.02 & 54\\
\hline
$-1$: Somewhat disagree & 0.19 & 0.02 & 72\\
\hline
1: Somewhat agree & 0.39 & 0.03 & 147\\
\hline
2: Strongly agree & 0.20 & 0.02 & 76\\
\hline
Missing & 0.00 & 0.00 & 0\\
\hline
I don't know & 0.06 & 0.01 & 24\\
\hline
\end{tabular}
\end{table}

\clearpage

{\renewcommand{\arraystretch}{1.3}\small
\begin{longtable}[ht]{|m{5cm}|c|c|c|m{2.6cm}|m{2.6cm}|}
\caption{\label{tab:sharingaspectsresearch} Sharing aspects of research (by demographic subgroup). Respondents were presented with three aspects of research randomly chosen from a list of six. For each aspect of research, they selected from six levels of openness (0 = it doesn’t matter; 1 = it’s completely up to the researchers to share or not to share; 2 = it’s preferred that researchers share but it’s not paramount that they do; 3 = researchers are encouraged to share; 3 = researchers need a very strong reason not to share; 4 = it must be shared every time). The table presents the mean response for each aspect of research by demographic subgroup, along with the associated standard error and sample size.}\\\hline
Aspect of research & Mean & \textit{SE} & \textit{N} & Demographic subgroup & Demographic subgroup type\\
\hline
\endfirsthead
\multicolumn{6}{c}%
{\tablename\ \thetable\ -- \textit{Continued from previous page}} \\
\hline
Aspect of research & Mean & \textit{SE} & \textit{N} & Demographic subgroup & Demographic subgroup type\\
\hline
\endhead
\hline \multicolumn{6}{r}{\textit{Continued on next page}} \\
\endfoot
\hline
\endlastfoot
\hline
High-level description of methods & 4.81 & 0.04 & 172 & All respondents & All\\
\hline
Detailed description of methods & 4.52 & 0.06 & 161 & All respondents & All\\
\hline
Results & 4.63 & 0.06 & 174 & All respondents & All\\
\hline
Code & 3.74 & 0.08 & 143 & All respondents & All\\
\hline
Training data & 3.54 & 0.08 & 173 & All respondents & All\\
\hline
Trained model & 3.46 & 0.10 & 143 & All respondents & All\\
\hline
Algorithm(s) & 4.30 & 0.07 & 153 & All respondents & All\\
\hline
High-level description of methods & 4.80 & 0.07 & 42 & US & Undergraduate country\\
\hline
Detailed description of methods & 4.46 & 0.12 & 42 & US & Undergraduate country\\
\hline
Results & 4.52 & 0.13 & 38 & US & Undergraduate country\\
\hline
Code & 3.40 & 0.15 & 35 & US & Undergraduate country\\
\hline
Training data & 3.30 & 0.15 & 46 & US & Undergraduate country\\
\hline
Trained model & 3.25 & 0.18 & 30 & US & Undergraduate country\\
\hline
Algorithm(s) & 4.37 & 0.15 & 31 & US & Undergraduate country\\
\hline
High-level description of methods & 4.79 & 0.12 & 19 & China & Undergraduate country\\
\hline
Detailed description of methods & 4.38 & 0.19 & 21 & China & Undergraduate country\\
\hline
Results & 4.62 & 0.13 & 21 & China & Undergraduate country\\
\hline
Code & 3.72 & 0.21 & 25 & China & Undergraduate country\\
\hline
Training data & 3.36 & 0.25 & 25 & China & Undergraduate country\\
\hline
Trained model & 3.67 & 0.29 & 18 & China & Undergraduate country\\
\hline
Algorithm(s) & 4.08 & 0.22 & 24 & China & Undergraduate country\\
\hline
High-level description of methods & 4.86 & 0.06 & 42 & Europe & Undergraduate region\\
\hline
Detailed description of methods & 4.62 & 0.12 & 37 & Europe & Undergraduate region\\
\hline
Results & 4.85 & 0.06 & 53 & Europe & Undergraduate region\\
\hline
Code & 3.94 & 0.17 & 33 & Europe & Undergraduate region\\
\hline
Training data & 3.67 & 0.17 & 33 & Europe & Undergraduate region\\
\hline
Trained model & 3.29 & 0.20 & 41 & Europe & Undergraduate region\\
\hline
Algorithm(s) & 4.38 & 0.15 & 34 & Europe & Undergraduate region\\
\hline
High-level description of methods & 4.81 & 0.07 & 44 & North America & Undergraduate region\\
\hline
Detailed description of methods & 4.52 & 0.11 & 47 & North America & Undergraduate region\\
\hline
Results & 4.58 & 0.11 & 44 & North America & Undergraduate region\\
\hline
Code & 3.40 & 0.16 & 40 & North America & Undergraduate region\\
\hline
Training data & 3.45 & 0.14 & 53 & North America & Undergraduate region\\
\hline
Trained model & 3.16 & 0.18 & 34 & North America & Undergraduate region\\
\hline
Algorithm(s) & 4.39 & 0.14 & 35 & North America & Undergraduate region\\
\hline
High-level description of methods & 4.78 & 0.07 & 55 & Asia & Undergraduate region\\
\hline
Detailed description of methods & 4.45 & 0.11 & 51 & Asia & Undergraduate region\\
\hline
Results & 4.59 & 0.11 & 51 & Asia & Undergraduate region\\
\hline
Code & 3.92 & 0.15 & 50 & Asia & Undergraduate region\\
\hline
Training data & 3.61 & 0.15 & 57 & Asia & Undergraduate region\\
\hline
Trained model & 4.04 & 0.15 & 45 & Asia & Undergraduate region\\
\hline
Algorithm(s) & 4.25 & 0.13 & 51 & Asia & Undergraduate region\\
\hline
High-level description of methods & 4.81 & 0.04 & 135 & Academic & Workplace type\\
\hline
Detailed description of methods & 4.47 & 0.08 & 115 & Academic & Workplace type\\
\hline
Results & 4.63 & 0.07 & 129 & Academic & Workplace type\\
\hline
Code & 3.77 & 0.10 & 103 & Academic & Workplace type\\
\hline
Training data & 3.60 & 0.09 & 131 & Academic & Workplace type\\
\hline
Trained model & 3.54 & 0.12 & 107 & Academic & Workplace type\\
\hline
Algorithm(s) & 4.35 & 0.08 & 111 & Academic & Workplace type\\
\hline
High-level description of methods & 4.80 & 0.08 & 44 & Industry & Workplace type\\
\hline
Detailed description of methods & 4.53 & 0.11 & 51 & Industry & Workplace type\\
\hline
Results & 4.65 & 0.11 & 48 & Industry & Workplace type\\
\hline
Code & 3.61 & 0.14 & 46 & Industry & Workplace type\\
\hline
Training data & 3.28 & 0.14 & 46 & Industry & Workplace type\\
\hline
Trained model & 3.05 & 0.20 & 42 & Industry & Workplace type\\
\hline
Algorithm(s) & 4.06 & 0.16 & 47 & Industry & Workplace type\\
\hline
\end{longtable}}

\begin{table}[ht] \centering 
\small
  \caption{Correlation between responses to the AI safety questions and support for pre-publication review. For all three models, the outcome variable is support for pre-publication review. Model 1 shows the bivariate relationship between familiarity with AI safety and support for pre-publication review. Model 2 shows the bivariate relationship between how much respondents thought AI safety research should be prioritized and support for pre-publication review. Model 3 includes responses to both AI safety questions as predictor variables. The Holm method was used to control the family-wise error rate.} 
  \label{tab:ai-safety-prepub-review} \renewcommand{\arraystretch}{1.2}
\begin{tabular}{@{\extracolsep{5pt}}p{6cm}p{1.8cm}p{1.8cm}p{1.8cm}} 
\\[-1.8ex]\hline & \multicolumn{3}{c}{Coefficient (\textit{SE})} \\ 
\\[-1.8ex] & (1) & (2) & (3)\\ 
\hline \\[-1.8ex] 
 (Intercept) & 0.403$^{***}$ & 0.397$^{***}$ & 0.396$^{***}$ \\ 
  & (0.083) & (0.083) & (0.082) \\ 
  Familiarity with AI safety & 0.153$^{**}$ &  & 0.149$^{***}$ \\ 
  & (0.045) &  & (0.029) \\ 
  How much AI safety must be proritized &  & 0.256$^{**}$ & 0.253$^{**}$ \\ 
  &  & (0.086) & (0.086) \\ 
  \hline
 \textit{N} & 257 & 257 & 257 \\ 
\textit{F}-Statistic & 11.29$^{**}$ \quad (\textit{df} = 1; 255) & 8.831$^{**}$ \quad (\textit{df} = 1; 255) & 18.48$^{***}$ \quad (\textit{df} = 2; 254) \\ 
\hline \\[-2.2ex] 
\multicolumn{4}{l}{$^{*}$p $<$ .05; $^{**}$p $<$ .01; $^{***}$p $<$ .001} \\ 
\end{tabular} 
\end{table} 

\clearpage

\begin{table}[ht] \centering 
\small
  \caption{Correlation between responses to the AI safety questions and how openly respondents think aspects of research should be openly shared. For all three models, the outcome variable is mean level of openness averaged across the three aspects of research respondents were randomly presented with. Model 1 shows the bivariate relationship between familiarity with AI safety and how openly respondents think aspects of research should be openly shared. Model 2 shows the bivariate relationship between how much respondents thought AI safety research should be prioritized and how openly respondents think aspects of research should be openly shared. Model 3 includes responses to both AI safety questions as predictor variables. The Holm method was used to control the family-wise error rate.
  } 
  \label{tab:ai-safety-sharing} 
\begin{tabular}{@{\extracolsep{5pt}}p{6cm}p{1.95cm}p{1.95cm}p{1.95cm}} 
\\[-1.8ex]\hline & \multicolumn{3}{c}{Coefficient (\textit{SE})} \\ 
\\[-1.8ex] & (1) & (2) & (3)\\ 
\hline \\[-1.8ex] 
 (Intercept) & 4.188$^{***}$ & 4.187$^{***}$ & 4.187$^{***}$ \\ 
  & (0.043) & (0.043) & (0.043) \\ 
  Familiarity with AI safety & $-$0.044 &  & $-$0.044 \\ 
  & (0.030) &  & (0.029) \\ 
  How much AI safety must be prioritized &  & 0.021 & 0.022 \\ 
  &  & (0.041) & (0.041) \\ 
  \hline
 \textit{N} & 256 & 256 & 256 \\ 
\textit{F}-Statistic & 2.101 & 0.2721 & 1.328  \\ 
& (\textit{df} = 1; 254) & (\textit{df} = 1; 254) & (\textit{df} = 1; 253) \\
\hline \\[-1.8ex] 
\multicolumn{4}{l}{$^{*}$p $<$ .05; $^{**}$p $<$ .01; $^{***}$p $<$ .001} \\ 
\end{tabular} 
\end{table} 

\end{document}